\documentclass[11pt]{article}

\pdfoutput=1

\usepackage{amssymb}
\usepackage{amsmath}
\usepackage{amstext}
\usepackage{graphicx,epsfig}
\usepackage{epsfig}
\usepackage{verbatim} 
\usepackage{fancybox}
\usepackage{color}
\usepackage{ulem}
\usepackage{enumitem}
\usepackage{subfigure}
\usepackage{bbm}
\usepackage{parskip}
\usepackage[numbers,sort&compress]{natbib}
\usepackage{tabularx}
\usepackage{makecell}

\newcommand*{\addheight}[2][.5ex]{%
  \raisebox{0pt}[\dimexpr\height+(#1)\relax]{#2}%
}

\newcommand{\Comment}[1]{{}}
\definecolor{darkblue}{rgb}{0.15,0.35,0.55}
\definecolor{reddish}{rgb}{0.65, 0.2, 0.2}
\usepackage[linktocpage=true]{hyperref}
\hypersetup{
colorlinks=true,
citecolor=darkblue,
linkcolor=reddish,
urlcolor=darkblue,
pdfauthor={},
pdftitle={},
pdfsubject={}
}

\newcommand{\be}{\begin{equation}}
\newcommand{\ee}{\end{equation}}
\newcommand{\bea}{\begin{eqnarray}}
\newcommand{\eea}{\end{eqnarray}}
\newcommand{\beas}{\begin{eqnarray*}}
\newcommand{\eeas}{\end{eqnarray*}}

\def\fr{\frac}
\def\mpl{M_{\rm Pl}}

\def\({\left(}
\def\){\right)}

\def\gsim{ \lower .75ex \hbox{$\sim$} \llap{\raise .27ex \hbox{$>$}} }
\def\lsim{ \lower .75ex \hbox{$\sim$} \llap{\raise .27ex \hbox{$<$}} }

\def\xyma{\xymatrix@M.7em}
\def\xymas{\xymatrix@M.1em}
\newcommand{\ba}{\begin{eqnarray}}
\newcommand{\ea}{\end{eqnarray}}

\title{}
\author{}

\numberwithin{equation}{section}

\begin{document}

\renewcommand{\thefootnote}{\fnsymbol{footnote}}
~
\vspace{1.75truecm}
\begin{center}
{\LARGE \bf{Emergence of the mass discrepancy-acceleration}}
\end{center} 
\vspace{.1cm}
\begin{center}
{\LARGE \bf{relation from dark matter-baryon interactions}}
\end{center} 
\vspace{.3cm}

\vspace{1truecm}
\thispagestyle{empty}
\centerline{{\Large Benoit Famaey,${}^{\rm a,}$\footnote{\href{mailto:benoit.famaey@astro.unistra.fr}{\texttt{benoit.famaey@astro.unistra.fr}}} Justin Khoury,${}^{\rm b,}$\footnote{\href{mailto:jkhoury@sas.upenn.edu}{\texttt{jkhoury@sas.upenn.edu}}} and Riccardo Penco${}^{\rm b,}$\footnote{\href{mailto:rpenco@sas.upenn.edu}{\texttt{rpenco@sas.upenn.edu}}}}}
\vspace{.5cm}
 
\centerline{{\it ${}^{\rm a}$Universit\'e de Strasbourg, CNRS, Observatoire astronomique de Strasbourg, }}
 \centerline{{\it  UMR 7550, F-67000 Strasbourg, France}} 
 \vspace{.25cm}

\centerline{\it $^{\rm b}$Center for Particle Cosmology, Department of Physics and Astronomy,}
\centerline{\it University of Pennsylvania, Philadelphia, PA 19104, USA}

\vspace{1.3cm}
\begin{abstract}
The observed tightness of the mass discrepancy-acceleration relation (MDAR) poses a fine-tuning challenge to current models of galaxy formation. We propose that this relation could arise from collisional interactions between baryons and dark matter (DM) particles, without the need for modification of gravity or ad hoc feedback processes. We assume that these interactions satisfy the following three conditions: $(i)$ the relaxation time of DM particles is comparable to the dynamical time in disk galaxies; $(ii)$ DM exchanges energy with baryons due to elastic collisions; $(iii)$ the product between the baryon-DM cross section and the typical energy exchanged in a collision is inversely proportional to the DM number density. As a proof of principle, we present an example of a particle physics model that gives a DM-baryon cross section with the desired density and velocity dependence.  For consistency with direct detection constraints, our DM particles must be either very light ($m \ll m_{\rm b}$) or very heavy ($m\gg m_{\rm b}$), corresponding respectively to heating and cooling of DM by baryons. In both cases, our mechanism applies and an equilibrium configuration can in principle be reached. In this exploratory paper, we focus on the heavy DM/cooling case because it is technically simpler, since the average energy exchanged turns out to be approximately constant throughout galaxies. Under these assumptions, we find that rotationally-supported disk galaxies could naturally settle to equilibrium configurations satisfying a MDAR at all radii without invoking finely tuned feedback processes. We also discuss issues related to the small scale clumpiness of baryons, as well as predictions for pressure-supported systems. We argue in particular that galaxy clusters do not follow the MDAR despite being DM-dominated because they have not reached their equilibrium configuration. Finally, we revisit existing phenomenological, astrophysical and cosmological constraints on baryon-DM interactions in light of the unusual density dependence of the cross section of DM particles.
\noindent
\end{abstract}

\newpage

\setcounter{tocdepth}{2}
\tableofcontents
\newpage
\renewcommand*{\thefootnote}{\arabic{footnote}}
\setcounter{footnote}{0}

\section{Introduction}
\label{intro}

The presence of mass discrepancies from the scales of dwarf galaxies to the largest scales in the Universe is nowadays backed by a plethora of astronomical data. After first hints for these mass discrepancies about 85 years ago ({\it e.g.},~\cite{oort1932force,Zwicky:1933gu}), the problem became manifest with the discovery of the asymptotic flatness of the rotation curves of disk galaxies~\cite{Rubin:1980zd,1981Bosma1,1981Bosma2}. Since then, numerous cosmological observations~\cite{Hinshaw:2012aka,Ade:2015xua,Abazajian:2008wr,Percival:2009xn} have confirmed the need for missing mass, dubbed dark matter (DM), on larger scales. In the present-day standard $\Lambda$CDM cosmological model, the missing mass is made of {\it cold dark matter} (CDM) particles, which are non-relativistic at decoupling and interact with each other and with baryons almost exclusively through gravity. In this context, the {\it fluid} of DM particles can be assumed to obey the {\it collisionless} Boltzmann equation, which is the fundamental equation of Galactic Dynamics for stars and DM particles. In the present contribution, we argue that, in order to explain certain (apparently) fine-tuned aspects of galaxy `rotation curves' phenomenology, it may be necessary to go beyond this collisionless approximation and consider collisional interactions between DM particles and baryons. Indeed, our viewpoint is that galaxy rotation curves provide us with important clues regarding the very nature of such interactions.

It is quite ironic that, despite being the first systems to have provided definitive evidence of mass discrepancies~\cite{Rubin:1980zd,1981Bosma1,1981Bosma2}, galaxies remain the most difficult systems to understand in the $\Lambda$CDM context~\cite{Bullock:2017}. This is in part due to the fact that, on galaxy scales, the physics of baryons can in principle play a major role in modifying the quasi-equilibrium configuration of CDM on secular time scales. Furthermore, despite the intrinsically stochastic nature of these putative feedback processes, baryons and DM in galaxies are observed to conspire in various {\it a priori} unexpected ways~\cite{Bullock:2017,vanAlbada1986,Persic:1995ru,Salucci:2007tm,Donato:2009ab,Gentile:2009,Famaey:2011kh}. The most well-known such conspiracy is the baryonic Tully-Fisher relation (BTFR)~\cite{McGaugh:2000sr,Papastergis:2016jqv,LelliBTFR}, relating the fourth power of the asymptotic circular velocity to the total baryonic mass of disk galaxies. Semi-empirical models with cored DM halos can reproduce the observed slope and normalization of such a relation, but not yet the small scatter~\cite{LelliBTFR,DiCintio:2015eeq,Desmond:2017}. 

An even more serious issue is that galaxies with the same asymptotic circular velocity---and hence the same total baryonic mass, according to the BTFR---display a very broad range of rotation curve {\it shapes} that is driven entirely by the broad range of surface density of the {\it baryons}~(see Fig.~15 of~\cite{Famaey:2011kh}). In other words, the rotation curve shapes of late-type spiral galaxies are all similar when expressed in units of disk scale length~\cite{Swaters:2009by}, and the DM cores actually correlate with scale-length~\cite{Donato:2004}. This presents another fine-tuning problem, perhaps the most severe one, since the {\it a priori} expectation would rather be that small, DM-dominated disk galaxies with similar asymptotic velocities display rotation curves with similar shapes given that they are embedded in similar DM halos~\cite{Oman:2015xda}. The observation that the central slope of the rotation curve correlates instead with the baryonic surface density means that feedback processes should be more efficient at removing DM from the central regions of galaxies with lower surface densities, which are often more gas-dominated. Hence, the feedback efficiency would have to increase with decreasing star formation rate~\cite{McGaugh:2011ac}, and be tightly anti-correlated with baryonic surface density (or correlated with scale-length at a given mass scale), independently of environment and accretion histories. This is arguably the most severe fine-tuning problem for galaxy formation models.

The BTFR, the variation in the shape of rotation curves with a given velocity scale, and their uniformity for a given baryonic surface density scale are all features that are captured by a single equation known as the {\it Mass Discrepancy-Acceleration Relation} (MDAR)~\cite{Sanders1990,McGaugh:2004aw}. This is a universal scaling relation between the total gravitational field $g$ and the Newtonian acceleration $g_{\rm b}$ generated by the observed distribution of baryons~\cite{McGaugh:2016leg}. The MDAR states that $g \simeq g_{\rm b}$ in the regime $g_{\rm b}\gg a_0$, and approaches the geometric mean $g \simeq \sqrt{a_0g_{\rm b}}$ whenever $g_{\rm b}\ll a_0$, where the characteristic acceleration is $a_0 \simeq 10^{-10}\,{\rm m}/{\rm s}^2$. The MDAR holds at any given radius within disk galaxies, and it implies the BTFR at large radii. The general properties of galaxy rotation curves as described by~\cite{Persic:1995ru,Salucci:2007tm} appear to be well in line with this scaling relation~\cite{Gentile:2008rv}, including the correlation between core size and scale-length~\cite{Donato:2004,Milgrom:2004ba}, and the relation can also be extended, with mixed success, to other types of galaxies~\cite{Lelli:2017vgz}. Some recent encouraging results have shown how the global shape of the MDAR might conceivably emerge in the context of $\Lambda$CDM~\cite{DiCintio:2015eeq,Keller:2017,Navarro:2017,Ludlow:2017,Read:2016}. However, the absence of residuals with radius and the observed small scatter, which could be solely driven by observational errors on the inclinations and distances of galaxies~\cite{McGaugh:2016leg,Lelli:2017vgz}, both remain unexplained~\cite{DiCintio:2015eeq,Desmond:2016azy}. 

Given the tight correlation between $g$ and $g_{\rm b}$, perhaps the simplest but most radical approach is to do away with DM, and assume that gravity is effectively modified in galaxies~\cite{Bekenstein:1984tv,Bekenstein:2004ne}. This paradigm, known as Modified Newtonian Dynamics (MOND), was first suggested more than 30 years ago by Milgrom~\cite{Milgrom:1983ca,Milgrom:1983pn} (see~\cite{Sanders:2002pf,Famaey:2011kh} for reviews). It can explain an impressive number of phenomena at galactic scales without DM, {\it e.g.}~\cite{Tiret:2007,Tiret:2008,Milgrom:2012,Milgrom:2013,Lughausen:2013,Thomas:2017}. In his pioneering work, Milgrom predicted the MDAR---which should therefore be called {\it Milgrom's relation}--- and all its rich phenomenology, well before being supported by precise observations. 

While exquisitely successful in rotationally-supported systems, the MOND paradigm as an {\it alternative to DM} seems increasingly unlikely, given the mounting evidence for DM behaving as a collisionless fluid
on cosmological and cluster scales. Indeed, the MOND law does not work on galaxy cluster scales~\cite{Gerbal:1992,Aguirre:2001fj,Sanders:2003,Clowe:2003tk,Clowe:2006eq,Pointecouteau:2005,Angus:2006qy,Angus:2007,Angus:2008}. Moreover, its relativistic extensions~\cite{Bekenstein:2004ne,Bruneton:2007si} cannot {\it a priori} explain the angular power spectrum of the Cosmic Microwave Background (CMB) without additional DM or unrealistically massive ordinary neutrinos~\cite{Skordis:2005xk}. Such theories are also highly constrained in the Solar System~\cite{Hees:2014kta,Hees:2015bna} (though see~\cite{Babichev:2011kq}) and have some problems at subgalactic scales~({\it e.g.,}~\cite{Ibata:2011ri}). 

A middle-ground interpretation of the MDAR is that it instructs us on the fundamental nature of DM, beyond the standard $\Lambda$CDM paradigm.
Minimal alterations to $\Lambda$CDM include tweaking the mass of DM particles~\cite{Bode:2001} or the strength of their self-interactions~\cite{Spergel:1999mh}. Interactions with photons~\cite{Schewtschenko:2015} or neutrinos~\cite{Boehm:2017} in the early Universe have also been considered. In the context of self-interacting DM, some recent encouraging results have shown how underdense halos could indeed be associated with extended baryonic disks~\cite{Kamada:2016,Creasey:2017}, in line with observations. However, getting a BTFR as tight as observed, as well as a MDAR at all radii remains challenging in this context. 
More radical approaches include Modified DM~\cite{Ho:2010ca,Ho:2011xc,Ho:2012ar,Edmonds:2017zhg} and Verlinde's emergent gravity~\cite{Verlinde:2016toy,Hees:2017}, both inspired by gravitational thermodynamics.

Another possibility is that the tight correlation between DM and ordinary matter embodied in the MDAR is the result of novel interactions between DM and baryons~\cite{Blanchet:2006yt,Blanchet:2008fj,Khoury:2014tka,Blanchet:2015sra,Blanchet:2015bia}. A recent prototypical example in this category is based on DM superfluidity~\cite{Berezhiani:2015pia,Berezhiani:2015bqa,Khoury:2016ehj,Khoury:2016egg,Hodson:2016rck,Berezhiani:2017tth}. In this scenario, the DM particles are axion-like, with masses of order~eV and strong self-interactions. They Bose-Einstein condense into a superfluid phase in the central regions of galaxy halos. The superfluid phonon excitations in turn couple to baryons and mediate an additional long-range force. For a suitable choice of the superfluid equation of state, this force can mimic Milgrom's relation. The framework thus marries the phenomenological success of the $\Lambda$CDM model on cosmological scales with that of the MDAR on galactic scales. 

The superfluid approach, while assuming Newtonian gravity, nevertheless relies on an additional long-range force between baryons to achieve the MDAR/MOND phenomena. In contrast with theories that propose to fundamentally modify Newtonian gravity, in this case the new long-range force mediated by phonons is an emergent property of the DM superfluid medium. 

In this paper, we propose a novel mechanism that instead primarily relies on DM-baryon {\it particle} ({\it i.e.}, short-range) interactions. Indeed, the most straightforward interpretation of the MDAR is that, given the baryonic density profile $\rho_{\rm b}(\vec{r})$ of a galaxy, the DM profile $\rho(\vec{r})$ can be uniquely predicted. The idea put forward here is that the desired DM profile may naturally emerge as the equilibrium configuration resulting from DM-baryon collisional interactions. The necessary properties of such interactions (such as the cross section) can then in principle be retro-engineered from our observationally-inferred knowledge of the MDAR. A brief heuristic overview of this retro-engineering and of the mechanism we are proposing can be found in Sec.~\ref{sec: heuristics}. The effects of such strong DM-baryon interactions have previously been considered in a number of different situations, {\it e.g.},~\cite{Starkman:1990nj,Dave:2000ar,Jacobs:2014yca}. The need for DM-baryon interactions in galaxies was emphasized recently in~\cite{Salucci:2017cet}. See also~\cite{Chen:2002yh,Sigurdson:2004zp,Boehm:2004th,Dolgov:2013una,McDermott:2010pa,Dubovsky:2001tr,Dubovsky:2003yn,Chuzhoy:2004bc,Hu:2007ai,Erickcek:2007jv,Albuquerque:2003ei,Natarajan:2002cw,Qin:2001hh,Mohapatra:1998nd,Cyburt:2002uw,Mack:2007xj,Dvorkin:2013cea,Munoz:2015bca,Munoz:2017qpy} for various cosmological and astrophysical constraints of strong DM-baryon interactions.

The detailed equations of our formalism will be presented in Secs.~\ref{setup} and \ref{sec:master} below. Then, under certain physical assumptions, we will find an equilibrium DM profile in rotationally-supported systems (Secs.~\ref{sec:rotational general} and \ref{exactnumerics}) such that $i)$ in regions where DM dominates the gravitational field---{\it i.e.}, in the `MONDian' regime---the DM profile gives rise to the deep-MOND law, $g \simeq \sqrt{a_0g_{\rm b}}$; in regions where baryons dominate, $g\simeq g_{\rm b}$, the DM profile contributes a sub-dominant, constant acceleration $g_{\rm DM}\simeq a_0\ll g_{\rm b}$. This corresponds to an approximately constant DM central surface-density, in good agreement with observations~\cite{Donato:2009ab}. In the MOND parlance, this matches the phenomenology of the `simple' interpolating function, which is well known to provide good fits to galaxy rotation curves~\cite{Famaey:2005fd,Gentile:2011}, as well as elliptical galaxies~\cite{Richtler:2011ec,Dabringhausen:2016klu,Chae:2017bhk}. 

It is important to stress that, unlike previous attempts to explain the MDAR with DM ({\it e.g.},~\cite{Kaplinghat:2001me}), our mechanism does {\it not} rely on the existence of a baryon-dominated region. It works equally well in low-surface brightness (LSB) galaxies, which are DM-dominated everywhere, as in high-surface brightness (HSB) galaxies, which do have a central baryon-dominated region. A limitation of our approach, however, is that, since we are working at equilibrium, we are unable at this stage to offer a dynamical explanation for the origin of the acceleration scale $a_0$. Instead, we regard $a_0$ simply as a parameter controlling the strength of DM-baryon interactions. Moreover, our equilibrium approach only allows us to demonstrate that the MDAR is necessarily obeyed at all radii once the BTFR is reached at large radii. The latter limitation can only be overcome by devising full-fledged simulations of galaxy formation in our framework.

There exist of course pressure-supported systems, such as galaxy clusters and certain dwarf spheroidals, that display a behavior distinct from the MDAR despite being dominated by DM. In Sec.~\ref{sec:pressure systems} we will show that this is easily explained by the fact that in such systems the relaxation time (determined by our DM-baryon interactions) can be larger than their age, and as such they have yet to reach their equilibrium configuration. 

For galaxy clusters, we will see that this is universally the case, which explains why such systems do not follow the MDAR~\cite{Gerbal:1992,Aguirre:2001fj}.
With dwarf spheroidals, the story is more nuanced. The standard MOND paradigm has a mixed track record when it comes to this kind of system~\cite{Famaey:2011kh}.
From our perspective, this is due to the fact that some dwarf spheroidals have already reached their equilibrium configuration whereas others have not.

\begin{table}
\caption{\label{observationssummary}Summary of observational consequences of our model}
\vspace{0.2cm}
\begin{tabularx}{\textwidth}{|X | p{0.55\textwidth}|}
    \hline\hline
\textsc{System} & \textsc{Behavior}\\
\hline\hline
\textbf{Rotationally-Supported Systems} & \\
 {\it Galaxy rotation curves} & \parbox[t]{0.55\textwidth}{Simple interpolating MOND, \\ no external field effect (EFE)} \\
 \hline
  \textbf{Pressure-Supported Systems} & \\
    {\it Galaxy clusters} & \parbox[t]{0.55\textwidth}{Not in equilibrium $\implies$ NFW profile \\ (for both dynamics and lensing)} \\
    &\\
      {\it Dwarf Spheroidals} & 
    	Some in equilibrium, others not. \\
    	&\\
    {\it Elliptical Galaxies} & \parbox[t]{0.55\textwidth}{Simple interpolating MOND \\ (subdominant DM acceleration $g_{\rm DM}\simeq a_0$)}\\
    &\\
    {\it Globular clusters} & No DM $\implies$ not MOND  \\
    \hline
  \textbf{Interacting Galaxies} & \\
    {\it Tidal dwarf galaxies} & No DM $\implies$ not MOND \\
    \hline
  \textbf{Galaxy-galaxy lensing} & Driven by NFW envelope $\implies$ not MOND \\
   \hline\hline
\end{tabularx}
\end{table}

When it comes instead to pressure-supported systems that are dominated by baryons, such as elliptical galaxies, 
we will argue that such systems are expected to have reached the equilibrium configuration. Our framework yields
a prediction for the subdominant DM contribution to the total acceleration that is consistent with the recent findings
of~\cite{Chae:2017bhk}. For the observationally-inclined reader, Table~\ref{observationssummary} summarizes the observational consequences of our model for various systems.

From the perspective of particle phenomenology, in order for our DM-baryon interactions to be effective we naturally need a strong cross section, of order $\frac{\sigma_{\rm int}}{m} \sim \frac{{\rm cm}^2}{{\rm g}}$ in the local neighborhood. Therefore, to satisfy direct detection constraints we are led to consider two limiting cases: the very heavy limit, $m \gg m_{\rm b}$, which
includes `macro' DM~\cite{Jacobs:2014yca}, and the very light limit, $m\ll m_{\rm b}$, which includes axion-like particles. As shown in Sec.~\ref{setup}, 
these correspond respectively to cooling and heating of DM by baryons. While in principle our mechanism can apply to either, the cooling case is technically
simpler since the average energy exchanged is approximately constant throughout the galaxy. Moreover, in the heating case one needs to be careful about possible Bose-Einstein condensation that may occur for $m \lesssim $ eV (see {\it e.g.},~\cite{Berezhiani:2015pia,Berezhiani:2015bqa}). For these reasons, most of our analysis in this paper will focus on the cooling/heavy DM case.

In this paper we will {\it not} offer a full-fledged example of a beyond-the-standard-model theory that realizes the desired interactions. 
Rather, we follow a `bottom-up' approach to determine, from kinetic theory arguments, what {\it kind} of particle physics interactions are necessary to obtain the MDAR, and to derive various astrophysical/cosmological consequences of such interactions. Most of the results of the paper are obtained by coarse-graining the baryonic and DM distribution functions over a typical scale of $\sim 10$ pc, therefore being independent of the clumpiness of baryons on small scales. We discuss in Sec.~\ref{sec: gas vs stars} how this assumption does not affect the heating rate but could affect detailed model-building to get the right scaling of the relaxation time.  Such a dedicated model-building effort is left for future work. Nevertheless, for illustrative purposes we present in Sec.~\ref{sec: particle physics models} an example of a particle physics model that gives the desired DM-baryon cross section dependence on density.

\section{Heuristic Overview} \label{sec: heuristics}

To gain some intuition, the basic physics behind the mechanism we will propose can be easily understood in the simple case of a spherically-symmetric system in the low acceleration regime. (In the language of Milgromian dynamics, this is called the `deep-MOND' regime.) Ignoring order unity factors controlling the ratio of velocity dispersion over circular velocity, we can roughly write $g \sim v^2/r$, where $v$ is the velocity dispersion of DM particles orbiting the halo (we will use this notation for the velocity dispersion to avoid confusion with the cross section throughout this paper), and $r$ is the spherical radius. The deep-MOND expression $g\simeq \sqrt{a_0g_{\rm b}}$ then implies
\be
v^4(r) \sim a_0GM_{\rm b}(r)\,.
\label{MDARsimple}
\ee
Thus the DM fluid temperature, $T \equiv mv^2$, is determined by the baryonic mass of the system. This strongly suggests that Eq.~\eqref{MDARsimple} results from integrating 
a {\it heat equation}, sourced by energy exchange with baryons.

In Sec.~\ref{setup} we will show that the desired heat equation at equilibrium (no time dependence) is
\be
\vec{\nabla}\cdot\left(\kappa \vec{\nabla} T \right)  \simeq n n_{\rm b} \sigma_{\rm int} v \epsilon\,,
\label{heatschematic}
\ee
where $\kappa$ is the thermal conductivity, $n$ and $n_{\rm b}$ are respectively the DM and baryon number densities, 
$\sigma_{\rm int}$ is the DM-baryon momentum-transfer interaction cross section, and $\epsilon$ is the typical energy exchanged per collision. Elastic interactions will be assumed. The left-hand side of this equation quantifies how efficiently heat is flowing through the DM medium, with the baryons {\it a priori} acting as the prime mediator of this heat flow (but noting that DM self-interactions could also play a role in it). It relies on the estimation of the relaxation time for DM particles. The right-hand side of the equation, on the other hand, quantifies the local rate of heating (cooling) that baryons inject into (remove from) the DM medium.

Our mechanism can in principle apply to either cooling ($\epsilon > 0$, with our sign convention) and heating ($\epsilon < 0$) of DM by baryons. For consistency with direct detection limits, these correspond respectively to very heavy ($m \gg m_{\rm b}$) and very light ($m\ll m_{\rm b}$) DM particles. The heating case would {\it a priori} appear more desirable, for instance to transform cusps into cores in the central regions of galaxy halos. However, as we will discuss in Sec.~\ref{setup}, the very heavy DM/cooling case is conceptually simpler, since in this case $\epsilon \sim m_{\rm b}v^2$, which is roughly constant for an approximate isothermal DM profile like the one we will derive in Sec. \ref{exactnumerics}. In that case, we will have to assume that cusps were already heated up and destroyed early in the history of the buildup of galaxy halos, and that the MDAR is then reached by cooling the DM fluid. Even though feedback might still be needed to erase the primordial cusps, our mechanism removes any need for fine tuning. The light DM/heating case, which might do away with feedback entirely, will be the topic of a follow-up paper.

To recover Eq.~\eqref{MDARsimple} as a solution to this heat equation, it is intuitively clear that: $i)$ the right-hand side of Eq.~\eqref{heatschematic} should depend  primarily on the baryon
density; and $ii)$ the characteristic acceleration $a_0$ should arise as some combination of the parameters. Both requirements are met by demanding
that the product $n \frac{\sigma_{\rm int}}{m_{\rm b}} \epsilon$, which has units of acceleration, be of order $a_0$:
\be
n \frac{\sigma_{\rm int}}{m_{\rm b}} \epsilon \sim a_0\,.
\ee
This {\it master relation} is critical to recover the MDAR. It also ensures, as shown in Sec.~\ref{sec: gas vs stars}, that all baryons contribute equally to the heating rate, regardless of whether they are distributed in a gas or they make up a star. Since $\epsilon \sim m_{\rm b} v^2$ in the cooling case of interest, the master relation requires $\sigma_{\rm int} \sim 1/nv^2$. 
The velocity dependence is quite natural, and arises, for instance, with Sommerfeld-enhanced interactions~\cite{Pospelov:2008jd,ArkaniHamed:2008qn,Iengo:2009ni,Feng:2010zp}
and charge-dipole interactions~\cite{Sigurdson:2004zp}. The density dependence is less familiar and suggests DM-baryon effective couplings that are sensitive to the
environment. We will show in Sec.~\ref{sec: particle physics models} how such a dependence on $n$ can arise naturally in a DM particle physics model. 

With the master relation, the heat equation~\eqref{heatschematic} simplifies to
\be
\vec{\nabla}\cdot\left(\frac{\kappa}{a_0} \vec{\nabla} T \right)  \sim v\rho_{\rm b}\,.
\label{heatschematic2}
\ee
Modulo a factor of $v$, the right-hand side depends on the baryon density, as desired.
The connection to MOND can be made even more transparent with the rough relation $v^2 \sim \Phi$, in which case Eq.~\eqref{heatschematic2} becomes
\be
\vec{\nabla}\cdot\left(\frac{\kappa}{a_0} \vec{\nabla} \Phi \right)  \sim v \rho_{\rm b}\,.
\ee
This is strikingly similar to the Bekenstein-Milgrom equation~\cite{Bekenstein:1984tv}, $\vec{\nabla}\cdot\left(\mu \left(\frac{|\vec{\nabla}\Phi|}{a_0}\right) \vec{\nabla} \Phi \right) = 4\pi G\rho_{\rm b}$,
with the thermal conductivity $\kappa$ effectively playing the role of the `interpolating function' $\mu$.
A key difference, however, is that, in the Bekenstein-Milgrom formulation of MOND, $\mu$ must have an ad hoc functional form, chosen by hand to satisfy the desired low- and high-acceleration behavior, whereas in our case $\kappa$ has a simple dependence on the DM density and temperature, as we are about to show.

A key requirement of our mechanism is that the typical relaxation time of DM particles is comparable to the dynamical time $t_{\rm dyn}$, {\it i.e.}, the time it takes for a DM particle to ``sample'' an appreciable fraction of the galaxy. This assumption, contrary to the heating rate, can in principle depend on the small scale clumpiness of baryons. The validity of this assumption on the relaxation time therefore relies on detailed model-building, which we will briefly discuss in Secs.~\ref{sec: gas vs stars} and~\ref{sec: particle physics models}. With this crucial assumption on the relaxation time, as shown in Sec.~\ref{sec:rotational general}, $\kappa$ is then simply given by
\be
\kappa \sim n \,r \,v \, .
\label{kappaintro}
\ee
In the kinetic theory of gases, this {\it Knudsen regime} applies to ultra-dilute gases, where molecules reach local thermal equilibrium by colliding with the walls of the container rather than among themselves. 
The thermal conductivity in that case is given by $\kappa \sim n\,L\,v$, where $L$ is the characteristic size of the container~\cite{Landau:1981mil}. 

With this expression for $\kappa$, we can integrate Eq.~\eqref{heatschematic2}, still in spherical symmetry and ignoring order unity factors, to obtain the Fourier's law, governing the temperature gradient of the DM fluid:\footnote{When integrating Eq.~\eqref{heatschematic2} we have treated the factor of $v$ on the right-hand side as an approximate power-law in~$r$.} 
\be
G\rho r \frac{{\rm d}v^2}{{\rm d}r} = a_0 g_{\rm b}\,.
\label{Fourierintro}
\ee
It is easy to show heuristically how the MDAR follows from this relation. Notice that the product $G \rho r$ is roughly of order the DM gravitational acceleration, $G \rho r \sim g_{\rm DM} = g - g_{\rm b}$.
Meanwhile, $\frac{{\rm d}v^2}{{\rm d}r}$ is very roughy of order of the total gravitational acceleration $g$. Putting these facts together, Eq.~\eqref{Fourierintro}
reduces to $(g-g_{\rm b})g = a_0 g_{\rm b}$, with solution
\be
g = \frac{g_{\rm b}}{2} \left(1 + \sqrt{1 + \frac{4a_0}{g_{\rm b}}}\right)\,.
\label{MONDsimple}
\ee
In MOND language, this is known as the `simple' interpolating function, which is well known to provide good fits to galaxy rotation curves~\cite{Famaey:2005fd,Gentile:2011}, as well as elliptical galaxies~\cite{Richtler:2011ec,Dabringhausen:2016klu,Chae:2017bhk}.\footnote{Of course the exact form of Eq.~\eqref{MONDsimple} should not be taken literally, since we have dropped various factors of order unity along the way. The important point is that it matches the `simple' function in both high- and low-acceleration limits.} In the low acceleration regime, $g_{\rm b}\ll a_0$, Eq.~\eqref{MONDsimple} correctly reproduces the deep-MOND relation $g\simeq\sqrt{a_0g_{\rm b}}$; in the high-acceleration regime, $g_{\rm b}\gg a_0$, it reduces to $g\simeq g_{\rm b} + a_0$, which includes a sub-dominant DM contribution $g_{\rm DM} \simeq a_0$.

With this heuristic approach in mind, we will now proceed with a more rigorous definition of the key equations of our model.

\section{Hydrostatic Equilibrium and Heat Transport}
\label{setup}

In the standard $\Lambda$CDM model, the DM component is comprised of particles which are non-relativistic at decoupling, and interact with each other and with baryons almost exclusively through gravity (apart from some extremely rare DM-baryon interactions looked for in direct detection experiments). Hence, they are governed by the {\it collisionless} Boltzmann equation, which states that the one-particle distribution function in phase space $f(t, \vec{r},\vec{v})$ obeys $\frac{{\rm d}f}{{\rm d}t}= 0$. Our approach generalizes this to a {\it Boltzmann transport equation}, in which case the right-hand side is non-zero and includes a collisional integral ${\cal I}[f,f_{\rm b}]$ to account for interactions between DM particles and baryons:
\begin{equation} 
\frac{\partial f}{\partial t} + \vec{v} \cdot \frac{\partial f}{\partial \vec r} + \vec{g}\cdot \frac{\partial f}{\partial \vec{v}} = {\cal I}[f,f_{\rm b}]\,.
\label{BTE}
\end{equation}
The (total) gravitational acceleration $\vec{g}$ is determined as usual by the Poisson equation
\be
\vec{\nabla}\cdot \vec{g}= -4\pi G\left(\rho + \rho_{\rm b}\right)\,,
\label{poisson}
\ee
where $\rho = m n \equiv m\int {\rm d}^3v \,f(\vec{x},\vec{v},t)$ and $\rho_{\rm b}$ are the DM and baryon densities, respectively.

To simplify the treatment in this first exploratory paper, it will suffice to take a Jeans approach~\cite{BinneyTremaine} and consider only the first few velocity moments of this equation. 
The zeroth, first and second velocity moments of Eq.~\eqref{BTE} respectively enforce mass, momentum and energy conservation:
\begin{subequations} \label{time dependent equations}
\bea
\label{continuity}
& \displaystyle \frac{\partial \rho}{\partial t} + \vec{\nabla}\cdot\left(\rho \vec{u}\right) = 0\,; & \\
\label{momentum}
& \displaystyle  \rho \left(\frac{\partial}{\partial t}  + \vec{u}\cdot\vec{\nabla}\right)u^i   +  \partial_jP^{ij} = \rho \vec{g}\,; &\\
\label{energy}
&\displaystyle  \frac{3}{2} \left(\frac{\partial}{\partial t}  + \vec{u}\cdot\vec{\nabla}\right) T +  \frac{m}{\rho} P^{ij}  \partial_i u_j + \frac{m}{\rho} \vec{\nabla}\cdot \vec q = \dot{{\cal E}}\,, &
\eea
\end{subequations}
where $\vec{u} \equiv \langle \vec{v} \rangle$ is the bulk DM velocity\footnote{We assume the standard notation $\langle A \rangle = \frac{1}{n} \int {\rm d}^3v \,f(\vec{x},\vec{v},t)A$.}, $P^{ij} \equiv \rho \left\langle \left(v^i - u^i\right)\left(v^j - u^j\right)\right\rangle$ is the pressure tensor,
$T \equiv \frac{m}{3} \langle |\vec{v}-\vec{u}|^2\rangle$ is the local DM temperature, and $\vec{q} \equiv \frac{1}{2}\rho \langle (\vec{v}-\vec{u})|\vec{v}-\vec{u}|^2\rangle$ is the
heat flux. The local heating rate, $\dot{{\cal E}}$, is due to interactions with baryons.

We will primarily be interested in equilibrium configurations with no spin of the DM halo. In this case, the DM bulk velocity can be set to zero, $\vec{u} = 0$, and the continuity equation~\eqref{continuity} is trivially satisfied. Furthermore, we will assume in this exploratory paper that the DM distribution is described to zeroth order by a local isotropic Maxwell-Boltzmann distribution. The pressure tensor hence becomes diagonal, $P^{ij} = \rho v^2\delta^{ij}$, where $v$ is the one-dimensional velocity dispersion,\footnote{The velocity dispersion is sometimes denoted by $\sigma$, but we reserve that symbol for the cross section.
The circular velocity will be instead denoted by $V$.} related to temperature by $v=\sqrt{T/m}$. The momentum equation~\eqref{momentum} thus reduces to
Jeans' equation, which in this case is equivalent to the condition of hydrostatic equilibrium:
\be
\vec{\nabla}\left(\rho v^2\right)  = \rho \vec{g}\,.
\label{jeans}
\ee
If we specialize to spherical symmetry in configuration space, as will be done later, then whenever $\rho$ and $v$ are power-laws in $r$ this equation implies that $v^2$
is proportional to the circular velocity squared, $V^2$. For instance, in the case of an isothermal sphere, for which $v$ is constant and $\rho \sim 1/r^2$, the proportionality factor is precisely one half.

To calculate the heat flux $\vec{q}$, we must perturb the Maxwell-Boltzmann distribution. To leading order in small deviations from local thermal equilibrium, the result is 
\begin{equation} \label{Fourierlaw}
	\vec q \simeq - \kappa \vec \nabla T \,.	
\end{equation}
This result should be thought of as the leading term in an expansion in derivatives of the temperature $T$.\footnote{For spherically symmetric distributions, higher order terms can be neglected provided ${\rm d} \log T / {\rm d} \log r$ is small. This criterion will indeed be satisfied by the equilibrium solutions obtained later.}
The thermal conductivity, $\kappa$, can be expressed in terms of microscopic quantities such as the characteristic distance $\ell$ ``sampled'' by a DM particle and the relaxation time $t_{\rm relax}$ it takes to lose memory of its initial velocity:
\begin{equation} \label{kappa}
	\kappa = \frac{3}{2} \frac{n \ell^2}{t_{\rm relax}}\,.
\end{equation}
Note that the thermal conductivity of the DM fluid is related in our case to its interactions with baryons, which act as a mediator of heat for the DM medium. Hence, in disk galaxies, the relaxation time will be fixed by the time it takes for a DM particle to find the baryonic disk, provided interactions with the baryons in the disk are efficient enough. We will come back to this issue in Secs.~\ref{sec:knudsen} and \ref{timescalechecks}, and briefly in Sec.~\ref{sec: gas vs stars}.

Near equilibrium, Eq.~\eqref{energy} enforcing energy conservation reduces to
\be
\vec{\nabla}\cdot \left(\kappa \vec \nabla (mv^2)\right) = - n\dot{{\cal E}}\,.
\label{heatbasic}
\ee
All that remains is to calculate the heating rate $\dot{{\cal E}}$ on the right-hand side, which depends on the baryon density and velocity dispersion. In order to simplify our analysis, we will assume at this stage that the probability distributions $f$ and $f_{\rm b}$ that appear in the Boltzmann transport equation \eqref{BTE} are obtained by coarse-graining over distances $\sim 10$ pc. This means that macroscopic quantities such as $\rho$, $v$ and $\vec u$ are all calculated by averaging over volumes that contain in general many stars as well as gas. By taking this viewpoint, we can describe baryons using a smoothly-varying mass density $\rho_{\rm b} = m_{\rm b} n_{\rm b}$ and local temperature $T_{\rm b} = m_{\rm b} v_{\rm b}^2$, with $m_{\rm b}\simeq 1~{\rm GeV}$. We will re-examine this working assumption in Sec.~\ref{sec: gas vs stars}.

Following~\cite{Munoz:2015bca}, we define the quantities 
\be
v_{\rm th} \equiv \sqrt{v^2 + v_{\rm b}^2}\,;\qquad\qquad \quad   y\equiv \frac{u_{\rm b}}{v_{\rm th}}\,,
\ee
where $u_{\rm b}$ is the baryon bulk velocity, and $y$ is thus the ratio of the baryon bulk velocity (typically, the rotational velocity) over the total velocity dispersion. We focus on elastic collisions, described by a transport (or momentum-transfer) cross section $\sigma_{\rm int}$ with power-law dependence on the relative baryon-DM velocity:\footnote{Recall that the transport cross section is related to the differential cross section by 
\be
\nonumber
\sigma_{\rm int} = \int {\rm d}\Omega\,\frac{{\rm d}\sigma}{{\rm d}\Omega}\,(1-\cos\theta)\,.
\ee
}
\be
\sigma_{\rm int} = \sigma_0 |\vec{v}-\vec{v}_{\rm b}|^{\alpha}\,.
\label{sigmapowerlaw}
\ee
Apart from $\alpha =0$, some physically-motivated values are $\alpha = -1$ (Sommerfeld enhancement away from resonance~\cite{Pospelov:2008jd,ArkaniHamed:2008qn,Iengo:2009ni}), $\alpha = -2$ (Sommerfeld enhancement at resonance~\cite{Cassel:2009wt,Slatyer:2009vg,Feng:2010zp} or charge-dipole interactions~\cite{Sigurdson:2004zp}) and $\alpha = -4$ (long-range mediator, {\it e.g.},~\cite{Dubovsky:2001tr,Dubovsky:2003yn,McDermott:2010pa}). As we will see, of particular interest to us will be the values $\alpha = 0$ and $-2$. Furthermore, for reasons already alluded to in Sec.~\ref{sec: heuristics}, it is critical for our mechanism that $\sigma_0$ have ``environmental" dependence on the local density $n$ and/or velocity dispersion $v$.

\begin{figure}[htp]
\centering
\begin{tabular}{c c}
       \addheight{\includegraphics[width=2.8in]{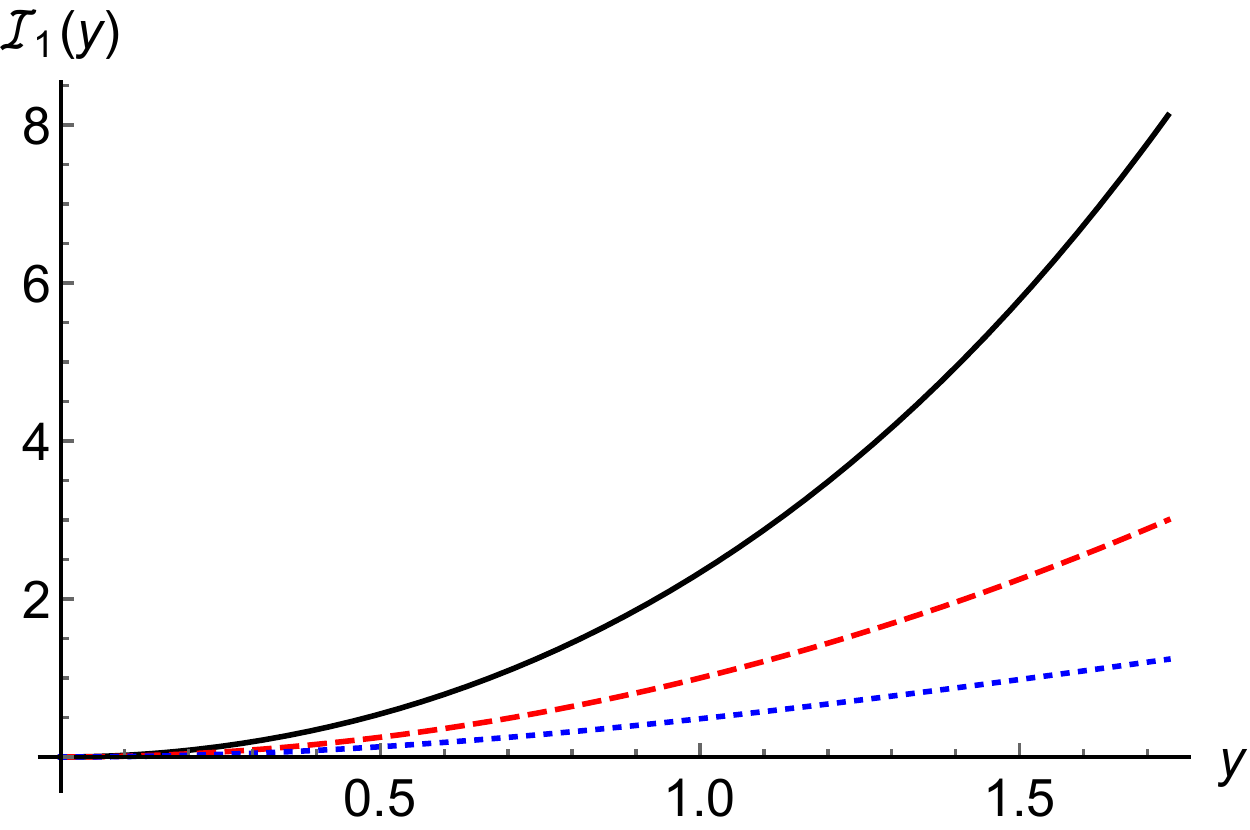}} &
      \addheight{\includegraphics[width=3.5in]{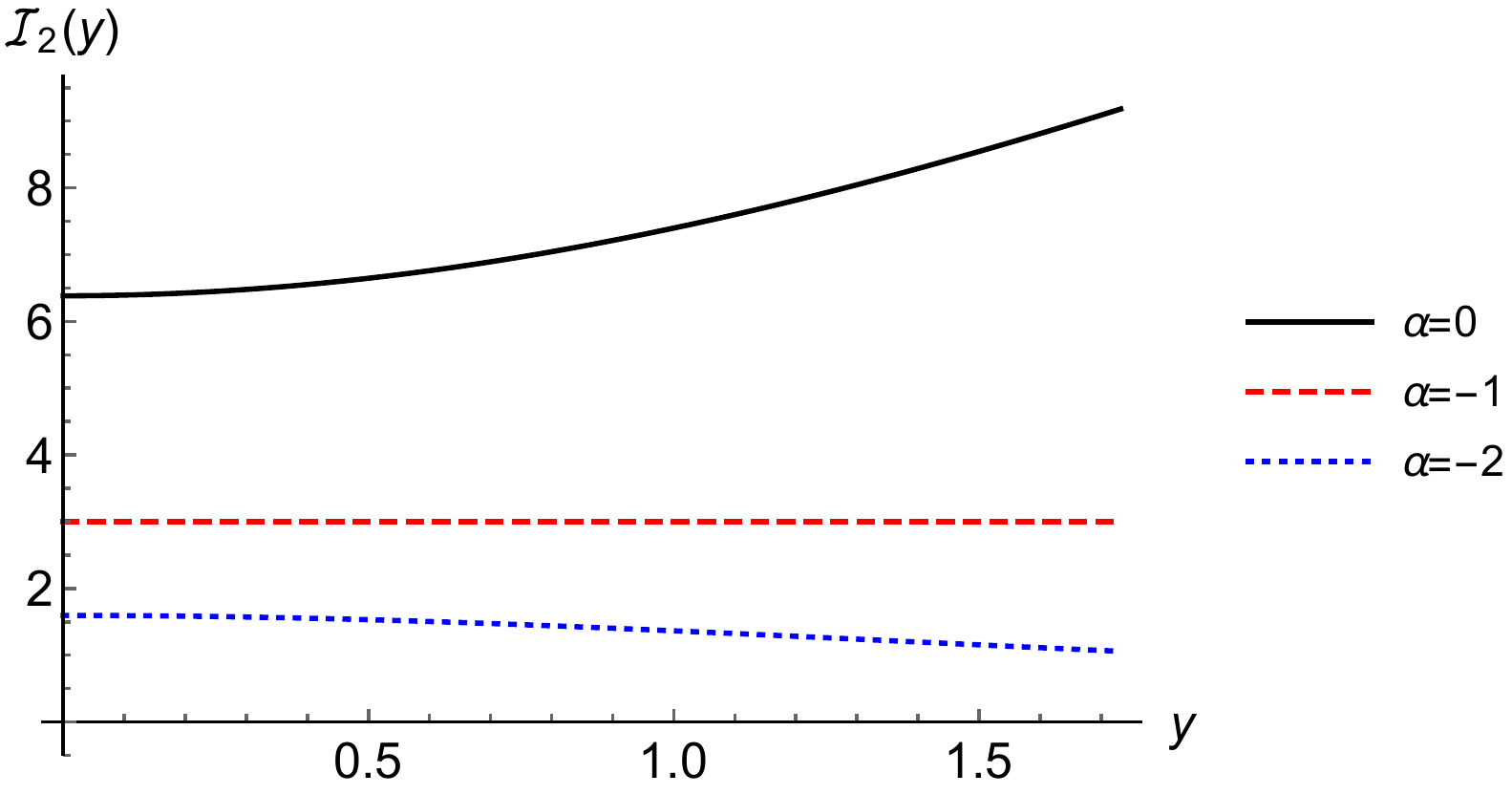}} \\
            \end{tabular}
      \caption{\small Functions ${\cal I}_1^{(\alpha)}(y)$ (left) and ${\cal I}_2^{(\alpha)}(y)$ (right) for a few values of $\alpha$.}
\label{Icoeffs}
      \end{figure}

Under these assumptions, the heating rate is given by~\cite{Munoz:2015bca}
\be
\dot{{\cal E}} = \frac{m m_{\rm b}}{(m+m_{\rm b})^2} n_{\rm b} \sigma_0v_{\rm th}^{3+\alpha} \left( m_{\rm b}{\cal I}_1^{(\alpha)}(y) - \frac{m v^2-m_{\rm b}v_{\rm b}^2}{v_{\rm th}^2} {\cal I}_2^{(\alpha)}(y)\right)\,,
\label{heatrate}
\ee
where $n_{\rm b}$ is the baryon number density, and we have introduced
\bea
\nonumber
{\cal I}_1^{(\alpha)}(y) &=& \frac{y^{\alpha+6}}{\sqrt{2\pi}} \int_{-\infty}^\infty {\rm d}x \,x e^{-\frac{1}{2}y^2x^2} \frac{(x-1)(\alpha+x+4)|x-1|^{3+\alpha}}{(\alpha+3)(\alpha+5)}\,;\\
{\cal I}_2^{(\alpha)}(y) &=&- \frac{y^{\alpha+6}}{\sqrt{2\pi}} \int_{-\infty}^\infty {\rm d}x \,x e^{-\frac{1}{2}y^2x^2} \frac{(x-1)^3\Big((\alpha+4)x + 1\Big)|x-1|^{\alpha+1}}{(\alpha+3)(\alpha+5)}\,.
\eea
The first term is positive definite and describes heating due to the relative bulk velocity. This term is non-zero whenever $u_{\rm b}\neq 0$, even if the two fluids have the same temperature. 
The second term is the usual heat flow proportional to the temperature difference. The functions ${\cal I}_1^{(\alpha)}$ and ${\cal I}_2^{(\alpha)}$ are plotted in Fig.~\ref{Icoeffs} for a few values of $\alpha$. 

Since our mechanism requires a healthy cross section between baryons and DM, to satisfy direct detection constraints we are led to consider two limiting cases: $m \gg m_{\rm b}$ and $m\ll m_{\rm b}$. As
we will see below, these correspond respectively to cooling and heating of DM by baryons.
\\

\begin{itemize}

\item {\bf Heavy DM, $m\gg m_{\rm b}$ (Cooling Case):} In this regime, we can safely assume that $m_{\rm b} v_{\rm b}^2 \ll m v^2$ in the second term of Eq.~\eqref{heatrate}. Furthermore, the bulk velocity term is 
suppressed by $m_{\rm b}/m$ and can be neglected. The heating rate thus reduces to
\be
\dot{{\cal E}} \simeq - n_{\rm b} \sigma_0 v_{\rm th}^{\alpha+1} {\cal I}_2^{(\alpha)}(y)\,m_{\rm b} v^2\,.
\ee
This quantity is manifestly negative, and thus describes {\it cooling}.

The heat equation~\eqref{heatbasic} boils down to an intuitively simple form,
\be
\vec{\nabla}\cdot\left(\kappa \vec{\nabla}(m v^2)\right)  \simeq n \Gamma_{\rm int}(v_{\rm th}) \epsilon\,,
\label{heatintuitive}
\ee
where $\Gamma_{\rm int}$ is the interaction rate
\be
\Gamma_{\rm int} = n_{\rm b}\sigma_{\rm int}(v_{\rm th}) v_{\rm th}\,,
\label{Gammaint}
\ee
and $\epsilon$ characterizes the typical energy exchanged per collision:\footnote{Our sign convention is such that positive $\epsilon$ corresponds to cooling.}
\be
\epsilon = {\cal I}_2^{(\alpha)}(y) m_{\rm b}v^2\,.
\label{epsiloncool}
\ee
As seen from Fig.~\ref{Icoeffs} (right panel), ${\cal I}_2^{(\alpha)}(y)$ is a slowly-varying function over the relevant range of $y$. For simplicity, we henceforth treat it as constant.
\\

\item {\bf Light DM, $m\ll m_{\rm b}$, (Heating Case):} In this regime, we instead have $m_{\rm b} v_{\rm b}^2 \gg m v^2$, and the heating rate becomes
\be
\dot{{\cal E}} \simeq n_{\rm b} \sigma_0 v_{\rm th}^{\alpha+1} \left({\cal I}_1^{(\alpha)}(y)  + \frac{v_{\rm b}^2}{v_{\rm th}^2}{\cal I}_2^{(\alpha)}(y)\right)m v_{\rm th}^2\,.
\ee
Both contributions are positive definite, hence this case describes {\it heating}.

Once again the heat equation can be cast in the form~\eqref{heatintuitive}, but with a characteristic energy exchanged now given by
\be
\epsilon = - \left({\cal I}_1^{(\alpha)}(y)  + \frac{v_{\rm b}^2}{v_{\rm th}^2}{\cal I}_2^{(\alpha)}(y)\right)m v_{\rm th}^2\,.
\label{epsilonheat}
\ee
Unlike the cooling case, however, $\epsilon$ cannot generally be treated as approximately constant. In disk galaxies, specifically, the prefactor varies
from $\ll 1$ in the central region to $\sim {\cal O}(1)$ in the outer region where the rotation curve flattens. 

To see this, first note that baryons have negligible velocity dispersion, $v_{\rm b} \ll v$, hence the second term, proportional to ${\cal I}_2$, is
suppressed by $v_{\rm b}^2/v_{\rm th}^2 \simeq v_{\rm b}^2/v^2 \ll 1$. Meanwhile, the baryon bulk velocity can be approximated by its circular velocity,
{\it i.e.}, $u_{\rm b} \simeq V_{\rm b}$. It follows that 
\be
y \simeq \frac{V_{\rm b}}{v}\,.
\ee
On the one hand, our hope is to reproduce the MONDian DM density profile, which is well-approximated by a cored isothermal
sphere with $v\simeq {\rm const.}$~\cite{Milgrom:2004ba}. On the other hand, $V_{\rm b}$ varies from 0 at the center to $\simeq \sqrt{2} v$ in the flat part
of the rotation curve. It follows that $y$ ranges from 0 to $\sqrt{2}$ and, correspondingly, we see from Fig.~\ref{Icoeffs} that ${\cal I}_1$ varies from
0 to order unity. Therefore, the prefactor in Eq. \eqref{epsilonheat} varies substantially and cannot be treated as constant in disk galaxies.\footnote{However, in pressure-supported systems
such as elliptical galaxies, DM and baryons have comparable velocity dispersion, $v_{\rm b} \sim v$, and negligible bulk velocity, $y\simeq 0$. Therefore,
the second term in Eq.~\eqref{epsilonheat} dominates and is everywhere approximately constant.} 
\\

\end{itemize}

The cooling and heating cases both result in a heat equation given by Eq.~\eqref{heatintuitive}. The main difference is that $\epsilon$ is positive and approximately constant in the cooling case, while it is negative and varies substantially in the heating case. For the sake of simplifying our analysis, we henceforth focus on the cooling case, leaving a detailed treatment of the heating case to future work.

To summarize, we henceforth assume that DM is heavy ($m\gg m_{\rm b}$), which results in baryons acting as a coolant.
The set of equations we will solve are
\begin{subequations}
\bea
\label{poissoncooling}
\vec{\nabla}\cdot \vec{g} &=& -4\pi G\left(\rho + \rho_{\rm b}\right)\,;\qquad \text{(Poisson)}\\
\label{jeanscooling}
\vec{\nabla}\left(\rho v^2\right)  &=& \rho \vec{g}\,;\qquad\, ~~~~~~~~~~~~~~~~\text{(Jeans)}\\
\label{heatcooling}
\vec{\nabla}\cdot \left(\kappa m\vec{\nabla}v^2\right) &=& n n_{\rm b} \sigma_{\rm int} v \epsilon\,; \qquad ~~~~~~~\text{(Heat)}\,.
\eea
\end{subequations}
Note that in the heat equation, for simplicity we have substituted $v_{\rm th}\simeq v$, a valid approximation in rotationally-supported systems.
(We will revisit this approximation in Sec.~\ref{sec:pressure systems}, when discussing pressure-supported systems.) The characteristic energy transferred
per collision, $\epsilon \sim m_{\rm b} v^2$, is positive and thus describes cooling. The proportionality  coefficient will be treated as approximately constant. 

Although we should in principle also include a heat transfer equation for the baryon component, for simplicity we will treat baryons as an input, whose
density profile $\rho_{\rm b}(\vec{x})$ is specified by observations. With the baryon distribution thus specified, we will solve the above equations to determine
the DM density and velocity profiles, $\rho$ and $v$.

\section{Acceleration Scale and Master Equation}
\label{sec:master}

At this stage, in order for this formalism to lead to the emergence of the MDAR, it is clear that an acceleration scale should appear in the heat equation. The reader familiar with the formalism of~\cite{Bekenstein:1984tv} will immediately recognize at this stage that, given the form of our heat equation as $\vec{\nabla} \cdot [\kappa(T) \vec{\nabla} T] = $ heat source, recovering the MDAR requires that the source term depend primarily on the baryon density. 

For this purpose we will only need to make a general assumption about the momentum-transfer interaction cross section. Specifically, we will assume that the product $n\frac{\sigma_{\rm int}}{m_{\rm b}} \epsilon$, which has units of acceleration, must be constant and of order the characteristic acceleration scale $a_0$:
\be
\sigma_{\rm int}  = \frac{C\,a_0m_{\rm b}}{n\,\epsilon}\,.
\label{master equation}
\ee
This simplifies our heat equation~\eqref{heatcooling} to
\be
\vec{\nabla}\cdot \left(\frac{\kappa}{a_0} m\vec{\nabla}v^2\right) = C\,v\rho_{\rm b}\,.
\label{heateqn1}
\ee
The dimensionless constant $C$ will be fixed later to be of order $10^{-1}$ (Sec.~\ref{sec:rotational general}). Since $\epsilon \sim m_{\rm b} v^2$, this {\it master relation}~\eqref{master equation} states that the cross section $\sigma_{\rm int}$ must scale like $1/nv^2$. 

We are of course used to cross sections depending on velocity (or energy), but density-dependent cross sections may be less familiar. It suggests that the strength of interactions between DM and baryons is sensitive to the {\it environment}. A number of well-known processes in nature exhibit environmental density dependence, such as the Mikheyev-Smirnov-Wolfenstein effect for solar neutrinos~\cite{Mikheyev:1985,Wolfenstein:1977ue} or photons acquiring an effective mass in plasmas~\cite{abrikosov2012methods}. Another example are phonons, whose interactions with themselves ({\it e.g.},~\cite{Endlich:2010hf}) as well as with other excitations ({\it e.g.},~\cite{Horn:2015zna,Nicolis:2017eqo}) are naturally suppressed by the density of the medium. We will show in Sec.~\ref{sec: particle physics models} that such a dependence on $n$ can arise naturally in DM particle physics models.

The $1/v^2$ dependence is more familiar. It arises, for instance, with Sommerfeld enhancement close to resonances~\cite{Pospelov:2008jd,ArkaniHamed:2008qn,Iengo:2009ni,Cassel:2009wt,Slatyer:2009vg,Feng:2010zp} and charge-dipole interactions~\cite{Sigurdson:2004zp}, both of which correspond to $\alpha = -2$. Another possibility is that the velocity dependence is also environmental, arising from DM-baryon
interactions being sensitive to the ambient DM velocity dispersion. In the explicit example of Sec.~\ref{sec: particle physics models}, in particular, the cross section will be inversely proportional to
the ambient DM pressure, {\it i.e.}, $\sigma_{\rm int} \sim 1/P \sim 1/nv^2$, corresponding to $\alpha = 0$.

Up until Sec.~\ref{sec: particle physics models}, we will remain agnostic as to the microphysical realization of Eq.~\eqref{master equation}.
We will treat it as an empirical requirement for the success of our mechanism and derive various model-independent consequences. Before focusing our attention on particular systems, in the rest of this section we will use our master equation to derive a few, very general results.

\subsection{Energy-exchange characteristic time}

In order to use the heat equation, we will need an explicit expression for the thermal conductivity Eq.~\eqref{kappa}. This will depend in general on the system under consideration. In particular, in order to correctly identify the relaxation time $t_{\rm relax}$ it is helpful to first introduce a characteristic time scale $\tau$ for energy loss to baryons. This is defined by the requirement that the energy
$E = \frac{3}{2} m v^2$ of a DM particle change by a relative factor $\Delta E / E \sim \mathcal O (1)$ in time $\tau$:
\be
\frac{{\rm d}(m v^2)}{{\rm d}t} = - \frac{2}{3} \Gamma_{\rm int} \epsilon = -\frac{2}{3} \rho_{\rm b} v \frac{C\,a_0}{n} \equiv - \frac{mv^2}{\tau} \,, 
\label{energy loss}
\ee
where in the next-to-last step we have used the master relation~\eqref{master equation}. It follows that
\be
\tau \simeq \frac{3}{2C} \frac{\rho}{\rho_{\rm b}} \frac{v}{a_0}\,.
\label{taudef}
\ee

When considering different systems in the next sections, we will compare $\tau$ to the age of the system, typically of order $H_0^{-1}$, and to the dynamical
time $t_{\rm dyn}$. The first comparison will allow us to deduce whether a given system has reached its equilibrium configuration (assuming that the energy loss due to self-interactions is subdominant). Using the value $a_0 \simeq cH_0/6$, we obtain
\be
H_0 \tau \simeq \frac{9}{C} \,  \frac{v}{c} \frac{\rho}{\rho_{\rm b}} \,.
\label{equilibriumcondition}
\ee
In order to justifiably claim a system is in equilibrium, one must check that $H_0\tau \;\lsim\; 1$. We will see that this is the case for disk galaxies, for instance, but not for galaxy clusters or certain dwarf spheroidals. 

The comparison between $\tau$ and the dynamical time $t_{\rm dyn}$ will determine the relaxation time, as discussed at length hereafter in Sec.~\ref{sec:rotational general}.
For a DM particle orbiting at a distance $r$ from the center of the galaxy, $t_{\rm dyn}$ is of order $\sim r/v$. Combining with Eq.~\eqref{taudef} we obtain
\be
\frac{\tau}{t_{\rm dyn}} \sim \fr{3}{2C} \frac{v^2}{a_0 r} \frac{\rho}{\rho_{\rm b}} \,.
\label{tautdynratio}
\ee
As we will see, our equilibrium solution requires $\tau \;\lsim\; t_{\rm dyn}$, within the galactic disk for rotationally-supported systems or throughout the halo for pressure-supported systems. We will demand however that $\tau \gtrsim t_{\rm dyn}$ in the halo of rotationally-supported systems. 

\subsection{Dynamics}
\label{dynamics}

While most of our discussion will focus on equilibrium configurations, it is worth at this stage to comment briefly on what can be expected dynamically, particularly during structure formation when an overdensity becomes non-linear and undergoes virialization. 

In $\Lambda$CDM, while continuously accreting matter from the cosmic web, halos undergo violent relaxation and phase-mixing at each step along the merger tree, until they reach a Navarro-Frenk-White (NFW)-like profile~\cite{NFW} in a few dynamical times. In our case, we expect that, at large enough scales in the early Universe after recombination, DM and baryons could still be mixed up with roughly the cosmological baryon fraction $\rho/\rho_{\rm b} \sim 6$. The first halos would then successively merge within this mixture until they give rise to the present-day mass halos. 

Let us estimate the ratio $\tau/t_{\rm dyn}$ at virialization. Substituting the cosmic fraction $\rho/\rho_{\rm b} \sim 6$ and the virial relation $v^2 = \frac{1}{3}\frac{GM}{R_{200}}$ (with $R_{200}$ the virial radius defined below), Eq.~\eqref{tautdynratio} becomes 
\be
\left.\frac{\tau}{t_{\rm dyn}}\right\vert_{\rm virial.} \sim \frac{3}{C}\frac{GM_{200}}{a_0 R^2_{200}}\,.
\label{tautdyn1}
\ee
This expression is nicely interpreted as the ratio of the acceleration at the virial radius, $GM/R^2_{200}$, over the critical acceleration $a_0$. Halos that are MONDian at virialization therefore have $\tau < t_{\rm dyn}$. The virial radius $R_{200}$ is defined as the radius at which the average density equals 200 times the present critical density. Assuming $H_0 = 70\;{\rm km}\,{\rm s}^{-1} {\rm Mpc}^{-1}$ for concreteness, it is given by
\be
R_{200}\simeq 203 \left(\frac{M_{200}}{10^{12}{\rm M}_\odot}\right)^{1/3}{\rm kpc}\,.
\label{R200def2}
\ee
The ratio of time scales in \eqref{tautdyn1} then reduces to
\be
\left.\frac{\tau}{t_{\rm dyn}}\right\vert_{\rm virial.} \simeq \frac{0.1}{C} \left(\frac{M_{200}}{10^{12}{\rm M}_\odot}\right)^{1/3}\,.
\ee
With $C \sim 10^{-1}$, we see that $\tau \;\lsim \; t_{\rm dyn}$ whenever $M_{200}\; \lsim \;10^{12} {\rm M}_\odot$. For smaller halo masses, the baryon fraction can be even lower, and the result still holds. This means that we do not expect our halos to reach a NFW profile early on. Later on, after baryons have collapsed at the center, the mergers with non-NFW profiles are still not expected to lead to NFW profiles. The very outskirts of halos could nevertheless still have a density dropping as a power-law with slope $-3$. However, since the NFW profile and its cold central cusp seems to be an attractor solution for cold enough initial conditions, it is not obvious that the DM cooling mechanism we are relying on would allow to reach the desired equilibrium configurations early on, especially at the center of halos, where  a constant density core is needed in most LSB galaxy halos. Therefore, some preheating of the central regions of DM halos might be needed before allowing them to reach equilibrium. In other words, even the most bound and cold DM particles at the center of halos could be in need of being strongly up-scattered before being allowed to cool again. Usual feedback processes could apply, but they would not need any sort of fine-tuning to reproduce the MDAR phenomenology. They could be very efficient, and severely heat the DM fluid in all halos, while the cooling mechanism resulting from interactions with baryons would then allow the profile to gently reach the desired equilibrium. 

It will also be important to check in simulations whether our cooling mechanism is not leading to too flattened halos, or too prominent dark disks, once the halos have an initial spin. In future work, we will look for equilibrium solutions in the heating case, which could self-consistently erase the cusps at the center of galaxy halos, and would not lead to flattening.

\section{Rotationally-Supported Systems: Analytical Results}
\label{sec:rotational general}

We first apply the framework discussed above to rotationally-supported systems, {\it i.e.}, primarily disk galaxies, since these are the systems for which the MDAR is particularly tight and well-defined.

\subsection{Knudsen regime in disk galaxies} \label{sec:knudsen}

From the perspective of a DM particle in a disk galaxy, we can distinguish three important time scales. The first one is the characteristic time $\tau_{\rm halo}$ associated with energy loss in the halo, far away from the baryons in the disk. The relevant collisions can be either DM self-interactions, or interactions with the few baryons that are not captured by the disk. This time scale is always comparable to (or larger than) the inverse interaction rate. (It is larger if multiple such collisions are needed to significantly change the energy of a DM particle.) The time scale $\tau_{\rm halo}$ will in general differ from the energy-loss time scale $\tau_{\rm disk}$ inside the disk, where collisions between baryons and DM are much more likely. Finally, the third important time scale is the dynamical time, $t_{\rm dyn}\sim r/v$.

The first crucial assumption we will make is that these three time scales satisfy the hierarchy
\be
\tau_{\rm halo} > t_{\rm dyn} > \tau_{\rm disk}\,. 
\label{timehierarchy}
\ee
We will justify this in Sec.~\ref{timescalechecks}, once we have derived the approximate density profile in both DM-dominated and baryon-dominated regimes. Under the assumption~\eqref{timehierarchy}, DM particles travel essentially unimpeded through the halo ($\tau_{\rm halo} > t_{\rm dyn}$) within the inner region where rotation curves are measured and take some time ($\sim t_{\rm dyn}$) to find the disk where most of the baryons are localized. However, once they do find the disk, interactions with baryons are fairly efficient ($\tau_{\rm disk} < t_{\rm dyn}$). In this scenario, the relaxation time that appears in Eq.~\eqref{kappa} is then set by the time it takes for a DM particle to find the disk, {\it i.e.}, 
\begin{equation}\label{relaxation time}
	t_{\rm relax} \sim t_{\rm dyn} .
\end{equation}

For a typical spiral galaxy, this relaxation time is much shorter than the age of the galaxy itself. Hence the system can relax fast, and even win over violent relaxation which typically takes a few dynamical times to be completed, but is halted as soon as an equilibrium is reached. We should emphasize however that the estimate in Eq.~\eqref{relaxation time} is applicable only for DM particles orbiting at distances not much larger than the size of the disk. DM particles orbiting much further away will take on average much longer to find the disk, hence for those particles $t_{\rm relax} \gg t_{\rm dyn}$. We will make this more precise in Sec.~\ref{NFWmatch}.

Using Eq.~\eqref{relaxation time} and the fact that $\ell \sim v \, t_{\rm dyn} \sim r$~\cite{Gnedin:2000ea} ({\it i.e.}, $r$ acts essentially as an infrared cutoff), the thermal conductivity given by Eq.~\eqref{kappa} for rotationally-supported disk galaxies reduces~to
\begin{equation} \label{kappa 2}
	\kappa \sim n \,r \,v \, .
\end{equation}
This optically-thin regime in which the relaxation and dynamical times are of the same order, also known as {\it Knudsen regime}, applies in the purely gravitational context to globular clusters~\cite{LyndenBell:1980}---see also Appendix B of~\cite{Gnedin:2000ea} for a related discussion in the context of self-interacting DM. In this regime the thermal conductivity, and hence the left-hand side of the heat equation~\eqref{heateqn1}, become insensitive to the microscopic details of baryon-DM interactions. The prototypical example of a system in the Knudsen regime is a very dilute gas, in which the mean free path of molecules is much larger than the size of the container. In this case the molecules reach local thermal equilibrium by colliding with the walls of the container rather than among themselves, and the heat conductivity is given by Eq.~\eqref{kappa 2}, where $r$ is now the size of the container~\cite{Landau:1981mil}. In our case, the disk effectively plays the role of the wall while the inner DM halo around the disk can be thought of as the container.

Substituting Eq.~\eqref{kappa 2} into the heat equation~\eqref{heateqn1}, we obtain
\be
\vec{\nabla}\cdot \left(\frac{\rho r v}{a_0} \vec{\nabla}v^2\right) = C\, v\rho_{\rm b}\,.
\label{heateqn2}
\ee
This equation, together with Poisson's equation~\eqref{poissoncooling} and the condition~\eqref{jeanscooling} for hydrostatic equilibrium ({\it i.e.}, the isotropic Jeans equation), form the complete set of equations to be integrated once the baryon density profile $\rho_{\rm b}$ is specified. 

\subsection{Scale invariance in DM-dominated regime}
\label{scaleinvarianceDM}

Consider a region where the gravitational acceleration is dominated by DM. The Poisson equation, Jeans equation, and heat equation become
\begin{subequations}
\bea
\vec{\nabla}\cdot \vec{g} \simeq -4\pi G \rho\,; \label{Poisson 2} \\
\vec{\nabla}\left(\rho v^2\right)  = \rho \vec{g}\,; \\
\vec{\nabla}\cdot \left(\frac{\rho r v}{a_0} \vec{\nabla}v^2\right) = C\, v\rho_{\rm b}\,.
\eea
\end{subequations}
These satisfy together, as a system of equations, a {\it space-time scale invariance}, which generalizes the scale invariance of the deep-MOND equation discussed in~\cite{Milgrom:2008cs}. Indeed, these equations are invariant under the anisotropic space-time rescaling $\vec{x}\rightarrow e^{\lambda} \vec{x}$, $t\rightarrow e^{z \lambda} t$ for an arbitrary value of $z$, provided the remaining quantities transform as follows:
\begin{subequations}\label{scalingequations}
\bea
v & \rightarrow & e^{\lambda(1-z)} v\, ; \\
\vec{g} & \rightarrow & e^{\lambda(1-2z)} \vec{g}\, ; \\
\rho & \rightarrow & e^{-2\lambda z} \rho\, ; \label{scaling rho}\\
\rho_{\rm b}& \rightarrow & e^{\lambda(1-4z)} \rho_{\rm b}\, . 
\eea
\end{subequations}

Much can be deduced about our predicted rotation curves from this scaling symmetry alone. Suppose we have a particular solution $v(\vec{x})$ in the
DM-dominated regime for a given baryon distribution $\rho_{\rm b}(\vec{x})$. Then, we immediately know that $\vec{v}(e^\lambda\vec{x})$ is a DM-dominated solution
for the rescaled distribution $e^{\lambda(1-4z)} \rho_{\rm b}(e^\lambda\vec{x})$. To the extent that $V^2 \sim v^2$, this implies that rotation curves for LSB galaxies (which are DM-dominated everywhere) are predicted to be self-similar when expressed in units of disk scale-length. The particular transformation with $z=1$, which leaves $v$ as well as the total baryonic mass invariant, has particularly striking implications. Consider two galaxies with the same total baryonic mass but different scale lengths $L_1$ and $L_2$. Then, their rotation curves should be related by $V_2(\vec{x}) = \vec{V}_1\left(\frac{L_1}{L_2}\vec{x}\right)$. Such behavior is of course also predicted by MOND, thanks to the same underlying scaling symmetry~\cite{Milgrom:2008cs}, and it is in fact confirmed by observations~\cite{Swaters2009}. 

It is interesting to point out that a subset of the scaling transformations above remains a symmetry even away from DM-dominated regions. In this case, one needs to add the baryonic density to the right-hand side of Poisson's equation \eqref{Poisson 2}. Our equations now remain invariant only if $\rho$ and $\rho_{\rm b}$ scale in the same way, which occurs for the particular value $z=1/2$. Incidentally, for this value the gravitational acceleration $\vec g$ is an invariant quantity. It would be very interesting to test the observational consequences of this symmetry. We should keep in mind though that the Jeans and heat equations for the baryonic component may break this symmetry explicitly. Conceivably, this breaking might play a larger role when the baryonic fraction is larger.

\subsection{Approximate analytic solution with spherical approximation}
\label{sec: Newton vs MOND}

To proceed analytically, it is necessary to make further simplifications. We will assume spherical symmetry, replacing the disk-like baryon distribution with an effective spherical distribution that reproduces the enclosed mass along the equatorial plane, keeping in mind that this is an effective method to derive results for disks, as the Knudsen regime does apply to flattened disks and not necessarily to this approximate spherical system. This spherical approximation, both in Newtonian and MONDian gravity~\cite{Brada:1994pk}, has been shown to be accurate within 20\% in velocity.  

With spherical symmetry, the isotropic Jeans equation~\eqref{jeanscooling} becomes
\be
\frac{1}{\rho}\frac{{\rm d}}{{\rm d}r}\left(\rho v^2\right) = - \frac{G\big[M(r) +M_{\rm b}(r)\big]}{r^2} \,,
\label{poisson2}
\ee
where $M(r)$ and $M_{\rm b}(r)$ are respectively the DM and baryonic mass enclosed within a sphere of radius $r$.
Meanwhile, the heat equation~\eqref{heateqn2} reduces to\footnote{Proportionality factors between $\ell$ and $r$ that have a geometric origin can always be absorbed in a redefinition of $C$.}
\be
\frac{1}{r^2}\frac{{\rm d}}{{\rm d}r} \left(\rho v r^3 \frac{{\rm d}v^2}{{\rm d}r}\right) = C\,a_0 v\rho_{\rm b}\,.
\label{heatspherical}
\ee

In order to gain intuition on the parametric dependence of the solution, we will first solve these two equations analytically by
making simplifying approximations and dropping order-unity coefficients along the way.
This `intuitive' analytic derivation will be carried out both in the DM-dominated (`deep-MOND') regime and in the baryon-dominated (`Newtonian') regime. We will then come back and integrate the equations numerically in Sec.~\ref{exactnumerics} for a simplified exponential profile for baryons. 

Starting with the more `intuitive' approach, we integrate the heat equation~\eqref{heatspherical} to obtain, up to an order-unity coefficient, 
\be
\rho \, r \frac{{\rm d}v^2}{{\rm d}r} \sim C\,a_0 \frac{M_{\rm b}(r)}{4\pi r^2}\,.
\label{Fourier}
\ee
This is our version of Fourier's law. This result is valid, for instance, if $\rho$ and $v$ are power-laws in~$r$. 
Next we must solve the Jeans equation~\eqref{poisson2}. We will consider separately the DM-dominated (`deep-MOND') and baryon-dominated (`Newtonian') regimes.\\

\noindent {\bf $\bullet$ DM-dominated regime:} Consider the region where the gravitational potential is dominated by DM, {\it i.e.}, $M(r)\gg M_{\rm b}(r)$. 
Assuming that $\rho$ and $v^2$ are power-laws in $r$, then Eq.~\eqref{poisson2} implies
\begin{equation}
	\rho (r) \sim \frac{v^2(r)}{2 \pi G r^2} \,.
\label{virialbis}
\end{equation}
Substituting this density profile into Fourier's law~\eqref{Fourier}, we find
\be
r \frac{{\rm d} v^4}{{\rm d}r} \sim C a_0 GM_{\rm b}(r) \,.
\label{MDAR1}
\ee
Considering $M_{\rm b}$ as a power-law, and up to a mild log-dependence when the power-index approaches zero, the solution to this equation is
\be
v^4(r) \sim C a_0 GM_{\rm b}(r) \,.
\label{BTFR from heat}
\ee

To compare this result to the BTFR, we need to translate the velocity dispersion into a circular velocity. In the flat part of rotation curves, where $M_{\rm b}(r)$ rises very slowly and can effectively be taken as constant, we can use
the isothermal relation $v^2 = V^2 /2$ to obtain
\be
V^4_{\rm flat} \simeq 4 C a_0 GM_{\rm b}\,.
\label{BTFRunnormalized}
\ee
This matches the normalization of the BTFR provided that $C \sim 1/4$. However this value cannot be taken too seriously, since we have dropped various $\mathcal{O} (1)$ factors along the way, especially within the logarithmic factor. A numerical integration of the equations in Sec.~\ref{exactnumerics} will show that a more accurate value is in fact $C \simeq 1/16$. 

It is worth emphasizing that, unlike MOND, our mechanism does not predict exactly flat rotation curves asymptotically, because of the logarithm that was dropped in passing from Eq.~\eqref{MDAR1} to Eq.~\eqref{BTFR from heat}. To shed light on this issue, we can solve analytically for the actual functional dependence of the velocity dispersion asymptotically far from baryons. However, note that this is also an idealization, since our assumption $t_{\rm relax} \sim t_{\rm dyn}$ is only valid a few scale lengths away from the galactic center. This criterion eventually breaks down sufficiently far away where DM particles take on average much longer than a dynamical time to find the disk, and should follow a classical NFW profile~\cite{NFW} with a {\it declining} rotation curve. We will revisit the matching to NFW in Sec.~\ref{NFWmatch}.

Ignoring this issue for the moment, and considering the asymptotic behavior of our equations, we can safely set $\rho_{\rm b} \simeq 0$ in the outskirts. In this case, an approximate solution to Eqs.~\eqref{poisson2} and~\eqref{heatspherical} is 
\be
v(r)\simeq v_\infty \log^{1/5}\frac{r}{r_0}\,; \qquad  \qquad \quad  \rho(r) \simeq \frac{v^2(r)}{2\pi G r^2}\,,
\label{asymptotic}
\ee
where $r_0$ is an arbitrary scale. This solution is valid only to leading order in the large logarithm $\log \frac{r}{r_0} \gg 1$. In this regime, the logarithm is approximately constant, and the DM density profile
is approximately isothermal. More generally, beyond the large-log approximation $v(r)$ is expected to be a function of logarithms and therefore slowly-varying. All this will be addressed more quantitatively in  Sec.~\ref{exactnumerics}.

The BTFR in Eq.~\eqref{BTFRunnormalized} relates the asymptotic circular velocity to the total baryonic mass. Perhaps the main result of this paper is that, if the normalization of Eq.~\eqref{BTFRunnormalized} through  the appropriate value of $C$ gives the BTFR at large radii, then in the DM-dominated regime Eq.~\eqref{BTFR from heat} {\it implies the MDAR at every point of the rotation curve}:
\be
\frac{V^4(r)}{r^2} \sim  \frac{a_0 V_{\rm b}^2(r)}{r} \,.
\label{MDAR2}
\ee
Equivalently, in term of accelerations,
\be
\frac{g(r)}{g_{\rm b}(r)} \sim \sqrt{\frac{a_0}{g_{\rm b}(r)}} \,,
\label{MONDlaw}
\ee
which is also the deep-MOND force law~\cite{Milgrom:1983ca,Milgrom:1983pn}. Incidentally, since we have assumed that DM dominates the gravitational potential, self-consistency of this approximation requires $g^2 \simeq a_0 g_{\rm b} < a_0g$, that is, $g < a_0$. This confirms that DM domination coincides with the low acceleration or `deep-MOND' regime (although it should be obvious that there is no modification of dynamics at all in our case). Of course there exist systems such as galaxy clusters that display a non-MOND behavior even though their gravitational potential is mostly due to DM. However, we will argue in Sec.~\ref{sec:galaxy clusters} that such systems have not yet reached equilibrium because their relaxation time is larger than $H_0^{-1}$.  As such, our result~\eqref{BTFR from heat} based on considerations at equilibrium does not apply. \\

\noindent {\bf $\bullet$ Baryon-dominated regime:} Let us now consider regions where the acceleration is dominated by baryons, {\it i.e.}, $M_{\rm b}(r)\gg M(r)$. Assuming once again that $\rho$ and $v^2$ are power-laws in $r$, which in turn requires $M_{\rm b}(r)$ to be a power-law, Eq.~\eqref{poisson2} implies
\begin{equation}
v^2(r) \sim \frac{G M_{\rm b}(r)}{r} \,.
 \label{virial theorem Newtonian}
\end{equation}
Substituting this result into Fourier's law~\eqref{Fourier}, we can express the DM density profile in terms of $v^2(r)$:
\be
\rho(r) \sim \frac{C\,a_0}{4 \pi G\,r} \,\frac{{\rm d} \log r}{{\rm d} \log v^2} \,.
\label{rho baryon domination}
\ee
To the extent that the logarithmic derivative on the right-hand side varies very slowly, we find that in the baryon-dominated regime the DM central surface density within radius $r$,\footnote{In Sec.~\ref{DMsurfacedensity} we will give a more precise definition of the DM central density as a column integral---see Eq.~\eqref{SigmaDMintegral}. The present estimate is closer to the definition of~\cite{Donato:2009ab}.} 
\be
\Sigma_{0}(r) \sim \rho(r)\, r \sim  \frac{a_0}{4 \pi G}\,,
\label{Sigma DM}
\ee
is approximately constant, in good agreement with the findings of~\cite{Donato:2009ab}. 

\begin{figure}
\centering
\includegraphics[width=3.2in]{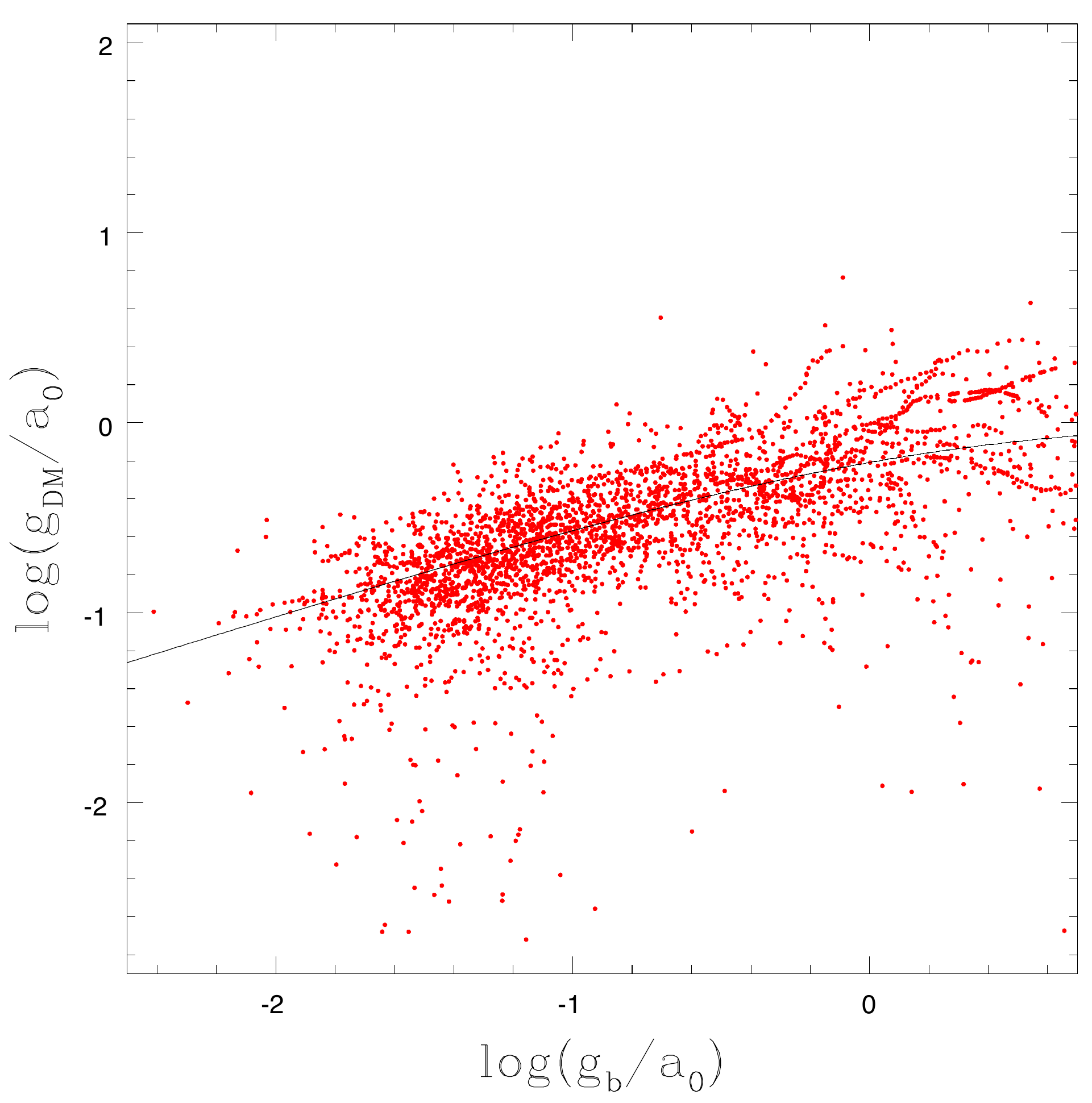}
\includegraphics[width=3.2in]{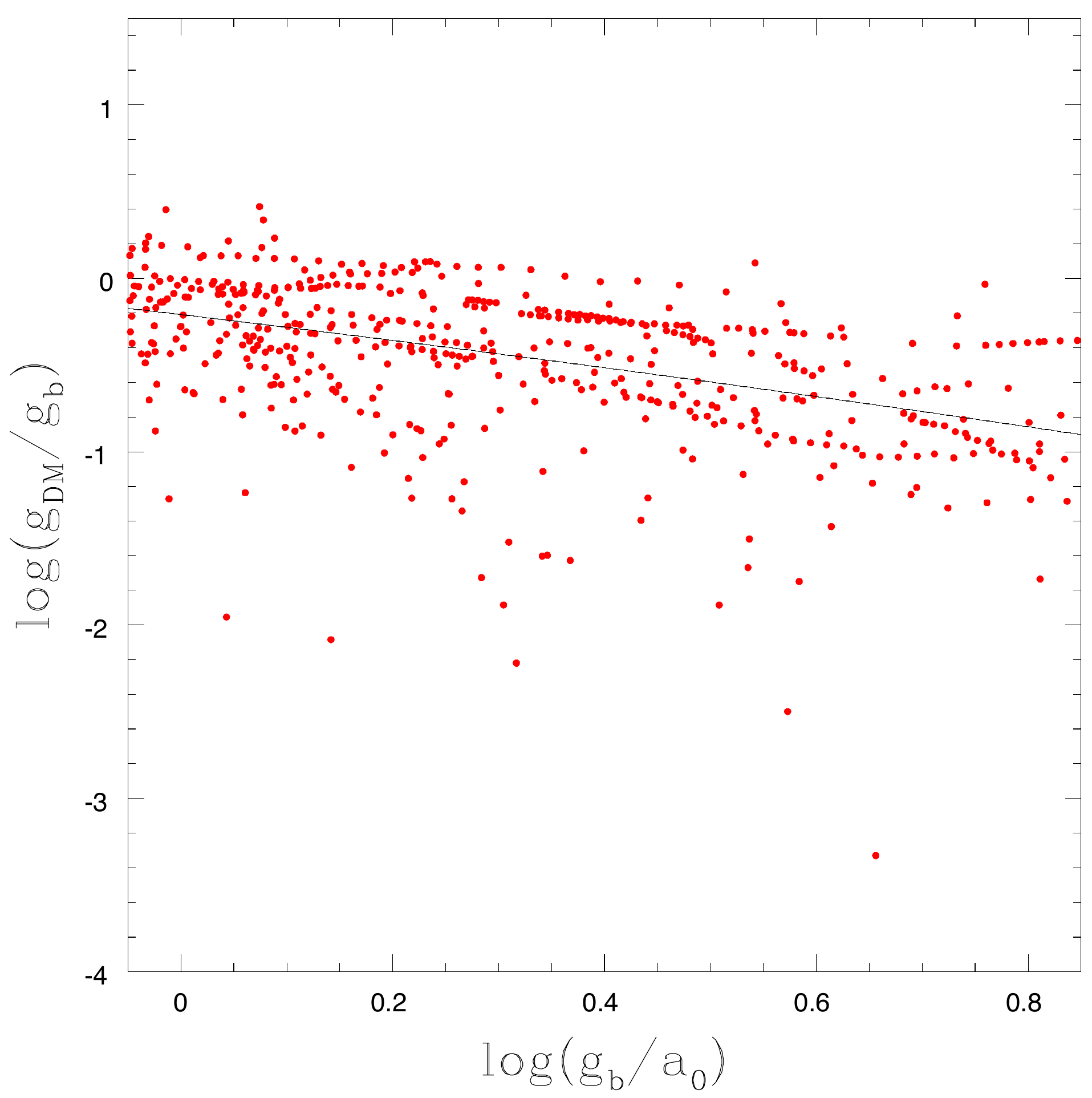}
\caption{\small {\it Left Panel}: Red points are the measured DM acceleration versus the baryonic one from~\cite{Lelli:2016zqa}, while the black line displays the prediction of the `simple' interpolating function. Note that, contrary to \cite{McGaugh:2016leg}, no quality criterion was applied. Even though the data become sparse in the high acceleration regime, it is clear that the `simple' interpolating function gives an overall good description of the scaling relation. {\it Right Panel:} The ratio $g_{\rm DM}/g_{\rm b}$ in the high acceleration regime (red points), together with the prediction from the `simple' interpolating function (black line), which coincides with our prediction of Eq.~\eqref{simplehighacc}. Note that the reported scatter on the MDAR should be taken with some caution given that each point contributing to the distribution is obviously not independent of the others on the above figures,  and because the radial sampling of each galaxy contributing to the relation is somewhat arbitary, typically set by the (angular) resolution of observations.} \label{sparcfig}
\end{figure}

The density profile~\eqref{Sigma DM} implies that DM contributes a constant acceleration,
\be
g_{\rm DM} \simeq a_0\,.
\label{simplehighacc}
\ee
In MOND terms, this matches the asymptotic behavior of the `simple' interpolating function, which is well known to provide good fits to galaxy rotation curves (see, {\it e.g.},~\cite{Famaey:2005fd,Gentile:2011}). We illustrate this in Fig.~\ref{sparcfig}, where we plot the DM acceleration versus the baryonic one from the sample of~\cite{Lelli:2016zqa}, as well as the ratio $g_{\rm DM}/g_{\rm b}$ in the high acceleration regime, compared to the `simple' interpolating function prediction. Incidentally, since we have assumed that baryons dominate the gravitational potential, self-consistency of this approximation requires $g_{\rm b} > g_{\rm DM} \simeq a_0$. This confirms that baryon domination coincides with the high acceleration or `Newtonian' regime.

\subsection{Consistency check: hierarchy of time scales}
\label{timescalechecks}

In order for our approach to be self-consistent, we should check that the assumed hierarchy of time scales, $\tau_{\rm halo} > t_{\rm dyn} > \tau_{\rm disk}$, is indeed satisfied. Assuming that $\tau_{\rm halo}$ and $\tau_{\rm disk}$ are both determined by interactions with baryons, we can use Eq.~\eqref{tautdynratio} to estimate the ratio away from the disk ($\tau_{\rm halo}/t_{\rm dyn}$) and inside the disk ($\tau_{\rm disk}/t_{\rm dyn}$).  

It is straightforward to argue that the first inequality, $\tau_{\rm halo} > t_{\rm dyn}$, is satisfied. Indeed, away from the disk the ratio $\rho / \rho_{\rm b}$ rises to typical values $\gtrsim \mathcal{O} (10^3)$. Using the fact that $C \sim 10^{-1}$, and that the MDAR is known empirically to be satisfied for $\frac{v^2}{a_0 r}\sim \frac{g}{a_0} \;\gsim\; 10^{-2}$, it follows that $\tau_{\rm halo} > t_{\rm dyn}$.

As for the second inequality, $\tau_{\rm disk} < t_{\rm dyn}$, we can show that it is true irrespective of whether DM or baryons dominate the gravitational potential. Consider first the DM-dominated regime. Substituting the density profile~\eqref{virialbis} in Eq.~\eqref{tautdynratio}, we find
\be
\frac{\tau_{\rm disk}}{t_{\rm dyn}} \sim \frac{3v^4(r)}{C a_0 G 4\pi r^3 \rho_{\rm b}} \simeq \frac{M_{\rm b}(r)}{(4/3)\pi r^3 \rho_{\rm b}}  \qquad (\text{DM-dominated}) \,,
\label{tauineqDM}
\ee
where in the next-to-last step we have used Eq.~\eqref{BTFR from heat}. Since, in the flattened disk of a real spiral galaxy, locally $\rho_{\rm b disk} \gg \rho_{\rm b \, sphericized}$, we have $\tau_{\rm disk} < t_{\rm dyn}$.

Next, let us consider the baryon-dominated regime. Using now the density profile~\eqref{rho baryon domination}, we obtain
\be
\frac{\tau_{\rm disk}}{t_{\rm dyn}} \sim \frac{3v^2}{G 8\pi r^2 \rho_{\rm b}} \simeq \frac{M_{\rm b}(r)}{(8/3)\pi r^3 \rho_{\rm b}}  \qquad (\text{baryon-dominated}) \,,
\label{tauineqbaryon}
\ee
where in the second step we have used the relation~\eqref{virial theorem Newtonian}. Again, since in the flattened disk of a real spiral galaxy, we locally have that $\rho_{\rm b disk} \gg \rho_{\rm b \, sphericized}$, we have $\tau_{\rm disk} < t_{\rm dyn}$. This estimate clearly relies on our assumption that most DM particles efficiently interact with baryons in the disk. We will revisit this assumption briefly in Sec.~\ref{sec: gas vs stars}.

To summarize the lessons of the `intuitive' derivation thus far, we have seen that the acceleration assumes a MONDian form $g \sim \sqrt{a_0g_{\rm b}}$ when DM dominates, and a Newtonian form $g \simeq g_{\rm b}$ when baryons dominate. According to the standard MOND paradigm, the transition between these two regimes is controlled by the value of the baryon's acceleration: systems are supposed to follow Newton's law for $g_{\rm b} \gg a_0$, and MOND's law for $g_{\rm b} \ll a_0$. We have argued though that these two seemingly different criteria are to a large extent equivalent to each other for practical purposes within our framework.

\section{Rotationally-Supported Systems: Numerical Results}
\label{exactnumerics} 

We now turn to the numerical integration of Eqs.~\eqref{poisson2} and~\eqref{heatspherical}. For this purpose, and to simplify the numerical treatment, we will again assume a sphericized exponential profile for the baryons,
\be
\rho_{\rm b}(r) = \frac{M_{\rm b}}{8\pi L^3}e^{-r/L}\,,
\label{rhobexp}
\ee
where $L$ is the characteristic scale-length of the baryon distribution (which in the astronomical literature is usually denoted as $h_R$ or $R_d$ within the disk), and $M_{\rm b}$ is the total baryonic mass. 
In future work, we plan to drop this spherical approximation for the baryons. Even more importantly, we plan to run simulations to follow the joint formation of DM halos and baryonic disks in a cosmological framework with our prescribed DM-baryon interaction cross section and energy exchange, in order to check that equilibrium configurations can actually be reached in nature. In the following, we concentrate only on equilibrium configurations within the above simplified distribution. Again, we stress that the Knudsen regime applies to flattened disks, and not necessarily to the sphericized distribution above, which is only used for convenience in solving the equations in this first exploratory paper.

\subsection{Numerical solution for exponential baryonic profile}

For the purpose of numerical integration, it is convenient to introduce the dimensionless radial coordinates
\be \label{x def}
x \equiv \frac{r}{L}\,,
\ee
as well as the dimensionless variables
\bea
\nonumber
& \hat{v}^2(x) \equiv \frac{v^2(r)}{\sqrt{C a_0G M_{\rm b}}}\,;\qquad \hat{q}(x)\equiv \frac{4\pi G \rho(r) r^2}{\sqrt{C a_0G M_{\rm b}}}\,;\qquad \hat{M}(x) \equiv \varepsilon \frac{M(r)}{M_{\rm b}}\,; & \\
& \displaystyle \hat{M}_{\rm b}(x) \equiv \frac{M_{\rm b}(r)}{M_{\rm b}} = 1 - \left(1 + \frac{x}{2}(x+2)\right)e^{-x} \,. & \label{dimensionless variables}
\eea
In defining $\hat M(x)$, we have introduced the dimensionless parameter
\be
{\varepsilon} \equiv \sqrt{\frac{G M_{\rm b}}{Ca_0L^2}}\,,
\label{epsilondef}
\ee
which characterizes whether the stellar distribution is typically in the DM-dominated/deep-MOND regime ($\varepsilon \ll 1$) or baryon-dominated/Newtonian regime ($\varepsilon \gg 1$). Typical HSB galaxies correspond to $\varepsilon \;\gsim\; 1$, whereas LSB galaxies correspond to $\varepsilon \;\lsim\; 1$. 

In terms of these variables, our equations reduce to the simple form
\begin{align}
\nonumber
\frac{x^2}{\hat{q}} \frac{{\rm d}}{{\rm d}x}\left(\frac{\hat{q}\hat{v}^2}{x^2}\right) &= - \frac{\hat{M} + \varepsilon \hat{M}_{\rm b}}{x^2}\,;\\
\frac{{\rm d}}{{\rm d}x}\left(\hat{q}\hat{v}^2 x \frac{{\rm d}\hat{v}}{{\rm d} x}\right) &= \frac{\hat{v}}{4} x^2e^{-x}\,;\\
\nonumber
\frac{{\rm d}\hat{M}}{{\rm d}x} &= \hat{q}\,.
\end{align}
As boundary conditions we demand regularity at the origin, and thus look for solutions to the above equations that can be expressed as a Taylor series around $x=0$. This requirement leaves us with a two-parameter family of solutions, which reads
\bea
\nonumber
\hat{v}^2(x) &=& \hat{v}_0^2 \left(1+\frac{x}{3R\hat{v}_0^2} - \frac{x^2}{8R\hat{v}_0^2}\left[1-\frac{7}{18R\hat{v}_0^2}\right] +\ldots\right) \,;\\
\label{expansionnearx=0}
\hat{q}(x) &=& \frac{R}{2}x^2 \left(1- \frac{x}{3R\hat{v}_0^2} + \frac{x^2}{8R\hat{v}_0^2}\left[1+\frac{1}{2R\hat{v}_0^2} - \frac{2}{3}R(R+\varepsilon)\right]  +\ldots\right) \,;\\
\nonumber \hat{M} (x) &=& \frac{R}{6}x^3 + \ldots
\eea
Note that regularity does not require ${\rm d}\hat{v}/{\rm d}x\rightarrow 0$ at the origin; what is constrained to vanish is instead the `heat current' $(\hat{q}\hat{v}^2/x) {\rm d}\hat{v}/{\rm d}x$.
The two free parameters in Eq.~\eqref{expansionnearx=0} have a simple physical interpretation: $\hat{v}_0$ is the value of the velocity dispersion at the origin, while $R$ fixes the
ratio between DM and baryon densities at the origin,
\be
\frac{R}{\varepsilon} =  \left. \frac{M(r)}{M_{\rm b}(r)}\right\vert_{r\rightarrow 0} = \left. \frac{\rho(r)}{\rho_{\rm b}(r)}\right\vert_{r\rightarrow 0}\,.
\label{density ratio}
\ee
Both $\rho$ and $\rho_{\rm b}$ are finite at the origin, corresponding to a DM core.
Once a solution for $\hat{v}$, $\hat{q}$ and $\hat{M}$ is found, we translate back to the physical variables $v$, $q$ and $M$.

\begin{figure}[htp]
\centering
\begin{tabular}{c c}
       \addheight{\includegraphics[width=3in]{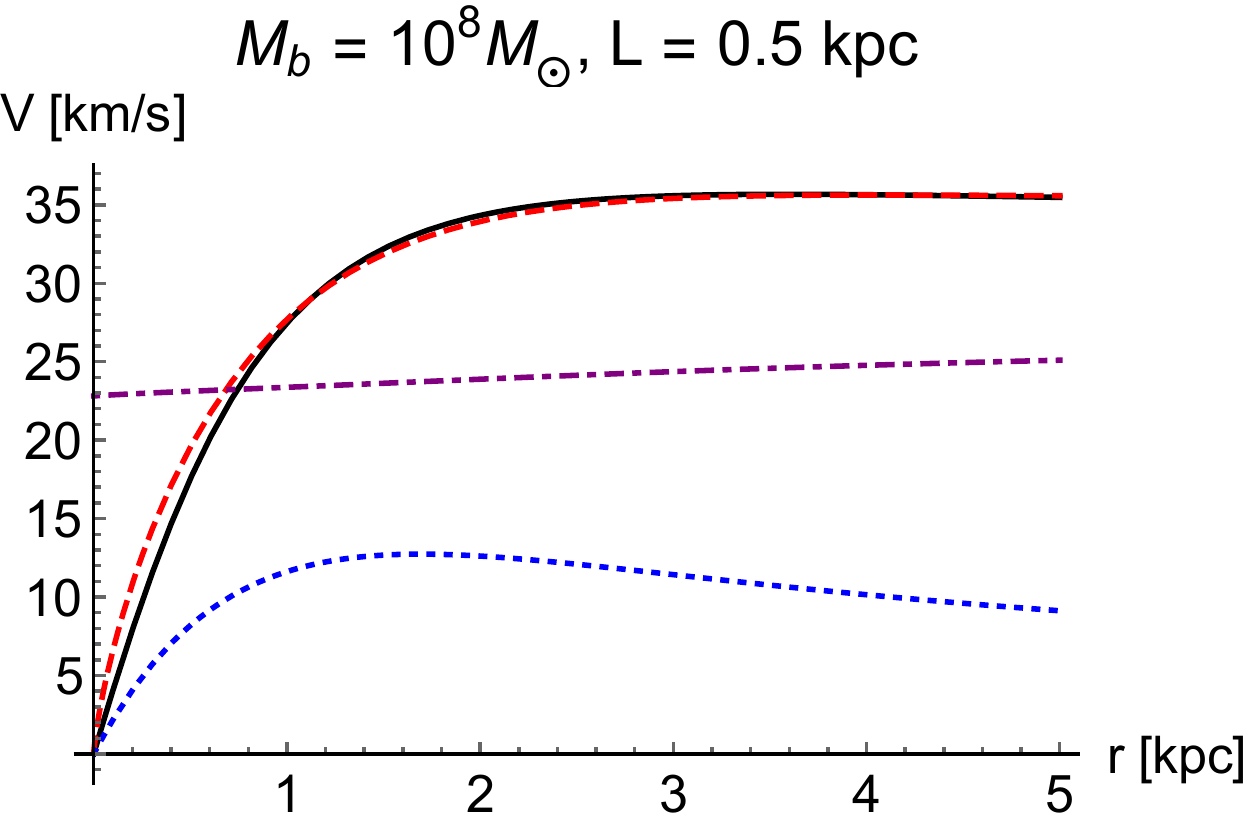}} &
      \addheight{\includegraphics[width=3in]{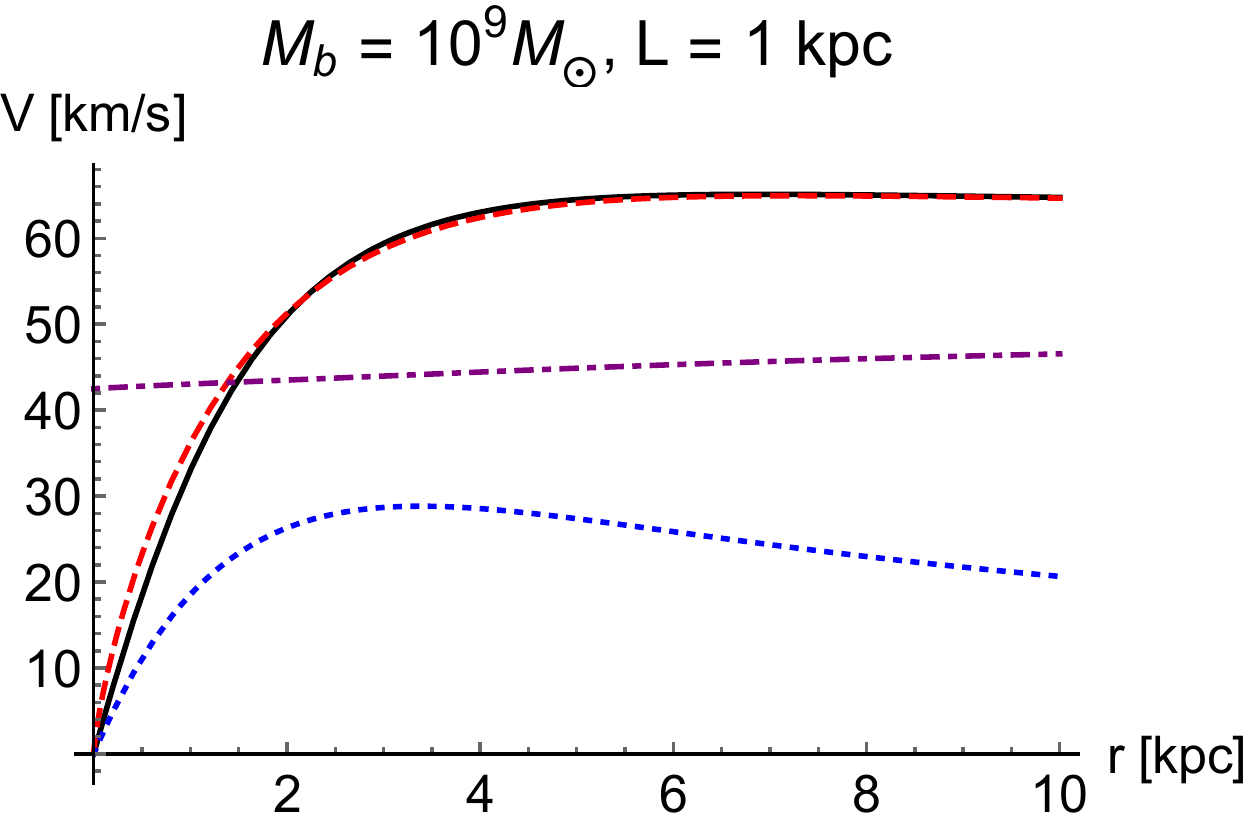}} \\
            \addheight{\includegraphics[width=3in]{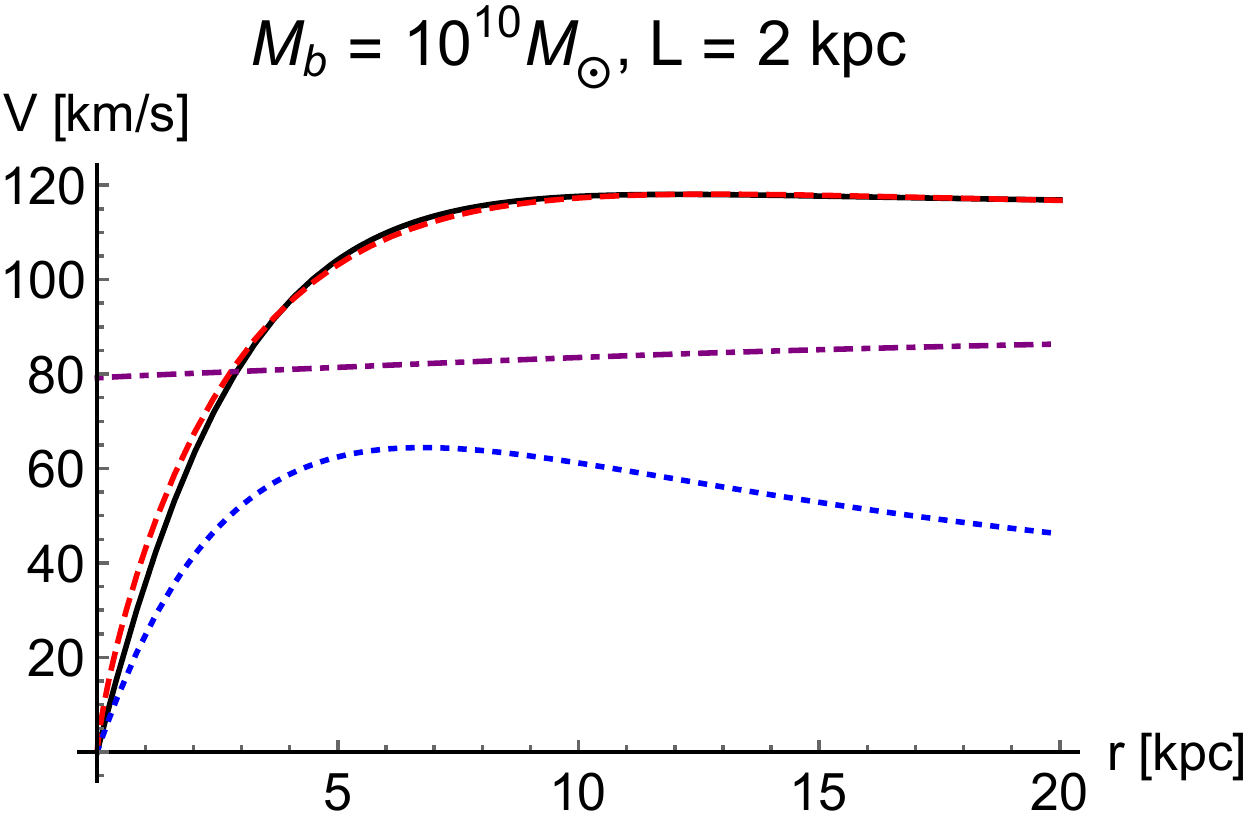}} & \\ 
            \end{tabular}
      \caption{\small Rotation curves for three sample galaxies, assuming a spherical exponential profile $\rho_{\rm b} \sim \exp(-r/L)$. The curves denote our fit (solid black), MOND with simple interpolating function (dashed red), and Newtonian gravity without DM (dotted blue). In all cases we have fixed $C = 1/16$. The values of $R$ and $\hat{v}_0^2$ are listed in Table~\ref{parameters}. The purple dash-dotted curve is the DM velocity dispersion, which is nearly constant everywhere.}
\label{rotationcurves}
\end{figure}

\begin{table}
\centering
\begin{tabular}{|c|c|c|c|c|c|}
      \hline
$M_{\rm b}[{\rm M}_\odot]$ & $L[{\rm kpc}]$ & $\varepsilon$ & $R$ & $\hat{v}_0^2$ & $\frac{\rho}{\rho_{\rm b}}\vert_{r = 0}$ \\
     \hline
$10^8 $ & $0.5$ & 2.7 & 5.9 & 1.7 & 2.2  \\
$10^9 $ & $1$  & 4.3 & 5.7 & 1.8 & 1.3  \\
$10^{10} $ & $2$ & 6.8 & 5.5 & 2.0 & 0.8  \\
     \hline
\end{tabular}
      \caption{\small Parameter values for the rotation curves shown in Fig.~\ref{rotationcurves}. The parameters $R$ and $\hat{v}_0^2$ were adjusted to obtain a good fit in each case. The parameter $\varepsilon$ is calculated from Eq.~\eqref{epsilondef}, while the density ratio $\rho/\rho_{\rm b}$ at $r= 0$ follows from Eq.~\eqref{density ratio}.}
      \label{parameters}
      \end{table} 

Since we have two free parameters ($R$ and $\hat{v}_0^2$) to find an equilibrium solution, one might argue that little has been achieved. However, as stated in the previous section, perhaps the main result of this paper is that once the BTFR is reached at large radii, then our framework imposes the MDAR at all radii. We thus choose to adjust these two parameters by $i)$ normalizing the asymptotic circular velocity at a fixed distance, $r = 10\,L$; and $ii)$ ensuring that the rotation is flat at that radial distance. By proceeding in this way, we have found that $C = 1/16$ generally allows for an overall reasonable fit. Fixing $C$ to this value, we then proceeded to adjust the parameters $R$ and $\hat{v}_0^2$ for various different cases.
 
Figure~\ref{rotationcurves} shows the rotation curves for three sample galaxies, spanning a range of typical values for $M_{\rm b}$ and $L$. Our model prediction is the solid black curve, the MOND prediction assuming the `simple' interpolating function is the red dashed curve, and the Newtonian baryons-only (no DM) is the blue dotted curve. The difference between our prediction for $V^4$ and the MOND result is in all cases less than 10\% over the entire range. The fitted values of $R$ and $\hat{v}_0^2$, and the physical parameters that derive from these, are listed in Table~\ref{parameters}. These plots confirm that our mechanism based on Newtonian gravity and DM-baryon particle interactions can reproduce exquisitely well the MOND phenomenology in disk galaxies, and therefore the MDAR.

\begin{figure}[t]
	\centering
    \includegraphics[width=3in]{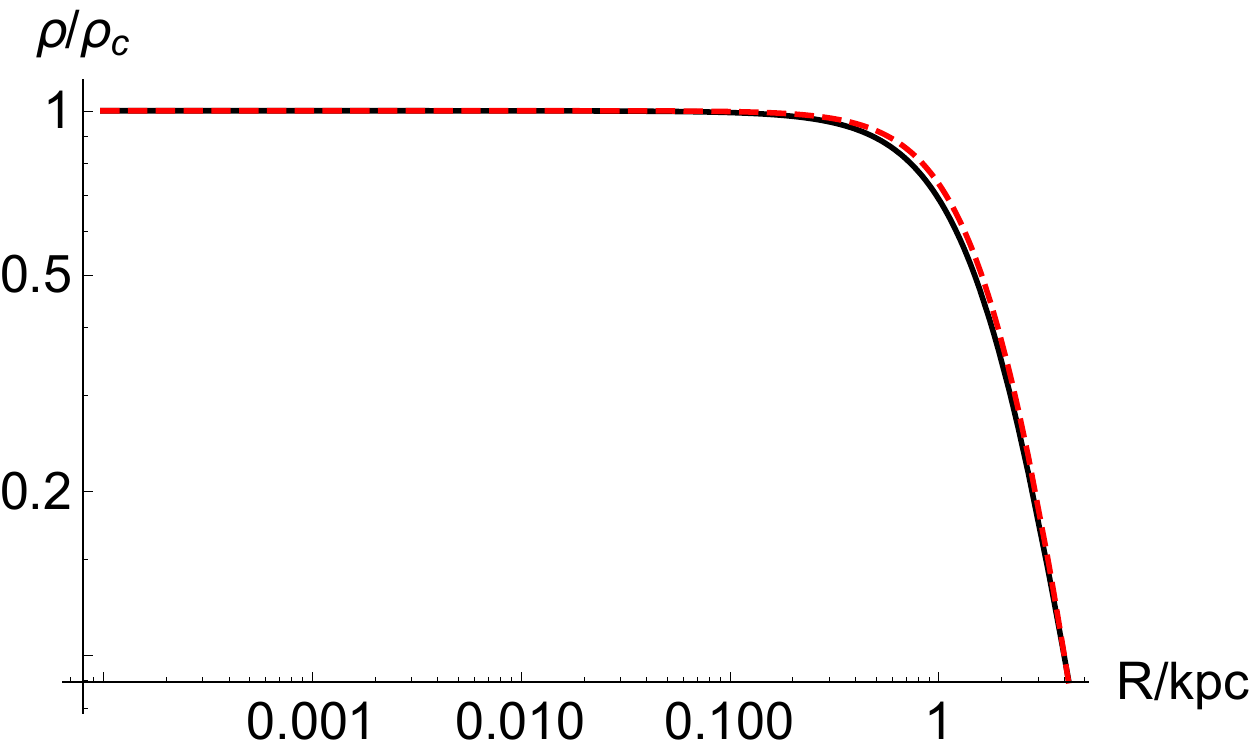}
    \caption{\small Comparison of our DM density profile (solid black) for the $M_{\rm b} = 10^9\; {\rm M}_\odot$, $L = 1$~kpc galaxy with a modified Hubble profile (Eq.~\eqref{modHubble}) with $r_{\rm core} \simeq 2.1~{\rm kpc}$ (dashed red), which in turn is a good approximation to a cored isothermal sphere within $2\,r_{\rm core}$~\cite{BinneyTremaine}.}
    \label{modifiedHubble} 
\end{figure}

The DM velocity dispersion, plotted as the purple dash-dotted curve, is nearly constant everywhere, with a small (almost linear) slope. This confirms that truncating Fourier's law at lowest order in the derivative expansion, as done in Eq.~\eqref{Fourierlaw}, is a self-consistent approximation.
Indeed, our DM profile is well approximated by a cored isothermal sphere~(see Sec.~4.3.3 of~\cite{BinneyTremaine}). Figure~\ref{modifiedHubble} compares the DM
profile for $M_{\rm b} = 10^9\;{\rm M}_\odot$,  $L = 1$~kpc with a modified Hubble profile
\be
\rho_{\rm Hubble}(r) = \frac{\rho_{\rm core}}{\left[1+\left(r/r_{\rm core}\right)^2\right]^{3/2}}\,. 
\label{modHubble}
\ee
The latter is well-known to approximate a cored isothermal sphere within $2\,r_{\rm core}$~\cite{BinneyTremaine}. To determine $r_{\rm core}$, we used the fact that for
a cored isothermal profile the velocity dispersion and the circular velocity cross each other at $r_{\rm core}/\sqrt{2}$. From the numerical solution, we thus obtained $r_{\rm core} \simeq 2.1~{\rm kpc}$.

\begin{figure}[t!]
\centering
\begin{tabular}{c c}
       \addheight{\includegraphics[width=3in]{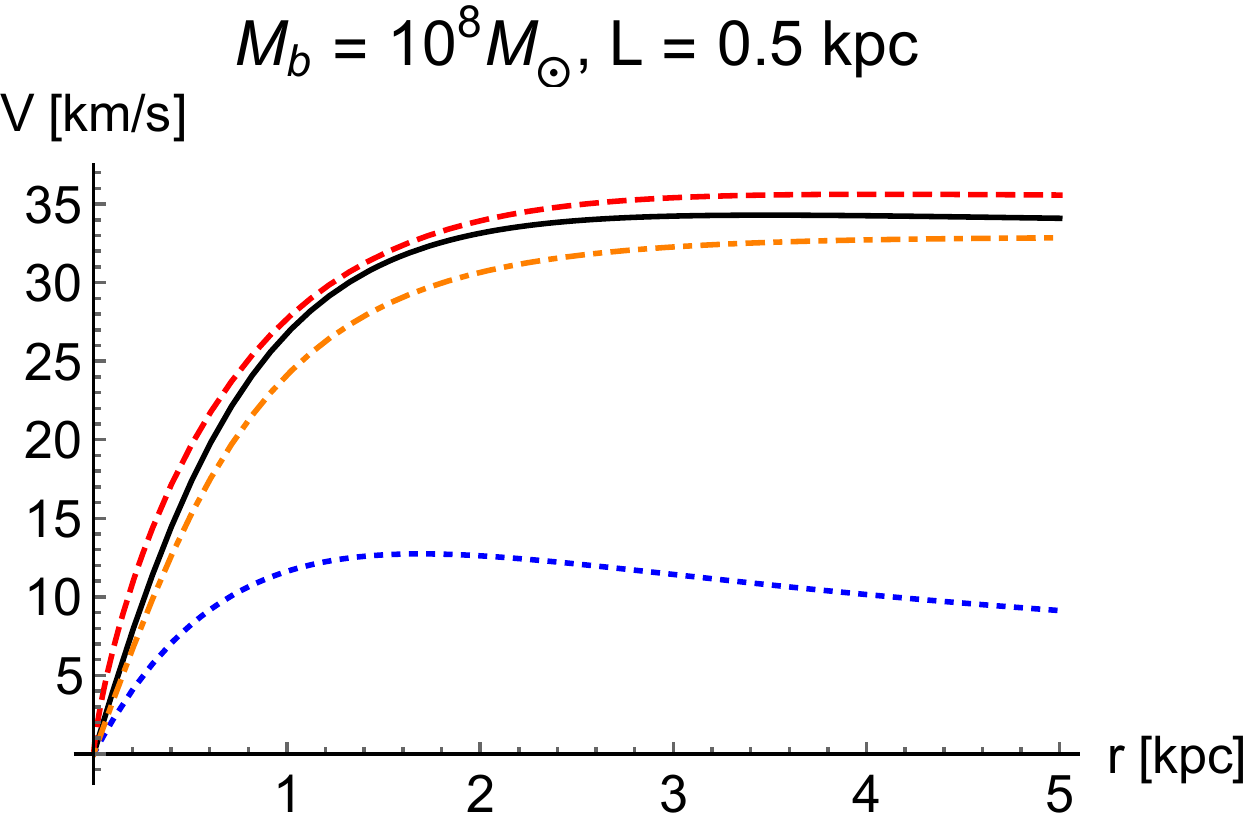}} &
      \addheight{\includegraphics[width=3in]{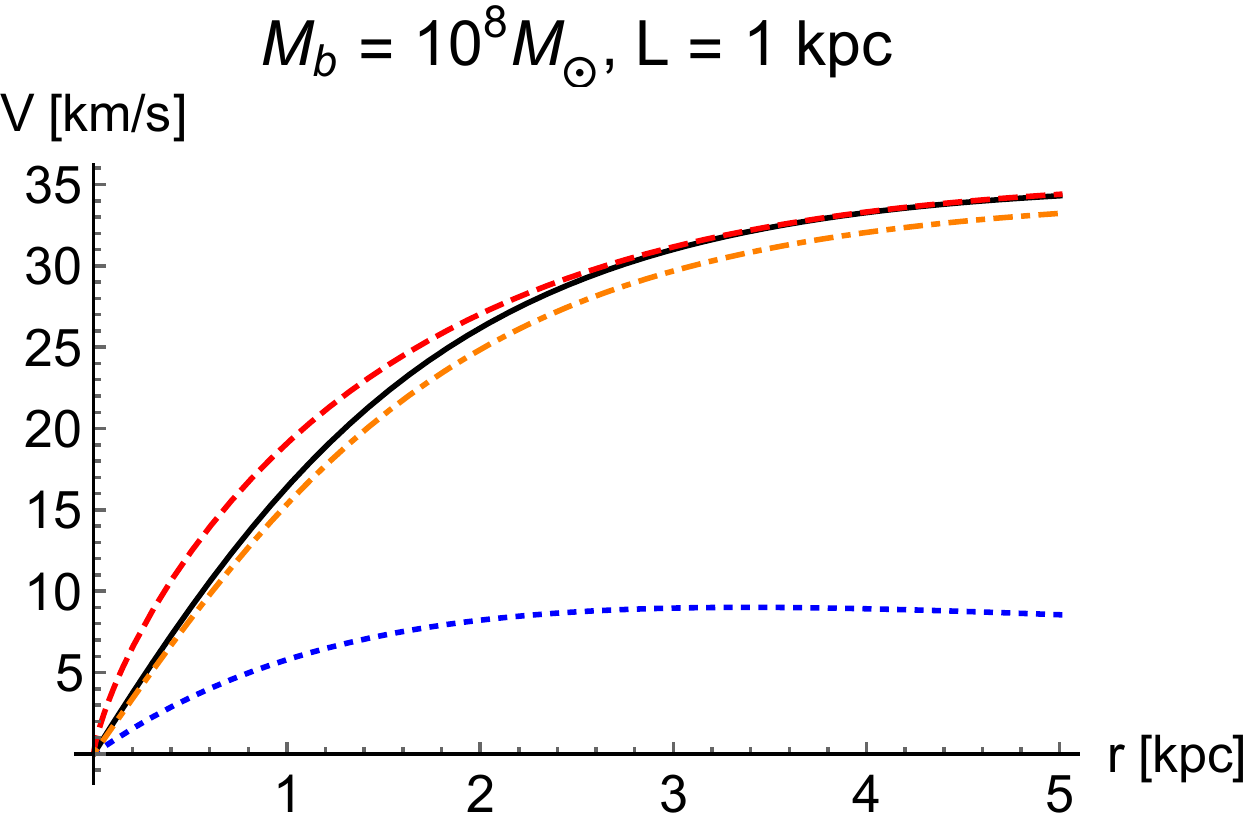}} \\ 
            \addheight{\includegraphics[width=3in]{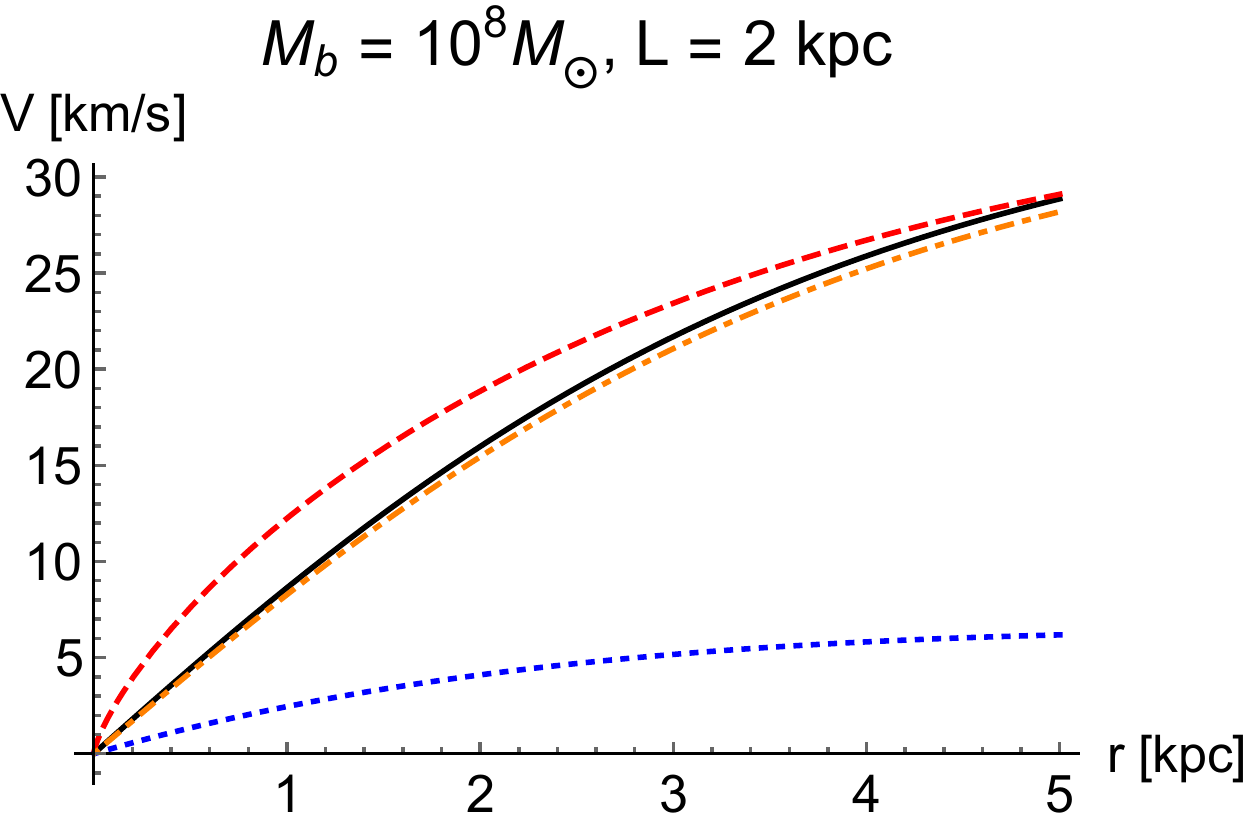}} & \\ 
            \end{tabular}
\caption{\small Rotation curves for $M_{\rm b} = 10^8 {\rm M}_\odot$ with scale lengths $L= 0.5$, 1 and 2~kpc. The curves denote the same components as in Fig.~\ref{rotationcurves}, with the addition of the DM-only contribution (orange dash-dotted). The parameters $R$ and $\hat{v}_0^2$ were adjusted to obtain a fit for the galaxy with largest scale length
      ($L = 2~{\rm kpc}$) only. The rotation curves for $L = 0.5$ and 1~kpc were then predicted by keeping $v_0^2$ fixed and scaling the central density ratio as
      $\frac{\rho}{\rho_{\rm b}} \rightarrow \frac{L}{2\,{\rm kpc}}\frac{\rho}{\rho_{\rm b}}$, in accordance with the scaling argument of Sec.~\ref{scaleinvarianceDM}. To highlight the `diversity' of
      predicted rotation curves, we have kept the range of $r$ fixed to 5~{\rm kpc} in all cases.}
\label{scalelengths}
      \end{figure}

Figure~\ref{scalelengths} shows instead the rotation curves for $M_{\rm b} = 10^8 {\rm M}_\odot$ with varying scale lengths ($L= 0.5$, 1 and 2~kpc). The curves denote the same components as before, with the addition of the DM-only contribution as the orange dash-dotted curve. In this case, we have adjusted $R$ and $\hat{v}_0^2$ to obtain a fit {\it for the largest scale length ($L = 2~{\rm kpc}$) only.} This galaxy has the lowest surface brightness and hence is the most DM-dominated in the center. In this regime of low surface brightness/high DM-baryon ratio, we expect the scaling symmetry of Sec.~\ref{scaleinvarianceDM} to be applicable. In accordance with the scaling argument for the particular value $z=1$, the rotation curves for $L = 0.5$ and 1~kpc were predicted by keeping $v_0^2$ fixed and scaling the central density ratio as $\frac{\rho}{\rho_{\rm b}} \rightarrow \frac{L}{2\,{\rm kpc}}\frac{\rho}{\rho_{\rm b}}$. 

To highlight the diversity of predicted rotation curves, however, we have kept the total range of 5~kpc fixed. It is clear that the rotation curves of small galaxies with small scale-lengths rise quite sharply, while those with large scale-lengths are gently rising, contrary to the $\Lambda$CDM expectation. Our formalism thus solves, at least in principle, the diversity of rotation curves problem~\cite{Oman:2015xda}, and the observed correlation of their shape with the disk scale-length (or baryonic surface density). However, let us note that some galaxies that tend to deviate from the MDAR are usually attributed in the MOND context to non-circular motions and morphological or kinematic disturbances \citep[e.g.][]{Gentile:2011}. Such galaxies should be understood in {\it any} theoretical context, through a full modelling of their 2D velocity field and possibly out-of-equilibrium dynamics. The above statement on solving the diversity of rotation curves problem thus holds only provided that such galaxies can be understood in the future.

In the analytical estimates of Sec.~\ref{sec:rotational general}, we showed heuristically how the BTFR could be a natural outcome of our formalism. The numerical solutions derived above reveal that, in fact, the BTFR cannot be uniquely predicted within our equilibrium treatment. Given our heuristic arguments, it is however tempting to speculate that the BTFR might be predicted by going beyond the equilibrium approximation, {\it i.e.}, by studying the dynamical evolution towards equilibrium. This would of course require numerical simulations of galaxy formation within our scenario, which is beyond the scope of the present analysis. It is worth pointing out that, even in the pure $\Lambda$CDM context, the slope and normalization of the BTFR can be reproduced, as they are a natural consequence of abundance matching ({\it e.g.},~\cite{Desmond:2017}). Only the scatter is slightly too large, a problem that our DM-baryon interactions might be able to solve within simulations. However, a much more serious issue is the diversity of rotation curve shapes at a fixed DM halo mass scale~\cite{Oman:2015xda} with a central slope correlated to the {\it baryonic} surface density (see Fig.~15 of~\cite{Famaey:2011kh}), or in other words, a MDAR obeyed at all radii, with no noticeable residual with the actual radius~\cite{Lelli:2017vgz}. What we have shown quantitatively is that, once the BTFR is attained at large radii, the MDAR at all radii necessarily follows from our formalism, hence removing any need for fine-tuned feedback to explain the diversity of rotation curve shapes at a fixed mass scale and their uniformity at a fixed baryonic density scale.  However, as before, we caution that this statement of course holds only provided that our formalism could also explain galaxies which apparently deviate from the MDAR, perhaps due to non-circular motions and morphological or kinematic disturbances, which should be understood through a full modelling of their 2D velocity field and possibly out-of-equilibrium dynamics in any theoretical context.

\subsection{DM central surface density in spiral galaxies}
\label{DMsurfacedensity}

It is also instructive to consider the DM central surface density as defined in~\cite{Lelli:2016uea}. In the spherically-symmetric case of interest, it is given by
\be
\Sigma_{\rm DM} \equiv 2 \int_0^\infty {\rm d}r\rho(r)\,.
\label{SigmaDMintegral}
\ee
The same quantity can be defined for baryons, which for our sphericized exponential profile in Eq.~\eqref{rhobexp} can be calculated explicitly:
\be
\Sigma_{\rm b} \equiv 2 \int_0^\infty {\rm d}r\rho_{\rm b}(r) = \frac{M_{\rm b}}{4\pi L^2}\,.
\ee
This is simply the total baryonic mass divided by the area of a disk of radius $L$.

\begin{table}
\centering
\begin{tabular}{|c|c|c|c|c|c|}
      \hline
$M_{\rm b}[{\rm M}_\odot]$ & $L$[kpc]& $\Sigma_{\rm b}[{\rm M}_\odot/{\rm pc}^2]$ & $\Sigma_{\rm DM}[{\rm M}_\odot/{\rm pc}^2]$ & $\sqrt{4 \Sigma_{\rm b}\frac{a_0}{2\pi G}}\;[{\rm M}_\odot/{\rm pc}^2]$& $\log\left(\frac{\Sigma_0}{{\rm M}_\odot {\rm pc}^2}\right)$  \\
     \hline
$10^6 $ & 0.125 & 5 & 57 & 53 & 1.6  \\
 $10^7 $ & 0.25 & 13 & 88 & 83 & 1.8  \\
$10^8 $ & 0.5 & 32 & 137 & 132 & 1.9  \\
$10^9 $ & 1 & 80 & 209 & 209 & 2.1  \\
$10^{10} $ & 2 & 199 & 313 & 330 & 2.3  \\
$10^{11} $ & 4 & 498 & 449 & 552 & 2.5 \\
$10^{12} $ & 8 & 1244 & 590 & 826 & 2.6 \\
     \hline
\end{tabular}
      \caption{\small Central surface densities for the six model galaxies and an additional one with $M_{\rm b} = 10^{12}{\rm M}_\odot$. See main text for the definitions of the quantities listed.}
      \label{surfacedensity}
      \end{table} 

The values of $\Sigma_{\rm b}$ and $\Sigma_{\rm DM}$ are listed respectively in the third and fourth columns of Table~\ref{surfacedensity}. We have augmented our three model galaxies with four additional ones, with $M_{\rm b} = 10^6$, $10^7$, $10^{11}$ and $10^{12}~{\rm M}_\odot$ to more clearly illustrate the trend as a function of baryonic mass. As expected, the less massive galaxies ($M_{\rm b}\leq 10^{10}{\rm M}_\odot$) have $\Sigma_{\rm b} < \Sigma_{\rm DM}$,
whereas the more massive ones ($M_{\rm b} = 10^{11}, 10^{12}{\rm M}_\odot$) have $\Sigma_{\rm b} > \Sigma_{\rm DM}$.

In the fourth column we list the values of the deep-MOND dynamical central surface density~\cite{Milgrom:2016ogb}:
\be
\Sigma_{\rm deep-MOND} \equiv \sqrt{4 \Sigma_{\rm b}\frac{a_0}{2\pi G}}\,.
\label{SigdeepMOND}
\ee
Less massive galaxies ($M_{\rm b} \leq 10^{10}{\rm M}_\odot$), those with $\Sigma_{\rm b} < \Sigma_{\rm DM}$, display good
agreement between $\Sigma_{\rm DM}$ and $\Sigma_{\rm deep-MOND}$. This makes sense---such galaxies are DM-dominated, and 
correspondingly agree with the deep-MOND expression. This agrees well with the observational results of~\cite{Lelli:2016uea}. The relation
$\Sigma_{\rm DM}\simeq \Sigma_{\rm deep-MOND}$ breaks down for more massive, baryon-dominated galaxies ($M_{\rm b} = 10^{11}, 10^{12}{\rm M}_\odot$),
where the deep-MOND expression is not expected to hold. 

The fifth column features the central DM surface density $\Sigma_0$ studied by Donato {\it et al.}~\cite{Donato:2009ab}. Because these authors assume a
Burkert density profile~\cite{Burkert:1995yz}, their definition of central surface density is related to ours by~\cite{Milgrom:2016ogb} $\Sigma_0 = \frac{2}{\pi} \Sigma_{\rm DM}$.
Donato {\it et al.}~\cite{Donato:2009ab} found that $\Sigma_0$ was nearly constant across a broad range of galaxies, with best-fit value
\be
\log\left(\frac{\Sigma_0}{{\rm M}_\odot {\rm pc}^{-2}}\right) = 2.15\pm 0.2\,.
\ee
In our case we see that galaxies in the intermediate mass range $10^7 {\rm M}_\odot \leq M_{\rm b} \leq 10^{11} {\rm M}_\odot$ agree
with this result within $2\sigma$, but there is a clear trend of mildly increasing $\Sigma_0$ with increasing $M_{\rm b}$.
As mentioned above, in the range $M_{\rm b} \leq 10^{10}{\rm M}_\odot$ our results agree well with the deep-MOND
behavior given by Eq.~\eqref{SigdeepMOND}, which implies $\Sigma_0 \sim \sqrt{\Sigma_{\rm b}}$.

\subsection{Stability}
\label{stability}

A full-fledged dynamical treatment would be needed to assess whether the equilibrium solutions we have found can indeed be reached dynamically. This is clearly beyond the scope of the present paper. At this stage, we will content ourselves with performing a preliminary check of the stability of these solutions.

To this end, we consider spherically symmetric, localized perturbations of our dimensionless variables,
\begin{equation} \label{perturbations}
	\hat q (\hat t, x) = \hat q (x) + \delta \hat q (\hat t, x)\,; \qquad \quad \hat M (\hat t, x) = \hat M (x) + \delta \hat M (\hat t, x)\,; \qquad \quad \hat v (\hat t, x) = \hat v (x) + \delta \hat v (\hat t, x) \ , 
\end{equation}
and let them evolve numerically while keeping the density of baryons fixed. The variable $\hat t$ appearing in Eq.~\eqref{perturbations} is a dimensionless time coordinate defined by $\hat t \equiv t \, (C a_0 G M)^{1/4} / L$. For concreteness, we will consider the solution with $M = 10^9 {\rm M}_\odot$ and $L = 1$ kpc.

\begin{figure}[t]
\centering
\includegraphics[width=3.2in]{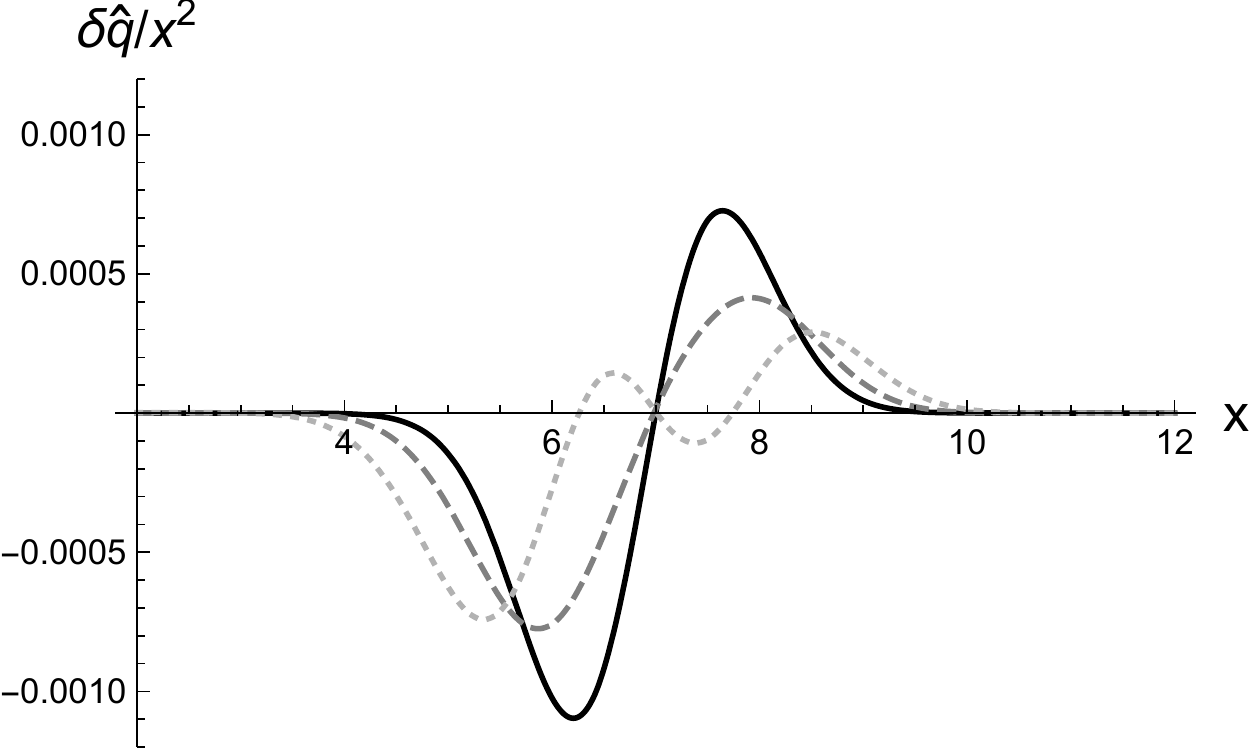}
\includegraphics[width=3.2in]{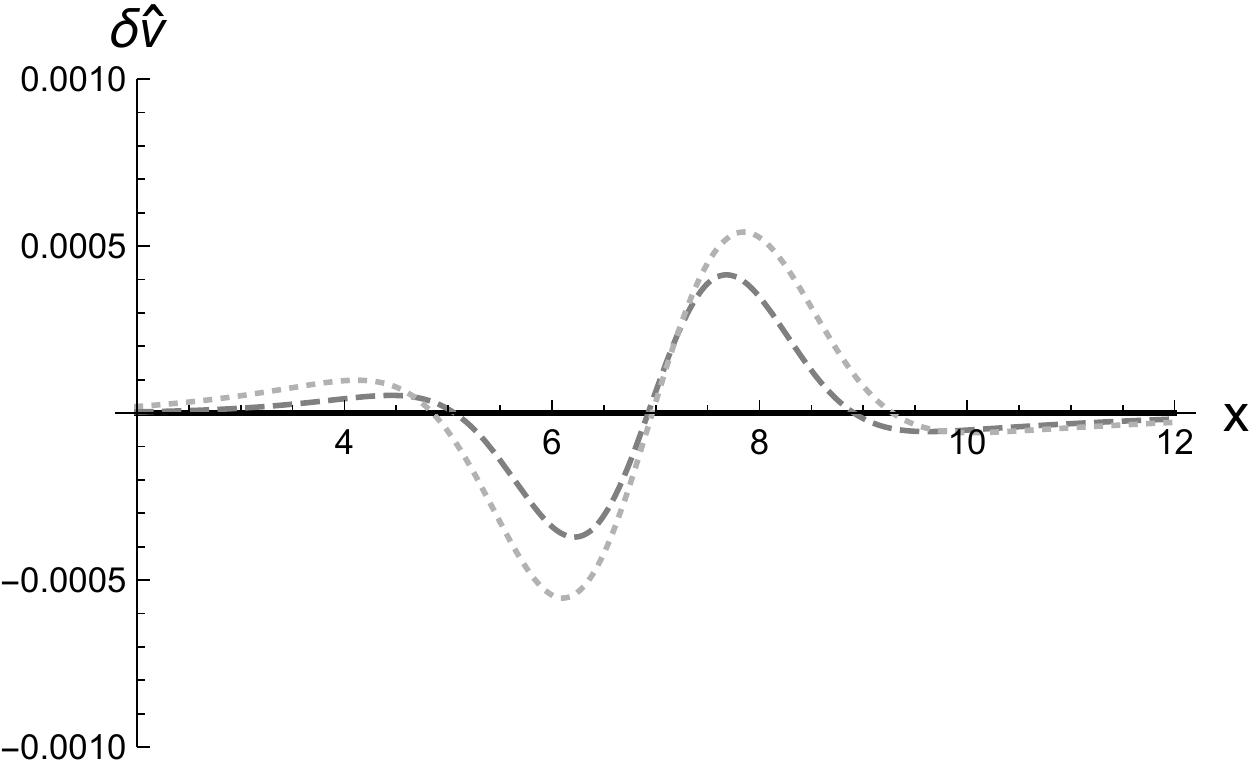}
\caption{\small {\it Left panel:} Evolution of a small DM density perturbation $\delta \hat q /x^2 = \delta \rho \, (4\pi G  L^2 / \sqrt{C a_0G M_{\rm b}})$ that preserves the total DM mass. {\it Right panel:} Evolution of the velocity dispersion perturbation $\delta \hat v = \delta v / (C a_0G M_{\rm b})^{1/4}$. In both cases, the solid black, dashed dark grey, and dotted light grey lines represent the profiles at times $\hat t = 0, \,   0.35, \, 0.6$ respectively. } \label{stability figure}
\end{figure}

Over time, these perturbations inevitably give rise to a bulk velocity perturbation in the radial direction, $\delta \hat u$, which therefore must be introduced for consistency. The linearized equations of motion for the perturbations $\delta \hat M, \delta \hat q, \delta \hat v$ and $\delta \hat u$ can be derived by rewriting Eqs.~\eqref{time dependent equations} in terms of the dimensionless variables introduced in Eqs.~\eqref{x def} and~\eqref{dimensionless variables} and expanding around our equilibrium solution. These partial differential equations should be supplemented with the constraint $ \partial_x \delta \hat M = \delta \hat q$ relating the perturbation of the total DM mass within a given radius with the DM density perturbation. The resulting set of partial differential equations is: 
\begin{subequations}
\begin{align}
& \partial_x \delta \hat{M} = \delta \hat{q}\ , \\
	& \partial_{\hat t} \delta \hat q + \partial_x (\hat q \delta u ) = 0 \ ,\\
	& \partial_{\hat t} \delta \hat u+ 2 \hat v \partial_x \delta \hat v +  \frac{\hat v^2 \partial_x \delta \hat q}{\hat q} + 2 \delta \hat v\left[ \hat v \left( \frac{\hat q'}{\hat q}- \frac{2}{x} \right) + \hat v' \right] - \frac{\hat v^2 \hat q'\delta \hat q}{\hat q^2} + \frac{\delta \hat M}{x^2} =0 \ ,\\
	& \partial_t \delta \hat v +\frac{\hat v \delta \hat q}{2 \hat q} - x \hat v \partial^2_x \delta \hat v- \partial_x \delta \hat v\left[ 4 x \hat v' + \frac {v (q + x q')}{q}\right] - \frac{x \hat v \hat v' \partial_x \delta \hat q}{\hat q} - \frac{\delta \hat v x(x e^{-x} - 8 \hat q \hat v^{\prime 2})}{4 \hat q \hat v} \\
	&\qquad \qquad \qquad \qquad \qquad \qquad \qquad \qquad \qquad \qquad \qquad \qquad \qquad  \nonumber -\frac{x \delta \hat q (x e^{-x}- 4 \hat v \hat q' \hat v')}{4 \hat q^2} = 0 \ .
\end{align}
\end{subequations}

Perhaps the most revealing evolution is that of density perturbations. In Fig. \ref{stability figure} we are showing a few snapshots of the initial time evolution of a localized density perturbation that preserves the total DM mass. We choose the other initial conditions to be $\delta \hat u = \delta \hat v =0$, making sure that the initial profile of $\delta \hat M$ is compatible with the constraint $\partial_x \delta \hat M = \delta \hat q$ at $\hat t = 0$. As we can see, the behavior appears to be damped and oscillatory, suggesting that our static solution should indeed be stable. 

We should however comment on the fact that $\delta \hat v$ initially grows negative (positive) in under-dense (over-dense) regions. This behavior is a direct consequence of the fact that the cooling rate is inversely proportional to the DM density, and is therefore larger (smaller) in under-dense (over-dense) regions. We expect this behavior to damp out as the density perturbation decays. Unfortunately, numerical issues prevent us at this stage from reliably following the time evolution of perturbations over much longer time scales to confirm this expectation.

\section{Matching to a NFW Profile}
\label{NFWmatch}

As mentioned earlier, our derivation of the MDAR in disk galaxies crucially relies on the fact that DM particles orbiting sufficiently close to the baryon distribution cross the disk on average once per dynamical time. Particles orbiting further away take on average much longer to find the galactic disk.  At sufficiently large distances, they experience less than one interaction per Hubble time $H_0^{-1}$, and therefore should behave to a good approximation like collisionless DM. We therefore expect our MONDian profile to interpolate to an NFW profile at large enough radii. In what follows we will crudely estimate the transition scale $r_*$ as the radius where the condition~\eqref{timehierarchy} $\tau_{\rm disk} < t_{\rm dyn}$ is saturated. A more realistic estimate would require investigating for reasonable distributions of orbits the radius at which DM particles interact with baryons once per Hubble time, including a more realistic geometry for the disk-halo configuration. We leave this to future work.

\subsection{Estimating the matching radius}

Following the above reasoning, the matching radius can be read off from Eq.~\eqref{tautdynratio} by demanding $\tau_{\rm disk} = t_{\rm dyn}$ at $r_*$:
\be
r_* \sim \frac{1}{C} \frac{v^2}{a_0} \frac{\rho}{\rho_{\rm b}} \,.
\ee
To proceed further, we will first assume, and then check {\it a posteriori}, that $r_*$ lies on the flat part of the rotation curve. In this region, the velocity dispersion follows the
BTFR,
\be
v^2 \simeq \frac{1}{2} \sqrt{a_0 G M_{\rm b}} = \frac{\sqrt{C}}{2}a_0\varepsilon L\,, 
\label{v^2BTFR}
\ee
where we have used Eq.~\eqref{epsilondef}. Accounting for the fact that DM particles have non-circular orbits in general,
we approximate the density ratio by its central value in Eq.~\eqref{density ratio}: $\frac{\rho}{\rho_{\rm b}} \simeq \frac{R}{\varepsilon}$.
From the results of our numerical analysis (Table~\ref{parameters}), $R$ was found to be nearly constant with value $\simeq 6$.
Putting everything together, and using $C \sim 10^{-1}$ we obtain
\be
r_* \simeq 10\,L\,.
\label{r*}
\ee
Remarkably, $r_*$ {\it is a fixed number of scale lengths.} This is consistent with the scaling symmetry of Sec.~\ref{scaleinvarianceDM}: under $L\rightarrow e^\lambda L$, the
density ratio transforms as $\frac{\rho}{\rho_{\rm b}}\rightarrow e^{-\lambda(1-2z)}\frac{\rho}{\rho_{\rm b}} $, while $v^2 \to e^{2\lambda (1-z)}v^2$. It follows that $r_* \rightarrow e^\lambda r_*$, {\it i.e.}, $r_*$ transforms like the scale length $L$. Incidentally, Eq.~\eqref{r*} confirms our starting assumption that $r_*$ lies on the flat part of the rotation curve. We will see below that upon matching to a NFW profile this gives a total halo mass consistent with the $\Lambda$CDM abundance matching prescription.

\subsection{Matching conditions}

To match our density profile to a NFW profile, we demand that both $\rho$ and $P= \rho v^2$ be continuous at $r_*$. Continuity of $\rho$ is obviously required, while continuity of $P$ 
ensures mechanical equilibrium. On the NFW side, the profile is
\be
\rho_{\rm NFW}(r) = \frac{\rho_0}{x(1+x)^2}\,;\qquad x \equiv \frac{r}{r_s}\,,
\label{rhoNFW}
\ee
where $r_s$ is the scale radius. This is related to the concentration parameter ${\rm c} = R_{200}/r_s$, with $R_{200}$ defined in Eq.~\eqref{R200def2}.
Meanwhile, on the MONDian side, for the sake of simplicity, we approximate the profile at large radii by a singular isothermal profile with velocity dispersion given by Eq.~\eqref{v^2BTFR}:
\be
\rho_{\rm MOND}(r_*) = \frac{1}{4\pi G r^2_*}\sqrt{a_0GM_{\rm b}}\,.
\label{MONDprofile}
\ee
Equating Eqs.~\eqref{rhoNFW} and~\eqref{MONDprofile} at $r_*$ gives
\be
4\pi G\rho_0 r_{\rm s}^2 =  \sqrt{a_0GM_{\rm b}} \frac{(1+x_*)^2}{x_*}\,,
\label{rho0rs}
\ee
with $x_* \equiv r_*/r_s$.

For pressure matching, note that on the MONDian side the pressure is straightforwardly obtained by multiplying
the density profile in Eq.~\eqref{MONDprofile} by the velocity profile in Eq.~\eqref{v^2BTFR}:
\be
P_{\rm MOND}(r_*) = v^2\rho_{\rm MOND} = \frac{a_0 M_{\rm b}}{8\pi r_*^2}\,.
\label{PMONDr*}
\ee
Meanwhile, on the NFW side the pressure can be computed by integrating Jeans equation~\eqref{poisson2},
\be
P_{\rm NFW}(r_*) = \int_{r_*}^\infty {\rm d}r \rho(r) \frac{GM(r)}{r^2}\,.
\label{Pintegral}
\ee
where we have neglected the baryonic contribution to the integrand---a reasonable approximation in the flat part of the rotation curve. 
The enclosed mass $M(r)$ for $r\geq r_*$ follows from integrating Eqs.~\eqref{rhoNFW} and~\eqref{MONDprofile}:
\bea
\nonumber
M(r) &=& \sqrt{\frac{a_0 M_{\rm b}}{G}} r_* + 4\pi \rho_0 r_{\rm s}^3 \left[\frac{1}{1+x} - \frac{1}{1+x_*}+ \log\left(\frac{1+x}{1+x_*}\right)\right]  \\
&=& \sqrt{\frac{a_0M_{\rm b}}{G}}r_* \left\{1+  \left(\frac{1+x_*}{x_*}\right)^2 \left[\frac{1}{1+x} - \frac{1}{1+x_*}+ \log\left(\frac{1+x}{1+x_*}\right)\right]\right\}\,,
\label{Menclosed}
\eea
where in passing from the first to the second line we have used Eq.~\eqref{rho0rs} to express $r_s$ in terms of~$x_*$.

The integral in Eq.~\eqref{Pintegral} can be carried out analytically, but the explicit expression is not particularly illuminating. The important point is that it is parametrically of the form
\be
P_{\rm NFW}(r_*) = \frac{a_0 M_{\rm b}}{8\pi r_*^2} {\cal F}(x_*)\,.
\ee
That is, the prefactor is identical to Eq.~\eqref{PMONDr*}. Equating the pressures therefore yields an equation for $x_*$, whose solution is
\be
x_*\simeq 0.65\,.
\label{x*}
\ee
Equations \eqref{r*}, \eqref{rho0rs} and \eqref{x*} determine the free parameters $\rho_0$ and $r_s$ of the NFW profile for any given  baryonic length scale $L$ and total baryonic mass $M_{\rm b}$.

\subsection{Mass-concentration relation}

We are now in a position to also derive expressions for the DM mass $M$ out to the NFW virial radius and the concentration~${\rm c}$. For concreteness we assume the fiducial relation $r_* = 10\,L$. Starting with the concentration parameter, by definition we can write $R_{200} = {\rm c} r_{\rm s} = \frac{{\rm c}}{x_*} r_* \simeq 15.4 \,{\rm c}L$. Equating
to~\eqref{R200def2} gives
\be
{\rm c} \simeq 13 \,\frac{\rm kpc}{L}\left(\frac{M}{10^{12}{\rm M}_\odot}\right)^{1/3} \,.
\label{Mc2}
\ee
Meanwhile, the total DM mass enclosed in the NFW virial radius is obtained by evaluating Eq.~\eqref{Menclosed} at $x = {\rm c}$:
\be
M = \sqrt{\frac{a_0M_{\rm b}}{G}}10 L \left\{1+  \left(\frac{1+x_*}{x_*}\right)^2 \left[\frac{1}{1+{\rm c}} - \frac{1}{1+x_*}+ \log\left(\frac{1+{\rm c}}{1+x_*}\right)\right]\right\}\,.
\label{Mc1}
\ee
For a galaxy of given $M_{\rm b}$ and $L$, these two equations can be solved numerically to obtain $M$ and~${\rm c}$. It is evident our model predicts that ${\rm c}$ should decrease with increasing $L$, the kind of anticorrelation anticipated by~\cite{Desmond:2016azy}.

\begin{table}
  \centering
\begin{tabular}{|c|c|c|c|}
      \hline
~$M_{\rm b}[{\rm M}_\odot]$~ &~$L [{\rm kpc}]$~  & ~$M[{\rm M}_\odot]$~ &~$M_{\rm AM}[{\rm M}_\odot]$~ \\
   \hline    
   $10^8 $ & $0.5$  & $1.1 \times 10^{10}$  & $10^{10}$ \\
  \hline
$10^9 $ & 1 & $6.3\times 10^{10}$ & $1.05\times 10^{11}$\\
   \hline
$10^{10} $ & 2  & $3.2 \times 10^{11}$ & $1.05\times 10^{12}$   \\
\hline
\end{tabular}
      \caption{\small Our predicted total DM mass for three sample galaxies, compared with the prediction of $\Lambda$CDM abundance matching prescription~\cite{Papastergis:2012wh,Behroozi:2012iw}.}
\label{AMcompare}
      \end{table} 

Table~\ref{AMcompare} lists the predicted values of $M$ for our three sample galaxies and compares the results to the $\Lambda$CDM abundance matching predictions~\cite{Papastergis:2012wh,Behroozi:2012iw}.
For this purpose we used the $M_{\rm b}-M_\star$ relation of~\cite{Papastergis:2012wh}, together with the $M_\star-M$ relation of~\cite{Behroozi:2012iw}. 
We see that our predicted halo masses are in good agreement with $\Lambda$CDM abundance matching. We tend to slightly underpredict the mass of
massive galaxies, which would thus give a light mass for the Milky Way~\cite{Gibbons:2014}.

\subsection{Velocity dispersions}

\begin{figure}[t]
	\centering
    \includegraphics[width=4in]{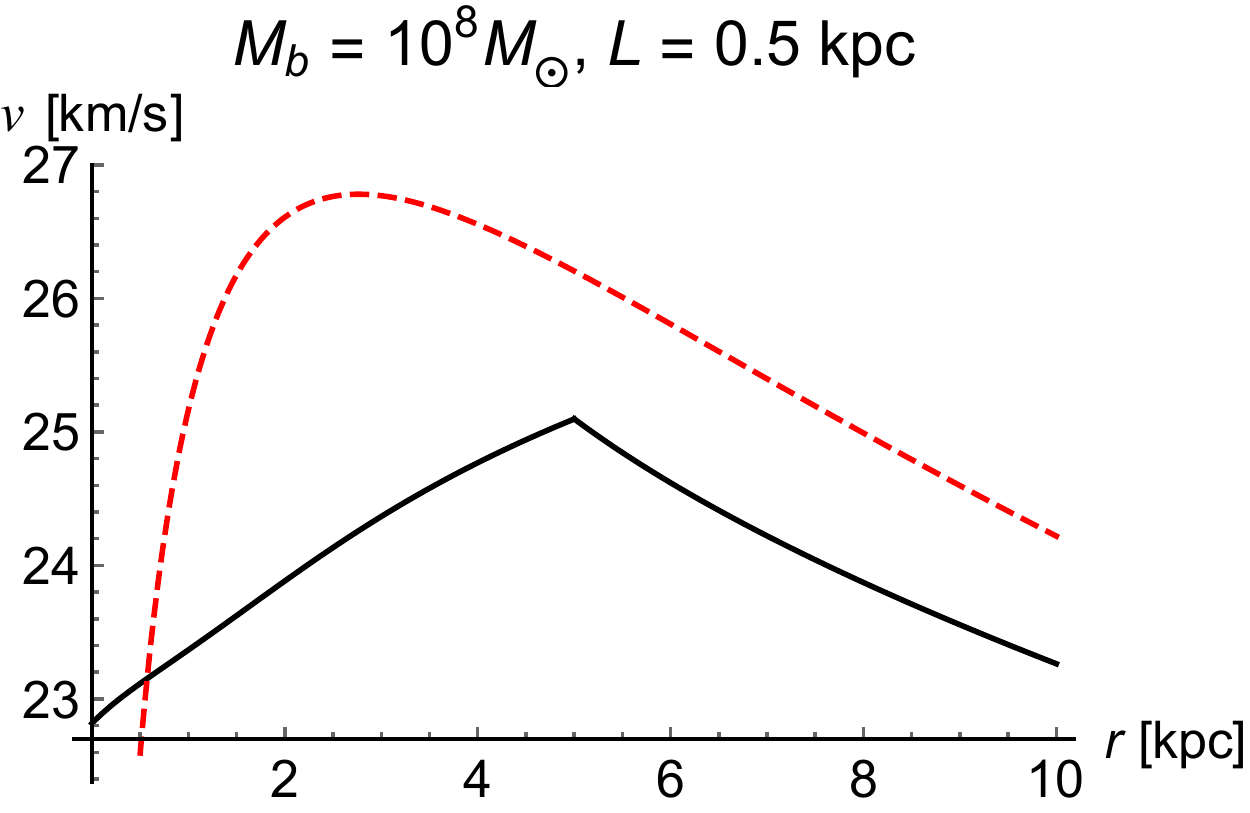}
    \caption{\small Our velocity dispersion profile (solid black curve) for $M_{\rm b} = 10^8 {\rm M}_\odot$ and $L = 0.5$ kpc, matched to a NFW profile at $r_* = 10\,L = 5~{\rm kpc}$. The total DM mass is $M = 1.1\times 10^{10}{\rm M}_\odot$, and the concentration parameter of the NFW envelope is ${\rm c} \simeq 12.6$. Also plotted for comparison is the velocity dispersion profile for a pure NFW profile with the same total DM mass and concentration (red dashed curve).} 
    \label{velocitycomparison} 
\end{figure}

It is instructive to compare the velocity dispersion profile $v(r)$ of our MOND/NFW halo with that of a pure NFW halo of the same mass and concentration. 
Figure~\ref{velocitycomparison} shows this comparison. The solid black curve represents the predicted $v(r)$ for a fiducial galaxy
with $M_{\rm b} = 10^8{\rm M}_\odot$ and $L = 0.5~{\rm kpc}$, obtained via the matching procedure outlined above. From Table~\ref{AMcompare} 
this galaxy has total DM mass $M = 1.1\times 10^{10}{\rm M}_\odot$. Equation~\eqref{Mc2} gives a concentration parameter for the NFW envelope of
${\rm c} \simeq 12.6$. The red dashed curve shows the velocity dispersion profile of a pure NFW with the same total mass and concentration~\cite{Lokas:2001}.

The velocity dispersions of our equilibrium DM halos are actually smaller than their NFW counterpart, apart from the very central region. This would be a natural result of our cooling mechanism: in our scenario, DM particles are slowed down and gently reach equilibrium around low surface brightness disk galaxies, without ever collapsing to a NFW shape in the central regions. However, this also needs the central velocity dispersion to be hotter than its NFW counterpart. Namely, the cold central cusp of the NFW profile expected in the absence of baryon-DM interactions should be replaced by a hotter constant density core at the very center of the halo. Our cooling mechanism of massive DM particles is not a natural way to obtain this.  Some preheating of the central regions of DM halos might therefore be needed before allowing the DM particles to reach equilibrium, {\it i.e.}, even the most bound and cold DM particles at the center of primordial halos should be strongly up-scattered before being allowed to cool again. Usual feedback processes could apply, but they would not need any sort of fine-tuning to reproduce the MDAR phenomenology. They could be very efficient, and severely heat the DM fluid in all halos hosting massive stars, while the cooling mechanism resulting from interactions with baryons would then allow the profile to gently reach the desired equilibrium. In future work, it will be important to also look for equilibrium solutions in the light DM heating case, which could self-consistently erase the cold central cusps at the center of galaxy halos.

\section{Pressure-Supported Systems}
\label{sec:pressure systems}

In the previous Section, we have argued that the MDAR emerges quite generically in flattened disk galaxies provided that the product of baryon-DM interaction cross section $\sigma_{\rm int}$ is inversely proportional to the local number density of DM particles and typical energy exchanged, as encoded in Eq.~\eqref{master equation} with $C \sim 10^{-1}$. By numerically integrating our equations, we were able to determine that good fits can be obtained for $C \simeq 1/16$.  

We will now broaden our discussion to include spheroidal stellar systems, in which the distribution of baryons is supported by velocity dispersion (`pressure') rather than by angular momentum. Unlike spiral galaxies, not all of these systems are observationally well described by the MDAR when dominated by DM ({\it e.g.},~\cite{Famaey:2011kh}). From our perspective here, we will see that these systems differ from disk galaxies because $i)$ baryons are not segregated in a disk, so that the relaxation time is no longer determined by how often DM particles interact with the disk, but instead by the characteristic time for energy loss due to interactions with baryons; and because $ii)$ some of these systems have yet to relax to their equilibrium configuration. 

Another important difference is that baryons and DM have comparable velocity dispersions in pressure-supported systems, {\it i.e.}, $v\simeq v_{\rm b}$. 
Recall that we simplified the heat equation~\eqref{heatcooling} by making the replacement $v_{\rm th} \rightarrow v$ in the interaction rate, which was a valid
approximation in rotationally-supported systems. In pressure-supported systems, however, we should instead use
$v_{\rm th}\simeq \sqrt{2}v$. This will be inconsequential for the analysis below, however, which focuses on the parametric dependence and ignores order-unity coefficients.

\subsection{Galaxy clusters} \label{sec:galaxy clusters}

Let us start by considering galaxy clusters. These systems do not follow the MDAR~\cite{Gerbal:1992,Aguirre:2001fj,Sanders:2003,Clowe:2003tk,Clowe:2006eq,Pointecouteau:2005,Angus:2006qy,Angus:2007,Angus:2008}. Within our approach, this is expected, due to the fact that clusters have a large relaxation time and have not reached their equilibrium configuration yet. This can be easily seen by comparing the characteristic time-scale for energy loss $\tau$ with an overestimate of the typical age of a galaxy cluster $\sim H_0^{-1}$, as given by Eq.~\eqref{equilibriumcondition}. Substituting $\rho \sim 10 \rho_{\rm b}$ and $v \sim 10^3\,\text{km/s} \sim 3 \times 10^{-3} c$ as typical values for DM density and velocity dispersion in a cluster, as well as $C = 1/16$, we~obtain
\begin{equation} \label{energy exchange time vs H0}
H_0 \tau \simeq \frac{9}{C} \,  \frac{v}{c} \frac{\rho}{\rho_{\rm b}}  \simeq 5 \,.
\end{equation}
Hence clusters have not completely relaxed yet to our equilibrium solution. Note that the age of galaxy clusters is actually smaller than $\sim H_0^{-1}$, making the argument even stronger. This is the reason why clusters should not obey a MDAR in our framework, despite the fact that their gravitational potential is dominated by DM. Instead DM in galaxy clusters should approximate a collisionless profile, such as the NFW profile~\cite{NFW}.

\subsection{Dwarf spheroidal galaxies}

Let us now turn our attention to dwarf spheroidal galaxies. The standard MOND paradigm has a mixed track record when it comes to this kind of system~\cite{Famaey:2011kh}. From our perspective, we will argue that this is due to the fact that some dwarf spheroidals have already reached their equilibrium configuration whereas others have not. To illustrate how this plays out, it is useful to rewrite the equilibrium condition~\eqref{equilibriumcondition} as
\be
H_0 \tau \simeq 5 \times 10^{-3} \frac{v}{10 \, \text{km/s}} \,\frac{\rho}{\rho_{\rm b}}  \,,
\label{dwarfequi}
\ee
where as before we have set $C = 1/16$. This quantity should be checked on a case-by-case basis. For instance, consider a prototypical dwarf galaxy with $v \sim 10 \, \text{km/s}$. Our formalism does {\it not} predict what the absolute dynamical mass ratio should be, depending on the history of each system. This should rely on a procedure like abundance matching, which could yield a rather scattered relation between baryonic and dynamical mass in the low halo mass range. But once this is settled for each dwarf, if it has a moderate mass-to-light ratio in the visible region, $\rho\; \lsim\; 10 \,\rho_{\rm b}$, such as Fornax, then it should be relaxed by now and thus should be amenable to our quasi-equilibrium analysis. If, instead, the dwarf has a larger mass-to-light ratio, $\rho \;\gsim\; 10^2 \rho_{\rm b}$, such as most ultra-faint dwarfs, then it did not have enough time to equilibrate.

Focusing on dwarf spheroidals that are in equilibrium, we now wish to solve our equations to obtain the DM density and velocity profiles for these objects, similarly to what we did for disk galaxies earlier. The Jeans equilibrium condition is unchanged and given by Eq.~\eqref{poisson2}. The main difference lies in the heat equation, and in particular in the expression for the thermal conductivity. Because there is no physical separation between baryons and DM ({\it i.e.}, baryons are not confined to a specific subregion, like in disk galaxies), the relaxation time of these systems is set directly by the energy-exchange~time:
\begin{equation} \label{t_relax dSph}
	t_{\rm relax} \sim \tau \, .
\end{equation}
Assuming that this energy-exchange time is smaller than (or of order) the dynamical time, there is no infrared cutoff as in the Knudsen regime of disk galaxies, and it follows that $\ell \sim \tau v$. Using Eq.~\eqref{taudef} we find that the thermal conductivity in Eq.~\eqref{kappa} becomes
\be
\kappa \sim n v^2\tau \sim \frac{v^3}{m C a_0} \frac{\rho^2}{\rho_{\rm b}}\,.
\ee
The heat equation~\eqref{heateqn1} in turn becomes 
\be
\frac{1}{r^2}\frac{{\rm d}}{{\rm d}r} \left( \frac{\rho^2}{\rho_{\rm b}} r^2  v^3\frac{{\rm d}v^2}{{\rm d}r}\right) \sim C^2 a_0^2 v\rho_{\rm b}\,.
\label{heatspherical_dSphs}
\ee
Akin to what we did in Sec.~\ref{sec: Newton vs MOND}, we will content ourselves with seeking an approximate analytic solution, focusing on the parametric dependence 
and ignoring order-unity coefficients. Thus, assuming again a power-law behavior, we can immediately integrate Eq.~\eqref{heatspherical_dSphs} to obtain Fourier's law:
\be
\rho^2 v^2\frac{{\rm d}v^2}{{\rm d}r} \sim C^2 a_0^2 \frac{M_{\rm b}}{4\pi r^2} \rho_{\rm b} \,.
\label{Fourierdwarfs}
\ee

Dwarf spheroidals are DM dominated, hence an approximate solution to the Jeans hydrostatic equilibrium condition is once again given by Eq.~\eqref{virialbis}, {\it i.e.}, $\rho (r) \sim \frac{v^2}{2 \pi G r^2}$. 
Fourier's law then reduces~to
\be
\frac{{\rm d}v^8}{{\rm d}r}\sim C^2 a_0^2 G^2 M_{\rm b} 4\pi r^2 \rho_{\rm b}\,,
\ee
which can be immediately integrated to yield
\be
v^8 \sim C^2 a_0^2 G^2 M_{\rm b}^2\,.
\ee
This is of course just the square of the MDAR---see Eq.~\eqref{BTFR from heat}. Hence, despite the different expression for the thermal conductivity, the end result for the velocity dispersion is parametrically
identical to that for rotationally-supported galaxies. 

For consistency, we should in principle check again that the time scale hierarchy $\tau \;\lsim\; t_{\rm dyn}$ is satisfied. However, this would be just a repeat of the analysis carried out in Sec.~\ref{timescalechecks}, with $\tau_{\rm disk}$ replaced by $\tau$ in Eq.~\eqref{tauineqDM}. Therefore we conclude immediately that
\be
\frac{\tau}{t_{\rm dyn}} \sim \frac{M_{\rm b}(r)}{(4/3)\pi r^3 \rho_{\rm b}}\,.
\ee
In this case, the sphericization is not artificial, meaning that $\tau \sim t_{\rm dyn}$. In our discussion of elliptical galaxies below, we will argue that the same holds true in the baryon-dominated regime. 

To summarize, our framework predicts that a dwarf galaxy which is not too DM-dominated, hence satisfying the equilibrium condition~\eqref{dwarfequi}, will typically follow the MDAR (or at least a rescaled version of the MDAR, with an effective rescaling of $a_0$) at every radius. For instance, irrespectively of the origin of its low DM-content, the recently discovered galaxy Crater 2, which has $\rho/\rho_{\rm b} \sim 25$ and $v \sim 3 \, {\rm km/s}$~\cite{Caldwell:2016} is predicted to have $H_0 \tau \, \lsim \, 5 \times 10^{-2}$, and hence to follow the MDAR. However, this would depend on whether the low DM content has been an ancient feature of the dwarf, or whether it is related to more recent DM stripping. If the dwarf has always had a low DM content, the fact that it would follow the MDAR is roughly consistent with~\cite{McGaugh:2016jwx}, but our mechanism does not predict any sort of external field effect.

\subsection{Elliptical galaxies}

Elliptical galaxies are another class of pressure-supported systems where we can test the predictions of our model. These objects are baryon-dominated, with relatively high velocity dispersion $v \sim {\cal O}(10^2)$~km/s. It is easy to show that they satisfy the equilibrium condition~\eqref{equilibriumcondition} and hence should be well-described by our formalism.

The analysis of ellipticals follows the same steps described above for dwarf spheroidals up until Fourier's law~\eqref{Fourierdwarfs}. Unlike dwarfs, which are DM-dominated, ellipticals are baryon-dominated. Hence in Fourier's law we should instead substitute the relation~\eqref{virial theorem Newtonian}, {\it i.e.}, $v^2 \sim G M_{\rm b}(r)/r$. This allows us to solve for the DM density profile and find
\be
\rho \sim \frac{a_0}{4\pi G r}\,,
\ee
which is virtually identical to the result~\eqref{rho baryon domination} for baryon-dominated disk galaxies. The subdominant DM component therefore contributes a nearly constant acceleration,
\be
g_{\rm DM} \simeq a_0\,.
\label{constantDMaccn}
\ee
This matches the Newtonian behavior of the `simple' interpolating function~\cite{Famaey:2005fd}.

Interestingly, this is consistent with strong gravitational lensing results, as shown in~\cite{Sanders:2008} (see their Figs.~2 and~5).
Moreover, a recent analysis~\cite{Chae:2017bhk} of $5619$ nearly-round elliptical galaxies from the Sloan Digital Sky Survey (SDSS) DR7 was able to constrain the subdominant DM contribution to the total acceleration in the range $a_0 < g < 30\,a_0$. The data are well-fitted by a power-law relation of the form
\be
\frac{g_{\rm DM}}{g_{\rm b}} \sim \left(\frac{g_{\rm b}}{a_0}\right)^{q}\,,
\ee
with $q \simeq - 1$ for $g_{\rm b} \gtrsim 10^{-9.5} \text{m}/\text{s}^2$. This implies that $g_{\rm DM} \simeq a_0$ in the high acceleration limit, in perfect agreement with the prediction of our model in Eq.~\eqref{constantDMaccn}. At the same time, the standard~\cite{Kent:1987zz}, exponential~\cite{McGaugh:2016leg} and Bekenstein~\cite{Bekenstein:2004ne} MOND interpolating functions, as well as the superfluid DM paradigm (which, with no renormalization of $a_0$, gives rise to Bekenstein's interpolating function)~\cite{Berezhiani:2015pia,Berezhiani:2015bqa,Khoury:2016ehj} and Verlinde's emergent gravity formula~\cite{Verlinde:2016toy}, all appear to be in tension with the data.

\section{Phenomenological and Cosmological Constraints}
\label{astroconstraints}

We now turn our attention to various phenomenological constraints, ranging from direct detection to cosmology.
In what follows we will remain as general as we can in deriving constraints, without committing to specific particle  physics realizations.

\subsection{Experimental constraints} \label{sec: experimental constraints}

\begin{figure}
\centering
\includegraphics[width=5in]{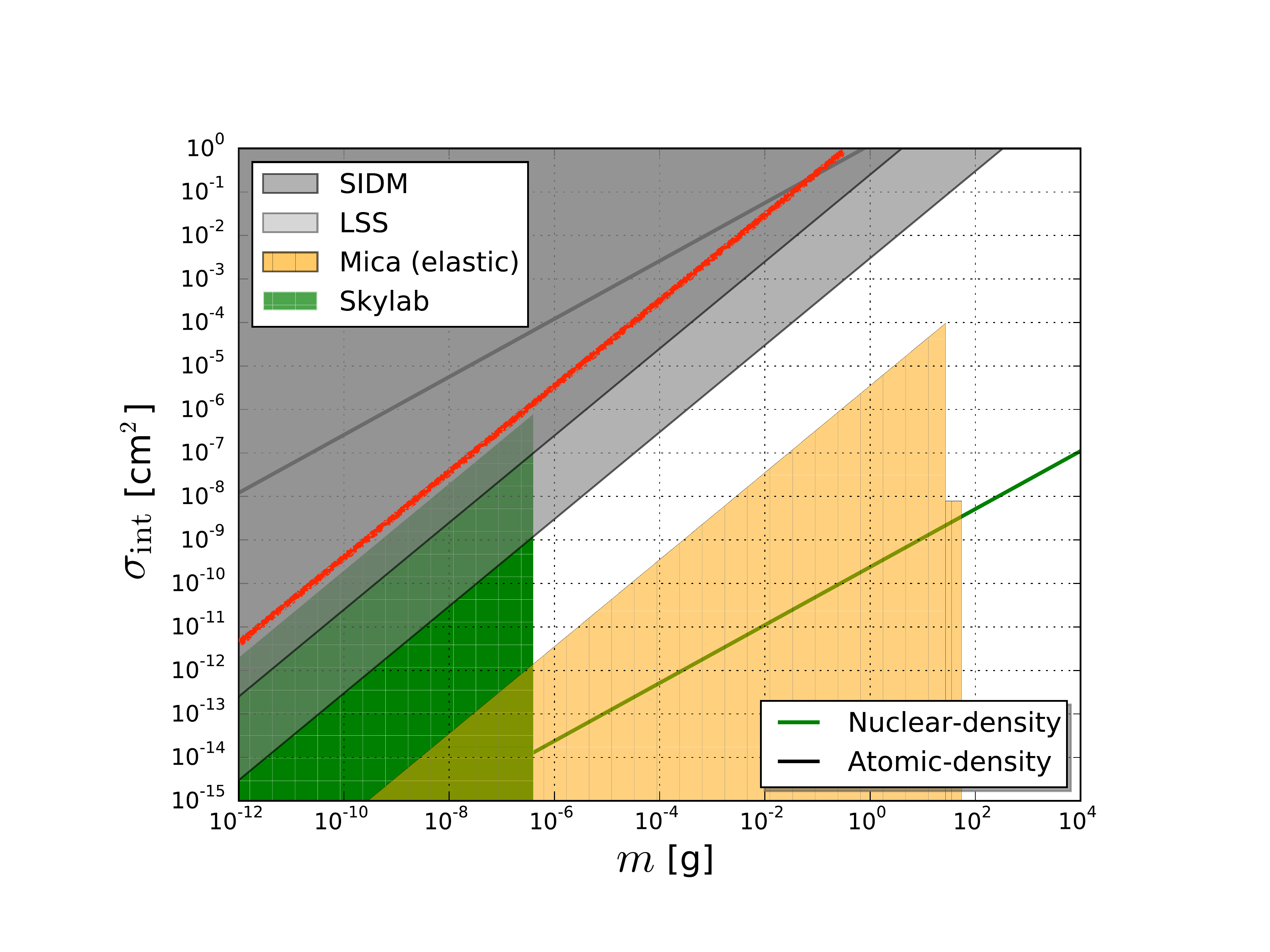}
\caption{\small Exclusion plot for Macro DM. Reproduced with permission from~\cite{Jacobs:2014yca}. The red line corresponds to the {\it local} value of $\frac{\sigma_{\rm int}}{m} \sim 5\,\frac{{\rm cm^2}}{{\rm g}}$ for our DM-baryon cross section. The grey contours seem to indicate that such a large cross section is ruled out by cosmological observations. However, those constraints cannot be immediately applied to our model with a density dependence of the cross section (see Sec.~\ref{sec:CMB & LSS constraints})} \label{macroexclusion}
\end{figure}

We must first check that the cross sections considered in the present paper are compatible with current direct detection experiment constraints. 
Using the fact that $\epsilon \sim m_{\rm b}v^2$ for the case of interest of heavy DM  ($m\gg m_{\rm b}$), the light DM case being the topic of future studies, our master relation~\eqref{master equation}
can be expressed as 
\be
\frac{\sigma_{\rm int}}{m} = \frac{C m_{\rm b} a_0}{\epsilon \rho}  \sim \frac{C a_0}{v^2\rho} \,.
\label{masterrepeat}
\ee
Substituting the local Solar neighbourhood values $\rho\simeq 10^{-24}~{\rm g}/{\rm cm}^3$ and $v \simeq \frac{220}{\sqrt{2}}~{\rm km}/{\rm s}$, as well as $C \sim 10^{-1}$, this gives
\be
\frac{\sigma_{\rm int}}{m} \sim \frac{C a_0}{v^2\rho} \sim 5~\frac{{\rm cm^2}}{{\rm g}}\,.
\label{masterrepeatelastic}
\ee
Figure~\ref{macroexclusion}, reproduced with permission from~\cite{Jacobs:2014yca}, shows the excluded region of $(\sigma_{\rm int},m)$ parameter space for $m \;\gsim\; 10^{-12}~{\rm g} \simeq 10^{11}~{\rm GeV}$. 
The local value of our cross section, $\frac{\sigma_{\rm int}}{m} \simeq 5~\frac{{\rm cm^2}}{{\rm g}}$, is shown as the red line. The grey contours seem to indicate that such a large cross section is ruled out by cosmological observations. However, those constraints cannot be immediately applied to our model with a density dependence of the cross section. Indeed, in Sec.~\ref{sec:CMB & LSS constraints} below we will revisit the analysis leading to light gray exclusion contour, which is based on cosmological linear perturbation theory. There, we will show that, once the density dependence of the cross section is taken into account, those constraints are lifted and are in fact compatible with our prediction for $\sigma_{\rm int} / m$. 

It is also worth noting that for $m \lesssim 10^{17}$~GeV our predicted cross section is barely allowed by the Skylab constraint (green region). Since we have neglected factors of $\mathcal{O}(1)$ that could eventually conspire to bring within the Skylab excluded region, to be on the safe side we are encouraged to consider scenarios like `macro' DM~\cite{Jacobs:2014yca}, where $m \gtrsim M_{\rm Pl} > 2 \times 10^{18}$~GeV.

Incidentally, we can also place a bound on the DM self-interaction cross section $\sigma$. Our assumption that DM be in the Knudsen regime in the halo of a disk galaxy requires that the mean free path $\lambda$ associated with self-interactions be such that $\lambda = \frac{1}{n\sigma} \gtrsim r$. In other words,
\be \label{inequality self interaction}
\frac{\sigma}{m} \lesssim \frac{1}{r\rho}\,.
\ee
Combining this result with Eq.~\eqref{masterrepeat} and using $\frac{v^2}{r} \sim g/2$, we obtain
\be \label{self interaction Knudsen}
\frac{\sigma}{\sigma_{\rm int}} \;\lsim\; \frac{1}{C}\,\frac{g}{2 a_0} \; \lsim \; 1\,,
\ee
where in the last step we have focused on the DM-dominated ($g < a_0$) regime for concreteness. What this shows is that DM self-interactions must be
somewhat weaker---though not necessarily orders-of-magnitude weaker---than DM-baryon interactions. Note that we have assumed up to now that the heat conductivity on the left-hand side of the heat equation was mostly driven by interactions of DM particles with baryons. However, if the inequality~\eqref{inequality self interaction} was saturated, then the relaxation time associated with DM self-interactions would also be of order $t_{\rm dyn}$, and this by itself would ensure the desired form of the thermal conductivity~\eqref{kappa 2}.

\subsection{CMB and large-scale structure} \label{sec:CMB & LSS constraints}

Dvorkin {\it et al.}~\cite{Dvorkin:2013cea} derived a constraint on the DM-baryon interaction rate from cosmological linear perturbation theory, including cosmic microwave background (CMB) and large-scale structure. They considered the case where the cross section depends on velocity, $\sigma_{\rm int}(v) = \sigma_0 v^n$, but is otherwise independent of density. The constraint comes predominantly from the redshift range $10^4 \;\lsim\; z \;\lsim\; 10^5$, during which DM dominates the expansion and begins to form structure while baryons are still tightly coupled to radiation. If baryons and DM interact significantly during this epoch, then baryons will transfer kinetic energy to DM, thereby jeopardizing the growth of DM perturbations.

Conveniently, the analysis of~\cite{Dvorkin:2013cea} focused on elastic scattering with $m\gg m_{\rm b}$. For $\sigma_{\rm int} \sim v^{-2}$, they
obtained (with $v$ in units of $c$)
\be
v^2 \frac{\sigma_{\rm int}}{m} \;\lsim\; 6\times 10^{-10}~\frac{{\rm cm}^2}{{\rm g}}\,.
\label{dvorkinbound}
\ee
Although this assumes a density-independent cross section, as a first pass it will suffice for our purposes to think of Eq.~\eqref{dvorkinbound}
as constraining our cross section in the relevant redshift range $10^4 \;\lsim\; z \;\lsim\; 10^5$. Using the cosmological density $\rho \simeq 2\times 10^{-30} (1+z)^3{\rm g}/{\rm cm}^3$, Eq.~\eqref{masterrepeat} gives
\be
v^2 \frac{\sigma_{\rm int}}{m} \simeq \frac{0.7}{(1+z)^3}~\frac{{\rm cm}^2}{{\rm g}}\,.
\ee
This comfortably satisfies Eq.~\eqref{dvorkinbound} for $10^4 \;\lsim\; z \;\lsim\; 10^5$.

It is natural to ask: at what redshift $z_\star$ would $\tau$ become comparable to $H^{-1}$? A simple calculation yields the value $z_\star \simeq 100$. At that redshift DM would thermalize both with itself and with baryons. Fortunately this is well after recombination, hence we expect negligible impact on the CMB but possible interesting signatures for 21-cm observations and the Lyman-$\alpha$ forest~\cite{Tashiro:2014tsa,Munoz:2015bca,Munoz:2017qpy}. We leave a detailed analysis of possible signatures of late-time thermalization to future work.

\subsection{Galaxy clusters}

Galaxy cluster mergers, such as the Bullet Cluster, constrain the DM self-interaction cross section per unit mass, $\frac{\sigma}{m} \;\lsim\; \frac{{\rm cm}^2}{{\rm g}}$~\cite{Markevitch:2003at,Randall:2007ph,Harvey:2015hha,Wittman:2017gxn}. The precise numerical value depends on the assumptions, ranging from $0.5~\frac{{\rm cm}^2}{{\rm g}}$~\cite{Harvey:2015hha} to $2~\frac{{\rm cm}^2}{{\rm g}}$~\cite{Wittman:2017gxn}.\footnote{Despite early claims to the contrary~\cite{MiraldaEscude:2000qt}, lensing-based studies of the shape of galaxy clusters yield comparable bounds~\cite{Peter:2012jh}.} More generally, what is constrained is the energy exchange characteristic rate
\be
\tau = \frac{m}{\rho \sigma v} \;\gsim\; 10^{16}~{\rm s}\,,
\label{mergingclusterbound}
\ee
where we have substituted the characteristic density $\rho \simeq 10^{-24} {\rm g}/{\rm cm}^3$ and velocity $v \simeq 10^3~{\rm km}/{\rm s}$.
For comparison, the time it takes the clusters to pass through each other is $\tau_{\rm passage} \sim \frac{{\rm Mpc}}{10^3~{\rm km}/{\rm s}}\sim 3 \times 10^{16}~{\rm s}$.
Thus Eq.~\eqref{mergingclusterbound} constrains each DM particle to have at most a few collisions per passage.

Although this bound was originally derived for DM self-interactions, we can equally well apply it to DM-baryon scattering, which in our case is
the dominant interaction. The expression for $\tau$ is once again given in Eq.~\eqref{taudef}, and using the fact that $\rho \sim 10 \rho_{\rm b}$ in clusters and $v \simeq 10^3~{\rm km}/{\rm s}$, we obtain
\be
\tau = \frac{\rho}{\rho_{\rm b}} \frac{v}{C a_0} \sim 10^{18}~{\rm s}\,.
\ee
This comfortably satisfies Eq.~\eqref{mergingclusterbound}. 

\subsection{Gas heating}

Elastic scattering with heavy DM particles can heat the gas in the central region of galaxy clusters, potentially resulting in a rising temperature profile towards the center, in conflict with observations~\cite{Qin:2001hh,Chuzhoy:2004bc,Hu:2007ai}. The resulting constraint for $\sigma_{\rm int} \sim v^{-2}$ is~\cite{Hu:2007ai}
\be
\frac{\sigma_{\rm int}}{m} \;\lsim\; 0.1 \left(\frac{v}{10^3~{\rm km}/{\rm s}}\right)^{-2}\frac{{\rm cm}^2}{{\rm g}}\,,
\label{clusterheating}
\ee 
assuming a central DM density of $\rho\simeq 0.02~{\rm M}_\odot/{\rm pc}^3 \simeq 1.5\times 10^{-24}~{\rm g}/{\rm cm}^3$. At that density, our cross section in Eq.~\eqref{masterrepeat} can be
expressed as
\be
\frac{\sigma_{\rm int}}{m} \simeq 0.08 \left(\frac{v}{10^3~{\rm km}/{\rm s}}\right)^{-2}\frac{{\rm cm}^2}{{\rm g}}\,,
\ee
which barely satisfies Eq.~\eqref{clusterheating}. It is also worth reminding ourselves that our estimated cross section in Eq.~\eqref{masterrepeatelastic} ignored a factor of order unity. Nevertheless, 
an order-of-magnitude improvement on the cluster bound would yield an interesting constraint on $C a_0$. 

The heating rate of baryonic gas is also constrained on galactic scales. Namely, the cooling rate of the interstellar medium in the galaxy should be higher than the
heating from DM interactions~\cite{Chivukula:1989cc}. The total cooling rate depends on the exact composition of the interstellar medium, but is observed to be roughly of the order of~({\it e.g.},~\cite{Pottasch:1979})
\be
\dot{\cal E}_{\rm cool} \simeq  8 \times 10^{-14}~\frac{{\rm eV}}{{\rm s}}\,.
\label{cooling}
\ee
On the other hand, our heating rate of baryons is
\be
\dot{\cal E}_{\rm heat} \simeq n \sigma_{\rm int} v \times m_{\rm b}v^2 = C a_0m_{\rm b}v \simeq 2 \times 10^{-14}~\frac{{\rm eV}}{{\rm s}}\,,
\ee
where in the last step we have substituted $v = \frac{220}{\sqrt{2}}~{\rm km}/{\rm s}$ and used $C = 10^{-1}$. This is somewhat smaller than the observed cooling rate (Eq.~\eqref{cooling}), but maybe uncomfortably close. In future work, we will include for completeness the Jeans and heat equations for baryons as well. The heat equation for the gas, in particular, would include both heating by DM and cooling from standard processes, as discussed above.

\section{Heating Rate on Small Scales: Gas versus Stars} \label{sec: gas vs stars}

Up to this point, we have taken a very phenomenological, bottom-up approach based on the first few moments of the Boltzmann transport equation and our master equation \eqref{master equation}. We also assumed that macroscopic quantities such as densities and velocity dispersions were defined by averaging over scales $\sim 10$ pc. This assumption enabled us not to specify the details of the baryon distribution at smaller scales, and in particular not to distinguish between gaseous and stellar components. However, in order to make contact with  particle physics model building---which will be the focus of the next section---it is important to take into account phenomena taking place on smaller scales. As a first step in this direction, we are now going to show that our expression for the heating rate $\dot{{\cal E}}$ is in fact largely independent of how baryons are distributed at small scales.

In the absence of long-range forces, this is purely a consequence of geometry, as we are now going to show. At smaller scales, the total baryon density is in general the sum of the density of gas and the density of~stars: 
\begin{equation}
	\rho_{\rm b} = \rho_{\rm gas} + \rho_{\rm stars}\,.
\end{equation}
The interaction rate with baryons in the gas can be immediately read off Eq.~\eqref{Gammaint}:
\begin{equation} \label{Gamma int gas}
	\Gamma_{\rm int} \simeq n_{\rm gas}\sigma_{\rm int}  v = \frac{\rho_{\rm gas}\sigma_{\rm int}  v}{m_{\rm b}}.
\end{equation}
Here and in what follows, we are using the fact that $v_{\rm th} \sim v$ in systems we are interested in. The interaction rate with baryons in stars is instead equal to the rate of interactions with stars, times the average number of interactions with baryons taking place inside a typical star:
\begin{equation} \label{Gamma stars * P int}
	\Gamma_{\rm int} = \Gamma_{\rm stars} \times N_{\rm int}\,.
\end{equation}

Let us consider for simplicity stellar populations defined by similar stellar mass $M_*$ and radius $R_*$. (The results can be trivially extended to all stellar populations.) The former factor is then equal to
\begin{equation} \label{Gamma stars}
	\Gamma_{\rm stars} = n_{\rm stars} \sigma_* v \simeq \frac{\rho_{\rm stars}}{M_*} \times \pi R_*^2 \times v \, ,
\end{equation}
where $\sigma_*$ is the cross section of a star of mass $M_*$ and radius $R_*$. The number of interactions $N_{\rm int}$ is instead given by the product of the interaction rate inside a star times the average time that a DM particle spends inside a star. Denoting with $n_*$ ($\rho_*$) the average number (mass) density of baryons in one star, and $\Delta t = R_*/v$ the typical time spent by a DM particle in the star, we find
\begin{equation} \label{P int}
	N_{\rm int} = n_* \sigma_{\rm int} v \times \Delta t  = n_* \sigma_{\rm int} R_*\, .
\end{equation}
Combining  the last three equations, and using the fact that $M_* \simeq \frac{4}{3} \pi \rho_* R_*^3$, we obtain, up to a factor of order unity,
\begin{equation} \label{Gamma stars 2}
	\Gamma_{\rm int} \simeq  \, \frac{\rho_{\rm stars}\sigma_{\rm int}  v}{m_{\rm b} } \,.
\end{equation}
This result has the same form as the interaction rate in Eq.~\eqref{Gamma int gas} with baryons in the gas, except for the replacement $\rho_{\rm gas} \to \rho_{\rm stars}$. Thus, the total interaction rate including gas and every stellar population is proportional to the total baryon density $\rho_{\rm b}$---and so is the heating rate $\dot{{\cal E}} = \Gamma_{\rm int} \epsilon$.

The situation becomes a bit more subtle in the presence of attractive long-range forces between baryons and DM particles, which necessarily include (but are not necessarily limited to) gravity. As we will see, the end result remains the same---the heating rate is still proportional to the total baryonic density---but {\it only due to the particular form of baryon-DM interactions we are considering.} The rate of interactions with the gaseous component, Eq. \eqref{Gamma int gas}, is not affected by additional long range forces. Thus, we need the same to be true for the stellar contribution to the heating rate, $\Gamma_{\rm stars} N_{\rm int} \epsilon$. At first sight this is not obvious, since the calculation of the latter is modified in several important ways.

First, the cross section of stars $\sigma_*$ is enhanced for small values of $v$ due to the Sommerfeld effect. Using $E = \frac{3}{2}mv^2$ as the typical energy of a DM particle asymptotically far from the star, the cross section becomes~\cite{Pospelov:2008jd,ArkaniHamed:2008qn}
\begin{equation} \label{Sommerfeld}
	\sigma_* \to \pi R_*^2 \left( 1 + \frac{v_{\rm esc}^2}{3v^2} \right)\,,
\end{equation} 
with $v_{\rm esc} \equiv (2 \alpha/ m R_*)^{1/2}$  the escape velocity of the star due to the long-range potential $\Phi \equiv -\frac{\alpha}{m r}$ (with $\alpha = G M_* m$ in the particular case of gravity). Second, the velocity factor in Eq.~\eqref{Gamma stars} gets boosted, $v\to v \sqrt{1 + \frac{v_{\rm esc}^2}{3v^2}}$, as DM particles accelerate while falling in the potential well of the star. Keeping these two effects into account, the rate of interactions with stars in Eq. \eqref{Gamma stars} changes as follows:
\be \label{Gamma int change}
\Gamma_{\rm stars} \to  \Gamma_{\rm stars}  \left( 1 + \frac{v_{\rm esc}^2}{3v^2} \right)^{3/2}\,.
\ee
Finally, the product $N_{\rm int} \epsilon$ is also affected by the long-range interaction. According to our master equation \eqref{master equation} and the expression for $N_{\rm int}$ in Eq. \eqref{P int}, we have
\begin{equation} \label{Pint epsilon}
	N_{\rm int} \epsilon = \frac{C\,a_0 m_{\rm b}n_* R_* }{n}
\end{equation}
When DM particles fall in the potential well generated by the star, they give rise to a slightly over-dense region. This local increase of the DM number density $n$ decreases in turn the product~$N_{\rm int} \epsilon$.

In order to quantify this effect, we extend the perturbative approach of~\cite{Hernandez:2008iq} to a non-linear analysis. We assume that the DM phase-space distribution in the absence of the star is the isotropic Maxwellian distribution, $f_0(u) \sim n_0\exp(-u^2/2v^2)$, which in the local neighborhood of the star can safely be approximated as homogeneous. Now, let's add a star. Its attractive potential $\Phi = -\frac{\alpha}{m r} \equiv - \frac{1}{2}v_{\rm esc}^2(r)$ will distort the original DM distribution to a new distribution, $f(u,r)$. This new, spherically-symmetric distribution can be determined by solving the stationary, collisionless Boltzmann equation, 
\be
u \frac{\partial f}{\partial r} =  \frac{v_{\rm esc}^2(r)}{2r} \frac{\partial f}{\partial u} \,.
\label{stationaryBoltzmann}
\ee
Substituting the locally Maxwellian ansatz $f(r,u) \sim n(r) \exp\left[-\frac{u^2}{2\left(v^2 + v_{\rm esc}^2(r)/3\right)}\right]$, and integrating over velocities,
we obtain an equation for the number density
\be
\frac{{\rm d}\ln n(r)}{{\rm d}\ln r} = - \frac{v_{\rm esc}^2(r)}{2\left(v^2 + \frac{v_{\rm esc}^2(r)}{3}\right)}  \,.
\ee
Demanding that $n$ approaches the homogeneous background density $n_0$ far away from the star, we can solve this equation to find
\be
n(r) = n_0 \left( 1 + \frac{v_{\rm esc}^2(r)}{3v^2} \right)^{3/2} \,.
\ee
This is exact and, in particular, agrees with the perturbative result of~\cite{Hernandez:2008iq} in the limit $v_{\rm esc}^2 \ll v^2$. 

This locally-enhanced number density is what should appear in the denominator in Eq.~\eqref{Pint epsilon}, resulting in a change
\be \label{N change}
N_{\rm int} \epsilon \to N_{\rm int} \epsilon \left( 1 + \frac{v_{\rm esc}^2}{3v^2} \right)^{-3/2}\,.
\ee
The extra factor in parentheses cancels exactly the one in Eq.~\eqref{Gamma int change}, so that the stellar heating rate $\Gamma_{\rm stars} N_{\rm int} \epsilon$ remains unchanged. Notice that this result relies crucially on our master equation~\eqref{master equation}. More generally, the stellar and gaseous components contribute equally to the heating rate $\dot{{\cal E}}$ only if baryon DM-interactions are such that $\sigma_{\rm int} \to \sigma_{\rm int} \left( 1 + \frac{v_{\rm esc}^2}{3v^2} \right)^{-5/2}$ when long-range effects are taken into account. 

Thus, we have demonstrated that the heating rate is independent of the clumpiness of gas and stars on small scales. In other words, the right-hand of the heat equation depends on the product $\Gamma_{\rm stars} N_{\rm int} \epsilon$, which is independent of any ``microscopic'' details.

For completeness, let us now briefly discuss also the left-hand side of the heat equation, and in particular the thermal conductivity. It is worth emphasizing that our analysis only relies on the assumption that $t_{\rm relax} \sim t_{\rm dyn}$, irrespective of the underlying microscopic mechanism that ensures this.  In this paper, we have assumed that the time scale for energy exchange with baryons in the disk is smaller than the dynamical time, so that the latter effectively sets the relaxation time scale. While we have shown that this assumption is self-consistent within our coarse-graining approach (see Sec. \ref{timescalechecks}), whether or not it is actually satisfied on smaller scales needs to be checked on a model-by-model basis. 

For instance, concrete models might  involve additional pc-range forces between DM and baryons, 
which would affect both the rate at which a single DM particle interacts with stars as well as the average number of collisions that such a particle experiences inside a star
as shown in Eqs.~\eqref{Gamma int change}~and~\eqref{N change}. Such a force could enhance the interaction rate with stars by orders of magnitude while decreasing the cross-section accordingly, so that the overall heating rate remains unaffected. In such a model, it would be expected that the contribution of this pc-range force to the heating rate would be negligible, and consistency with post-Newtonian constraints in the Solar system should be checked. 

One should also keep in mind the possibility that DM self-interactions, while in principle not required by our scenario, could  also affect the relaxation time, as mentioned in Sec. \ref{sec: experimental constraints}. At this stage, we delay detailed model-building to future work, and only provide in the next section an example giving the right dependence of the cross section on density, as a proof of principle.

\section{Particle Model Building} \label{sec: particle physics models}

Our key requirement for the emergence of the MDAR is that the product of the DM-baryon interaction cross section $\sigma_{\rm int}$
and the typical energy exchanged $\epsilon$---which for elastic collisions depends on the DM and/or baryon velocity---is inversely proportional to the ambient DM number density, as 
stated in Eq.~\eqref{master equation}. We are of course used to cross sections depending on velocity (or energy). For instance, inverse powers of velocity can be easily obtained if the dominant interaction between baryons and DM is mediated by a light particle, as in a plasma~\cite{Sigurdson:2004zp, Landau:1981mil}, or boosted by Sommerfeld enhancement (see {\it e.g.},~\cite{ArkaniHamed:2008qn,Iengo:2009ni,Feng:2010zp}).  However, density-dependent cross sections may be less familiar. In this Section we will show that such environmental dependence can in fact arise naturally in particle physics models. Our main goal at this stage is not to propose fully-fledged models to be tested against precise observations, especially since it is not clear what the heavy `macro DM' would be composed of, but rather to discuss possible physical mechanisms that would {\it generically} give rise to the correct density and velocity dependence, and to provide the reader with a proof-of-principle toy model. We expect to conduct a more detailed phenomenological analysis of specific models in future work. 

Perhaps the most straightforward interpretation of our master equation \eqref{master equation} is that the strength of interactions between DM and baryons is somehow inversely proportional to the DM density. This feature is strikingly reminiscent of phonons, whose interactions with themselves (see, {\it e.g.},~\cite{Endlich:2010hf}) as well as with other excitations (see, {\it e.g.},~\cite{Horn:2015zna,Nicolis:2017eqo}) are naturally suppressed by the density of the medium. Thus, it would seem worthwhile to entertain the possibility that baryons interact predominantly with phonons in the DM sector rather than with individual DM particles. After all, it was already proposed elsewhere~\cite{Berezhiani:2015pia,Berezhiani:2015bqa,Khoury:2016egg,Berezhiani:2017tth} that, if DM was in a superfluid state, phonons could give rise to the MDAR (at least for a certain choice of the superfluid equation of state) by mediating an additional long-range force between baryons. However, our requirement that DM be in the Knudsen regime necessarily prevents phonons from playing any significant role in our context. This is because phonons are collective excitations that are well defined only at scales much larger than the mean free path of the medium, which for self-interactions coincides with the typical energy-loss length-scale. Since we want the latter to be larger than $r$ (or at least of that order), we immediately conclude that phonons cannot be excited at the scales of interest. 

Here, we will describe instead a different mechanism that gives rise to the correct density {\it and} velocity dependence in one fell swoop. In the case of heavy DM (the ``cooling'' case), the typical energy exchanged in a single elastic collision is $\epsilon \simeq 2 m_{\rm b} v^2$, and our master equation~\eqref{master equation} reduces to
\begin{equation}
	\sigma_{\rm int} \simeq \frac{C a_0}{n v^2} \ . \label{master eq model building}
\end{equation}
The toy model we will write down is based on the simple observation that for a non-relativistic, thermal distribution of DM particles, the product $n v^2$ that appears in the denominator is proportional to the DM pressure $\bar P = m n v^2$ at equilibrium. 

Given that the typical velocities at play in galaxies are of $\mathcal{O}(10^{-3})$ (relative to $c$), it is certainly a good approximation to work with a non-relativistic field theory. We will therefore consider the following non-relativistic action:
\begin{align}
	S &= \int {\rm d}t {\rm d}^3 x \left\{ \gamma \, \frac{e^{-3 q \pi}}{2} (\vec \nabla \pi)^2 + \varepsilon \,e^{- 5 q\pi} + e^{ q \pi} \left[ \frac{i}{2} \left(\psi^*  \partial_t\psi - \psi\partial_t\psi^*\right)- \frac{|\vec \nabla \psi|^2}{2m} + \mu |\psi|^2 \right. \right. \label{toy model Lagrangian} \\
	& \qquad \qquad \qquad \qquad \qquad \qquad \qquad  \left. \left.  -\frac{q}{m} (\psi^* \vec \nabla \psi + \psi \vec \nabla \psi^*) \cdot \vec \nabla \pi - 2\frac{q^2}{m} |\psi|^2 (\vec \nabla \pi)^2 \right] \right\} \ , \nonumber 
\end{align}
where $\psi$ is the non-relativistic DM field, and $\pi$ mediates DM self-interactions as well as DM-baryon interactions. (we will write down the coupling between $\pi$ and baryons in a moment.) The parameter $\varepsilon$ has units of energy density, $\mu$ is the DM chemical potential, $\gamma = \pm 1$ depending on whether $\pi$ mediates repulsive or attractive interactions, and $q$ has dimensions of $[L]^{1/2} [T]^{1/2}$.

A few comments are in order at this point. First, notice that the combination of terms involving $\psi$ within the bracket in the first line of Eq.~\eqref{toy model Lagrangian} is precisely the DM pressure $P$. This can be inferred from the fact that this combination is just the non-relativistic limit of the finite density relativistic Lagrangian for a massive complex scalar field. The latter can be easily shown to be equal to the pressure of the system by calculating the relativistic stress-energy tensor with standard methods. Second, the form of our non-relativistic Lagrangian in Eq.~\eqref{toy model Lagrangian} is dictated by the requirement that in vacuum ({\it i.e.}, when $\mu = 0$) it is invariant under the Schr\"odinger group~\cite{Hagen:1972pd}, which contains Galilean transformations as a subgroup. In particular, it is easy to check that Eq.~\eqref{toy model Lagrangian} is invariant under the following non-relativistic scale transformation:
\begin{subequations} \label{scale transf}
\begin{align}
	t &\to e^{2\lambda} t \,; \\
	\vec x &\to e^{\lambda} \vec x\,; \\
	\psi &\to  e^{-2\lambda} \psi \,;\label{toy scaling psi} \\
	\pi &\to \pi + \lambda / q \ . 
\end{align}
\end{subequations}
The field $\pi$ plays the role of a non-relativistic dilaton, realizing non-linearly both the scale transformations above as well as Galilean special conformal transformations---which mandate the terms in the second line of Eq.~\eqref{toy model Lagrangian}. It is known that the non-linearly realized Schr\"odinger symmetry forbids time derivatives in the kinetic term for $\pi$~\cite{Arav:2017plg}. This should not be a source of concern, however, since our model should be thought of as a non-relativistic effective theory which ultimately admits some relativistic UV completion. The Schr\"odinger invariance could be an accidental symmetry which emerges for small velocities but does not have an obvious relativistic counterpart---as is the case for a free non-relativistic field. Hence, the field $\pi$ should not necessarily be interpreted as the non-relativistic limit of some relativistic dilaton.  

At this point, the reader might be wondering whether there is any relation between the scaling transformation above and the scaling invariance  discussed in Sec. \ref{scaleinvarianceDM}. There, we showed that, in the DM-dominated regime, the combined systems of Poisson, Jeans and heat equations for DM are invariant under the anisotropic rescaling in Eq.~\eqref{scalingequations}. The rescaling of the coordinates we are considering here clearly corresponds to the particular case $z=2$. The scaling dimension of $\psi$ we are postulating in Eq.~\eqref{toy scaling psi} is such that $\rho = m |\psi|^2 \to e^{-4 \lambda} \rho$, which reproduces the transformation rule in Eq.~\eqref{scaling rho} with $z=2$.

It is also easy to understand why the term proportional to $\mu$ breaks the scale invariance in Eq.~\eqref{scale transf}. Adding a chemical potential to the Lagrangian via the replacement $\partial_t \to \partial_t - i \mu$ is just a trick for taking into account the fact that one is considering fluctuations around finite density states rather than the vacuum~\cite{Kapusta:1981aa}. Because these states are such that $ n = \langle |\psi|^2 \rangle \neq 0$, they {\it spontaneously} break the scale invariance. This spontaneous breaking translates into an {\it explicit} breaking once the chemical potential is added at the level of the Lagrangian, as in Eq.~\eqref{toy model Lagrangian}.

Finally, in the presence of a non-zero density of DM, we expect the combination of the first terms involving $\psi$ in the bracket, and corresponding to the DM pressure, to give a non-trivial contribution to the tree-level effective potential for $\pi$. The latter can be calculated by integrating out $\psi$ in a finite-temperature path integral, which can be done exactly since the Lagrangian is quadratic in $\psi$. To this end, we restrict ourselves to constant field configurations of $\pi$, in which case the effective potential is given by
\begin{equation}
	\exp \left[ - \frac{1}{\hbar}\int_0^{\hbar \beta} {\rm d} \tau\int {\rm d}^3 x \, V_{\rm eff} (\pi) \right] = \exp \left[ \frac{1}{\hbar} \int_0^{\hbar \beta} {\rm d} \tau\int {\rm d}^3 x \, \varepsilon \, e^{- 5 q\pi} \right] \int \mathcal{D} \psi \mathcal{D} \psi^* e^{- S_{\rm E}[\psi,\pi]/\hbar} \ , \label{effective potential}
\end{equation}
with 
\begin{equation}
	S_{\rm E} [\psi,\pi] = \int_0^{\hbar \beta} {\rm d} \tau\int {\rm d}^3 x \, e^{ q \pi} \left[ \frac{\hbar}{2} \left(\psi^*  \partial_\tau \psi - \psi\partial_\tau\psi^*\right) + \frac{\hbar^2}{2m}|\vec \nabla \psi|^2 - \mu |\psi|^2 \right] \ .
\end{equation}

We have reintroduced explicitly all factors of $\hbar$, since this will enable us to calculate the effective potential by borrowing standard results in statistical field theory. Notice in fact that a constant $\pi$ can always be eliminated from the path integral in Eq. \eqref{effective potential} by performing the rescaling
\begin{subequations}
\begin{align}
	\hbar &= e^{-q \pi/3} \hbar' \ ;\\
	\tau &= e^{-q \pi/3} \tau' \ ;\\
	\vec x &= e^{-q \pi/3} \vec x' \ .
\end{align}
\end{subequations}
Then, the path integral in Eq.~\eqref{effective potential} reduces to the grand-canonical partition function for a free, non-relativistic field with a rescaled Planck constant and rescaled coordinates. Using a standard result in statistical mechanics~\cite{huang1987statistical}, this partition function can be expressed in terms of volume, temperature and pressure as follows:
\begin{equation}
	\int \mathcal{D} \psi \mathcal{D} \psi^* e^{- S_{\rm E}[\psi,\pi]/\hbar}  = e^{\beta V' \bar P} = \exp \left( \frac{1}{\hbar'} \int_0^{\hbar' \beta} d\tau' {\rm d}^3 x' \, \bar P \right) = \exp \left( \frac{1}{\hbar} \int_0^{\hbar \beta} d\tau {\rm d}^3 x \,  \bar P \, e^{q \pi} \right) \ ,
\end{equation}
where we have used the fact that at equilibrium the pressure is constant and, in the last step, we have reverted to the original coordinates. By combining this result with Eq.~\eqref{effective potential}, we conclude that the effective potential  for $\pi$ is
\begin{equation}
	V_{\rm eff} (\pi) = - \varepsilon \, e^{- 5 q\pi} - \bar P \, e^{q \pi} \ . \label{effective potential 2}
\end{equation}
The extremum of this potential is located at a value $\pi_*$ such that
\begin{equation}
	e^{3 q \pi_*} = \sqrt{\frac{5 \varepsilon}{\bar P}} \ .
\end{equation}

One may worry about instabilities, given that this is actually a local maximum of the effective potential rather than a minimum. The (square of the) characteristic momentum scale $k_*$ associated with this instability can be derived by expanding the effective potential around $\pi =\pi_*$ and reading off the coefficient in front of the term quadratic in $\delta \pi$. The most conservative approach is to require that $k_*$ be outside the present cosmological horizon, which yields the constraint
\begin{equation}
	k_*^2 \equiv 30 \, \varepsilon q^2 e^{-2 q \pi_*} < H_0^2/c^2 \ . \label{long range}
\end{equation}
When this constraint is satisfied, baryon-DM and DM-DM interactions mediated by $\pi$ are still short-range because DM is derivatively coupled to $\pi$. 

In order to calculate the cross section for baryon-DM interactions, we will supplement the action given by Eq.~\eqref{toy model Lagrangian} with the following coupling between baryons and $\pi$:
\begin{equation} \label{toy model baryon interaction}
	\Delta S = \int {\rm d}t {\rm d}^3 x \, q_{\rm b} \pi |\psi_{\rm b}|^2 \ .
\end{equation}
With this coupling, $\pi$ mediates long-range interactions among baryons whenever the condition in Eq.~\eqref{long range} is satisfied. It is important to stress though that this fifth force will not play any role in what follows, and that we are emphatically {\it not} relying on any long-range modification of gravity. In fact, we will explicitly demand below that interactions among baryons mediated by $\pi$ be completely negligible compared to gravitational interactions---see Eq. \eqref{Newton toy model}. This can always be achieved if $q_{\rm b}$ is sufficiently small. Hence, even though the interaction in Eq.~\eqref{toy model baryon interaction} breaks explicitly the Schr\"odinger symmetry, the latter remains a very good approximate symmetry. Moreover, the smallness of $q_{\rm b}$  helps to ensure that the finite density of baryons yields a negligible correction to the effective potential for $\pi$ in Eq. \eqref{effective potential 2}.

The tree-level scattering amplitude for baryon-DM interactions is easily calculated to be
\begin{equation}
	i \mathcal{M} = i q_{\rm b} \times \frac{-i e^{3 q \pi_*}}{\vec Q^2} \times \frac{i q \vec Q^2}{3m} = \frac{iqq_{\rm b}}{3m} \sqrt{\frac{5 \varepsilon}{\bar P}} \ ,  \label{baryon-DM amplitude}
\end{equation}
where $\vec Q$ is the momentum exchanged, and we have simplified the DM vertex using the DM dispersion relation $\omega_k = k^2 /2 m - \mu$.  The corresponding scattering cross section reads
\begin{equation}
	\sigma_{\rm int} \simeq \frac{1}{4\pi} \left( \frac{2m_{\rm b} q q_{\rm b}}{3 m} \right)^2 \,\frac{5\varepsilon}{m n v^2} \ ,
\end{equation}
where we have used $\bar P = m n v^2$. By comparing this result with our master equation \eqref{master eq model building}, we can immediately identify
\begin{equation}
\varepsilon q^2 q_{\rm b}^2 \left(\frac{m_{\rm b}}{m} \right)^3 \simeq \frac{9 \pi C m_{\rm b} a_0}{10} \ .  \label{master eq toy model}
\end{equation}

Similarly, the amplitude for DM self interaction is equal to the one shown in Eq.~\eqref{baryon-DM amplitude} with the factor $i q_{\rm b}$ replaced by $i q \, \vec Q^2/(3m)$. The corresponding scattering cross section averaged over a thermal distribution of initial momenta is therefore
\begin{equation} 
	\langle \sigma \rangle = \frac{3 m^2}{2 \pi} \left( \frac{4qv}{3} \right)^4 \frac{5 \varepsilon}{m n v^2}\ .
\end{equation}
This means in particular that our requirement that $\langle \sigma \rangle / \sigma_{\rm int} \lesssim 1$ in order for DM to be in the Knudsen regime---see Eq. \eqref{self interaction Knudsen}---amounts to  
\begin{equation} \label{Knudsen toy model}
	q^2 \lesssim \frac{3}{128} \frac{1}{v^4} \frac{q_{\rm b}^2 m_{\rm b}^2}{m^4}	\ .
\end{equation}
Finally, as mentioned earlier, $\pi$ mediates also a conventional long-range force between baryons. By demanding that this additional force yields a correction to Newton's constant that is smaller than the current experimental error, $\left\vert\frac{\Delta G}{G}\right\vert \;\lsim\; 10^{-6}$~\cite{Rosi:2014kva,Jaffe:2016fsh}, we find
\begin{equation} \label{Newton toy model}
	 \frac{q_{\rm b}^4 \varepsilon}{\rho v^2} \lesssim \mathcal{O} (10^{-12}) \frac{m_{\rm b}^4 c^2}{M_{\rm Pl}^4} \ .
\end{equation}

Equations~\eqref{master eq toy model}, \eqref{Knudsen toy model} and \eqref{Newton toy model} are the main relations that constrain the parameters of our toy model. These can be combined to eliminate $q, q_{\rm b}$ and $\varepsilon$, and obtain a bound on the DM mass, which in natural units reads
\begin{equation} \label{toy bound on m}
	\left( \frac{m}{m_{\rm b}} \right)^7 \lesssim \frac{\mathcal{O}(10^{-12})}{v^2} \frac{\rho}{3 H_0^2 \mpl^2} \frac{m_{\rm b}}{\mpl} \frac{H_0}{\mpl} \sim 10^{-77} \ ,
\end{equation}
where the estimate in the last step was obtained by using the fiducial values $1/ v^2 \sim 10^8$ and $\frac{\rho}{3 H_0^2 \mpl^2} \sim 10^5$. Unfortunately, this inequality implies $m / m_{\rm b} \lesssim 10^{-11}$, which is incompatible with the mass hierarchy $m \gg m_{\rm b}$ we have been assuming all along. 
  
This mechanism seems rather to favor the opposite mass hierarchy, $m \ll m_{\rm b}$ ({\it i.e.}, the ``heating'' case). In this case the energy exchanged has a more complicated dependence on the velocity dispersion $v$, as discussed in Sec. \ref{setup}. This in turn affects the expression for the baryon-DM cross section we get from our master equation. However, we can still derive an order-of-magnitude bound on $m$ by using $\epsilon \simeq \mathcal{O}(1) m v^2$. Repeating the analysis above for the light DM case, we end up with a bound similar to the one in Eq. \eqref{toy bound on m} with the LHS replaced by $(m / m_{\rm b})^2$. This implies $m \lesssim 10^{-29}$ eV, which corresponds to a DM candidate with a Compton wavelength of $\mathcal{O}(\text{Mpc})$ much larger than the size of a galaxy. Taken at face value, such a light mass would be incompatible with the cosmological bound $m \gtrsim 10^{-26}$ eV that follows from the observed large scale structures (see, {\it e.g.},~\cite{Hui:2016ltb}). It is worth mentioning though that this bound is usually derived without taking into account baryon-DM interactions. Given the strength of the interactions required by our framework, this bound would probably need to be revisited.

Regardless, the particular model we have discussed here is clearly too simple to be phenomenologically viable. The main point of this Section, however, was to illustrate how the required density and velocity dependence could emerge naturally in a particle physics model. From this perspective, we see no fundamental reason that prevents a variation of the mechanism described here (or a completely different mechanism, for that matter) from being successfully implemented in a more realistic model. 

\section{Conclusions}

Of all the oft-discussed small-scale tensions with $\Lambda$CDM~\cite{Bullock:2017}, the MDAR (i.e., Milgrom's relation) stands out as arguably the richest phenomenologically, because it tightly connects the DM and baryon distributions in disk galaxies. In this paper we have proposed a novel mechanism for the origin of the MDAR through strong collisional DM-baryon interactions, rather than a consequence of modification of gravity
or finely-tuned feedback. The idea put forward here is that the observed particle DM profile in galaxies may naturally emerge as the {\it equilibrium} configuration resulting from DM-baryon interactions.

For simplicity we have focused on elastic collisions, though in principle our analysis can be generalized to inelastic interactions as well. In order for our scenario to work while being consistent
with direct detection constraints, our DM particles must be either very heavy ($m \gg m_{\rm b}$) or very light ($m\ll m_{\rm b}$), corresponding respectively to cooling and heating of DM by
baryons. We have focused here on the heavy DM/cooling case, both to streamline the presentation and because this case is conceptually simpler, as the average energy
exchanged is approximately constant over the velocity range of interest. In future work we plan to extend our analysis to the heating case.

Our approach has been thoroughly ``bottom-up". We used the MDAR as a guide towards determining, from kinetic theory arguments, what {\it kind} of particle physics interactions
are necessary. What we have discovered is that obtaining an equilibrium MDAR configuration in disk galaxies requires $i)$ a relaxation time of the order of the dynamical time; $ii)$ a DM-baryon cross section that is inversely proportional to the DM number density and velocity squared: $\sigma_{\rm int} \sim 1/nv^2$. Such a dependence of the cross section on density and velocity comes from the fact that the MDAR involves an acceleration constant, and that the product of cross section, number density and velocity squared has the desired dimensionality. The $1/v^2$ dependence is quite natural, and occurs for instance with Sommerfeld enhancement, as well as charge-dipole interactions. The $1/n$ dependence is less familiar and suggests environmental dependence of DM-baryon couplings. We believe in fact that our scenario opens up a new avenue for model building which is at the same time grounded in astrophysical observations and yet largely untapped. 

The $1/n$ dependence of $\sigma_{\rm int}$ is critical, and plays three independent roles:

\begin{itemize}

\item It implies that the heat source depends primarily on the baryon density, which is key in deriving the MDAR from the heat equation. 

\item It implies that the interaction rate of a baryon with DM targets, $\sim n \sigma_{\rm int} v$, is {\it independent of density}. This greatly weakens any constraint
from early universe cosmology, such as from the CMB and structure formation.

\item Remarkably, as shown in Sec.~\ref{sec: gas vs stars}, it implies that the contribution to the heating rate from interactions with gas and stars is identical.
This ensures that the MDAR applies equally well to star-dominated {\it vs} gas-dominated galaxies, {\it i.e.}, that only the total baryonic density plays a role, as required by observations.

\end{itemize}

While it is remarkable that the heating rate is independent of the clumpiness of baryons on small scales, this is less evident for the relaxation time being of the order of the dynamical time, which is a crucial assumption of the model. Here we coarse-grained the baryonic and DM distributions over a typical scale of $\sim 10$ pc, but at the microscopic level, and in the absence of additional pc-range forces, most DM particles would take too long to interact with any star. We pointed out in Sec.~\ref{sec: gas vs stars} that a pc-range force as well as DM self-interactions would help to naturally realize the desired relaxation time. We delay such a detailed investigation to future model-building efforts, and took here this assumption on the relaxation time as a core hypothesis for the viability of the model.

With the above assumptions, we argued heuristically in Sec.~\ref{sec:rotational general} that the Poisson, Jeans and heat equations admit an equilibrium DM profile in rotationally-supported systems that parametrically matches the deep-MOND relation of Milgrom $g \simeq \sqrt{a_0g_{\rm b}}$ wherever DM dominates the gravitational field, and gives a sub-dominant, constant acceleration
$g_{\rm DM}\simeq a_0\ll g_{\rm b}$ wherever baryons dominate the gravitational field. This behavior matches that of the MOND simple interpolating function, which does in general an excellent job at fitting galaxy rotation curves~\cite{Famaey:2005fd,Gentile:2011}. This `intuitive' derivation, however, assumed spherical symmetry and ignored order-unity prefactors. 

In Sec.~\ref{exactnumerics} we refined the argument and solved numerically for the DM profile, still within the spherical approximation (but using a thermal conductivity appropriate for a flattened disk) and assuming a toy exponential profile for baryons. We found that solutions that are regular at the origin are described by two parameters ($R$ and $\hat{v}_0^2$), which physically characterize the central values of DM density and temperature. The freedom in specifying those two parameters thus implies that   the  BTFR cannot be uniquely predicted by our mechanism, at least within the equilibrium treatment. Nevertheless, what we have shown quantitatively is that, {\it once the BTFR is attained at large radii, the two parameters can be fixed and the MDAR at all radii necessarily follows from our formalism}, hence removing any need for fine-tuned feedback to explain the diversity of rotation curve shapes at a fixed mass scale and their uniformity at a fixed baryonic surface density scale. With these parameters thus set, the resulting DM profile is well-approximated by a cored isothermal sphere and matches well the MOND simple-interpolating prediction at all distances.

It remains to be seen whether the BTFR can also be predicted by our mechanism, once we go beyond equilibrium and study dynamical evolution of galaxy formation. Even if a dynamical analysis falls short of explaining the BTFR, it is worth emphasizing that already within $\Lambda$CDM  the slope and normalization of the BTFR can be reproduced, as they are a natural consequence of abundance matching. Only the scatter is too large, a problem that our DM-baryon interactions might be able to solve within simulations.

Our model makes a number of distinguishing predictions from standard MOND and/or $\Lambda$CDM cosmology:

\begin{enumerate}

\item While our mechanism reproduces the MOND/MDAR phenomenology in rotationally-supported systems, it distinguishes itself in pressure-supported systems, such as 
galaxy clusters and certain dwarf spheroidals, which are well-known to be problematic for standard MOND. These are DM-dominated systems, and yet do not
abide to deeply-MONDian dynamics. In our framework, this is naturally explained by the fact that such systems have not had time to relax to the equilibrium configuration. This is
universally the case for galaxy clusters, in particular. We instead predict that DM in galaxy clusters should assume an approximately collisionless profile, such as NFW, in
good agreement with observations.

The situation with dwarf spheroidals is more nuanced. Some dwarf spheroidals are expected to have already reached their equilibrium configuration, whereas others have not.
The necessary condition -- Eq.~\eqref{dwarfequi} -- for a dwarf spheroidal to have reached equilibrium must be checked on a case-by-case basis. Generally, we can say that a dwarf spheroidal with prototypical
velocity dispersion $v \sim 10 \, \text{km/s}$ and moderate mass-to-light ratio, $\rho\; \lsim\; 10 \,\rho_{\rm b}$, such as Fornax, should be relaxed by now and thus amenable to our quasi-equilibrium analysis. On the other hand, a dwarf galaxy with comparable velocity dispersion but much larger mass-to-light ratio, $\rho \;\gsim\; 10^2 \rho_{\rm b}$,
such as most ultra-faint dwarfs, should not have had enough time to equilibrate.

Elliptical galaxies are another important class of pressure-supported systems, which are dominated by baryons. Such objects are expected to have reached the equilibrium configuration. The constant, subdominant acceleration contribution from DM, $g_{\rm DM}\simeq a_0\ll g_{\rm b}$, is consistent with the recent analysis of Chae {\it et al.}~\cite{Chae:2017bhk}.

\item An essential aspect of MOND phenomenology is the external field effect (EFE). Consider a subsystem with subcritical internal acceleration, $g_{\rm int} \ll a_0$, in the presence of a large homogeneous external acceleration $g_{\rm ext}\gg a_0$. In General Relativity (GR), $g_{\rm ext}$ does not have physical consequences and can be removed by a coordinate transformation, in accordance with the equivalence principle. In MOND, however, $g_{\rm ext}$ is physical and renders the subsystem Newtonian. In our scenario, baryon-DM interactions are local and there is no EFE. The properties of a system that determine whether it is effectively `MONDian' (DM-dominated) or `Newtonian' (baryon-dominated) are all {\it local} quantities, such as the local DM density and velocity dispersion. 

\item In our scenario, like in $\Lambda$CDM, objects that are {\it a priori} expected to be devoid of DM, such as globular clusters (if formed in a DM-free environment) and tidal dwarf galaxies, should not exhibit a mass discrepancy. In the context of standard MOND, globular clusters that are sufficiently far from the Milky Way ({\it e.g.}, NGC~2419~\cite{Ibata:2011ri} and Pal~14~\cite{Jordi:2009aw}), such that the external field effect is negligible, are problematic. This is not the case in our scenario. 

\item As argued in Sec.~\ref{astroconstraints}, on large scales our scenario matches the $\Lambda$CDM expansion history and linear growth of perturbations. Indeed, because of our cross section scales as $1/nv^2$, DM-baryon interactions are suppressed in early universe, ensuring consistency with the CMB and structure formation. One potentially important difference with $\Lambda$CDM is the prediction made at the end of Sec.~\ref{sec:CMB & LSS constraints} that DM-baryon interactions should become significant around $z \sim 100$. This is well after recombination, so the CMB should be unaffected, but there might be interesting signatures for future 21-cm observations~\cite{Tashiro:2014tsa,Munoz:2015bca}.

\end{enumerate}

Our scenario must be further developed on a number of fronts. Some of the directions we plan to pursue in future works include:

\begin{itemize}

\item {\bf Model building:} An obvious priority is to develop a full-fledged example that realizes the desired $1/nv^2$ dependence while being
consistent with phenomenological constraints. For illustrative purposes, we constructed in Sec.~\ref{sec: particle physics models} an explicit particle physics model,
based on a non-relativistic (Schr\"odinger) scaling symmetry. The model generates the desired cross section $\sigma_{\rm int} \sim 1/nv^2$, with both velocity and density dependence generated environmentally at once. In its current guise, however, the model fails on phenomenological grounds, and for this reason should be considered no more than a proof of principle. 

An important handle in model building may lie in the scaling symmetry uncovered in Sec.~\ref{scaleinvarianceDM}. There we found that, in the DM-dominated regime, our equations
are invariant under the anisotropic space-time rescaling $\vec{x}\rightarrow e^{\lambda} \vec{x}$, $t\rightarrow e^{z \lambda} t$, for arbitrary value of $z$. More remarkably,
away from DM domination, our equations are invariant under $z = 1/2$ scaling transformations. It will be interesting to see if an explicit model with this scaling symmetry can be constructed. As already pointed out, realizing the core assumption on the relaxation time being of the same order as the dynamical time might also depend on model-building details such as additional pc-range forces between DM and baryons or DM self-interactions. Finally, one could also investigate whether there may be a dynamical explanation (for instance linked to $H_0$) for the origin of the acceleration scale $a_0$, which in our mechanism controls the strength of DM-baryon interactions.

\item {\bf Studying the light DM/heating case:} As shown in Sec.~\ref{setup}, the characteristic energy exchanged with baryons in the case of light DM cannot be treated as approximately constant in disk galaxies, as it depends strongly on the circular velocity of baryons which is sharply varying in the central regions. We thus plan to perform a similar analysis to the one conducted here, by carefully taking into account this subtlety in the heating case.

\item {\bf Equilibrium solution away from spherical symmetry:} For simplicity, much of our analytical and numerical derivations relied on spherical symmetry. It is imperative to generalize the analysis
to a realistic geometry of disk galaxies, still within the static/equilibrium approximation. A warm-up problem along these lines would be to consider a toy baryonic disk profile, such as the Kuzmin disk, whose gravitational potential is that of a point mass above and below the disk~\cite{BinneyTremaine,Brada:1994pk}.

\item {\bf Numerical simulations of galaxy formation:} It would be very instructive to extend the equilibrium analysis of this paper to a fully dynamical study of galaxy formation
in our scenario. Because our mechanism relies on standard gravity and simple particle interactions, it should be straightforward to modify existing hydrodynamical codes to include
a DM-baryon cross section with the appropriate density and velocity dependence, with a coarse-graining on a 10~pc scale. Such a study would allow us to check, among other things, $i)$ that our equilibrium solution is indeed stable; $ii)$ that the outskirts of galaxy disks are not too severely perturbed by the interactions with DM; $iii)$ that the clumping of DM in subhalos does not affect the results (we expect the smallest DM subhalos to be tidally destroyed close to the disk, and therefore our smooth assumption for the DM distribution to hold); $iv)$ that our equilibrium solution can be reached dynamically on the predicted time scale; $v)$ whether the BTFR is a natural outcome of the dynamical evolution as expected from our heuristic arguments, and if so, with what scatter; $vi)$ whether an evolution of the acceleration scale $a_0$ (for instance with $H_0$), which in our mechanism controls the strength of DM-baryon interactions, could lead to the desired equilibrium.

\end{itemize}

\paragraph{Acknowledgements:} We thank Yacine Ali-Ha\"imoud, Lasha Berezhiani, James Bullock, Kyu-Hyun Chae, Fran\c{c}oise Combes, Harry Desmond, Xavier Hernandez, Federico Lelli, Stacy McGaugh, Vinicius Miranda, Ira Rothstein, Paolo Salucci, Ravi Sheth and Glenn Starkman for helpful discussions. We are also very grateful to Fran\c{c}oise Combes, Harry Desmond and Paolo Salucci for helpful feedback on an early version of this manuscript. We would like to thank the anonymous referee for helpful suggested improvements. B.F. acknowledges the financial support from the ``Programme Investissements d'Avenir'' (PIA) of the IdEx from the Universit\'e de Strasbourg. J.K. is supported in part by NSF CAREER Award PHY-1145525, US Department of Energy (HEP) Award DE-SC0017804, NASA ATP grant NNX11AI95G, and the Charles E. Kaufman Foundation of the Pittsburgh Foundation. R.P. acknowledges the hospitality of the Sitka Sound Science Center during the final stages of this work, and is supported by the US Department of Energy (HEP) Award DE-SC0013528.

\renewcommand{\em}{}
\bibliographystyle{hieeetr}
\bibliography{heat-v20.bib}

\end{document}